\documentclass[11pt,letterpaper]{article}
\pdfoutput=1
\usepackage{jheppub}
\usepackage{color}
\usepackage{graphicx}
\usepackage{tabularx}
\usepackage{xspace}
\usepackage{empheq}
\usepackage{verbatim}
\usepackage{amsmath}
\usepackage{amssymb}
\usepackage[caption=false]{subfig}
\usepackage{url}
\usepackage{bbold}
\usepackage{slashed}
\usepackage{array}
\usepackage{aas_macros}

\usepackage{multirow}
\usepackage{threeparttable}
\usepackage{paralist}
\usepackage{xspace}
\usepackage{upgreek}
\usepackage{lipsum}
\usepackage[T1]{fontenc}
\usepackage{scalerel}

\usepackage[framemethod=TikZ]{mdframed}
\usepackage{setspace}
\usepackage{mathtools}

\allowdisplaybreaks

\newcommand{\be}{\begin{equation}}
\newcommand{\bea}{\begin{eqnarray}}
\newcommand{\ee}{\end{equation}}
\newcommand{\eea}{\end{eqnarray}}

\def\nn{{\nonumber}}

\DeclareMathAlphabet\mathbfcal{OMS}{cmsy}{b}{n}
\mathchardef\mhyphen="2D 

\newcommand{\aW}{\alpha_{\scriptscriptstyle W}}
\newcommand{\aNA}{\alpha_{\scriptscriptstyle \textrm{NA}}}
\newcommand{\TaW}{\tilde{\alpha}_{\scriptscriptstyle W}}
\newcommand{\mW}{m_{\scriptscriptstyle W}}
\newcommand{\mZ}{m_{\scriptscriptstyle Z}}
\newcommand{\gW}{g_{\scriptscriptstyle W}}
\newcommand{\thetaW}{\theta_{\scriptscriptstyle W}}
\newcommand{\sW}{s_{\scriptscriptstyle W}}
\newcommand{\cW}{c_{\scriptscriptstyle W}}

\newcommand{\mo}{\mathcal{O}}

\begin{document}

\preprint{\hbox{MIT-CTP/5615; CERN-TH-2023-168}}

\title{The Quintuplet Annihilation Spectrum}

\author[1]{\small Matthew Baumgart,}
\author[2]{\small Nicholas L. Rodd,}
\author[3]{\small Tracy R. Slatyer,}
\author[4]{\small and Varun Vaidya}

\affiliation[1]{\footnotesize Department of Physics, Arizona State University, Tempe, AZ 85287, USA}
\affiliation[2]{\footnotesize Theoretical Physics Department, CERN, 1 Esplanade des Particules, CH-1211 Geneva 23, Switzerland}
\affiliation[3]{\footnotesize Center for Theoretical Physics, Massachusetts Institute of Technology, Cambridge, MA 02139, USA}
\affiliation[4]{\footnotesize Department of Physics, University of South Dakota, Vermillion, SD 57069, USA}

\emailAdd{Matt.Baumgart@asu.edu}
\emailAdd{nrodd@cern.ch}
\emailAdd{tslatyer@mit.edu}
\emailAdd{Varun.Vaidya@usd.edu}

\abstract{We extend the Effective Field Theory of Heavy Dark Matter to arbitrary odd representations of SU(2) and incorporate the effects of bound states.
This formalism is then deployed to compute the gamma-ray spectrum for a $\mathbf{5}$ of SU(2): quintuplet dark matter.
Except at isolated values of the quintuplet mass, the bound state contribution to hard photons with energy near the dark-matter mass is at the level of a few percent compared to that from direct annihilation.
Further, compared to smaller representations, such as the triplet wino, the quintuplet can exhibit a strong variation in the shape of the spectrum as a function of mass.
Using our results, we forecast the fate of the thermal quintuplet, which has a mass of $\sim$13.6 TeV.
We find that existing H.E.S.S. data should be able to significantly test the scenario, however, the final word on this canonical model of minimal dark matter will likely be left to the Cherenkov Telescope Array (CTA).
}

\maketitle
\setcounter{page}{2}

\section{Introduction}

A fundamental difficulty in the search for the particle nature of dark matter (DM) is the enormous array of well motivated candidates that exist.
There are many ways to forge a path through the vast model space.
One guiding principle is a bottom-up notion of simplicity---a preference is given for the minimal modification to the Standard Model (SM) consistent with observations.
As originally emphasized in Ref.~\cite{Cirelli:2005uq}, with the mantra of minimality, few models are simpler than quintuplet DM.
Within this model, the SM is augmented by a single new field, a Majorana fermion transforming in the $\mathbf{5}$ or quintuplet representation of SU(2), and as a gauge singlet under SU(3)$\times$U(1).
This representation sits at a ``sweet spot.''
It is large enough that no additional symmetries are needed to make it cosmologically stable, with decay operators only appearing at dimension-6.
Simultaneously, it is small enough that the SU(2) Landau pole remains above the GUT scale.
After electroweak symmetry breaking, the five Majorana fermions of the quintuplet reorganize themselves into a neutral Majorana fermion, $\chi^0$ -- the DM candidate -- as well as singly and doubly-charged Dirac fermions, $\chi^+$ and $\chi^{++}$, which will play important roles in the phenomenology.
The coupling between the DM and the SM is fixed by measurements of the electroweak coupling $\aW$, leaving the DM mass, $M_{\chi}$, as the single free parameter in the model.
Assuming a conventional thermal origin of the DM, even the mass becomes fixed to $M_{\chi} = 13.6 \pm 0.8~{\rm TeV}$~\cite{Mitridate:2017izz,Bottaro:2021snn}.\footnote{This value includes the contribution of bound states to the relic abundance.
The value without these effects is closer to 9.6 TeV~\cite{Cirelli:2007xd}.
The value is, however, computed with the LO rather than NLO potentials of Refs.~\cite{Beneke:2019qaa, Urban:2021cdu} that we will make use of throughout, as described in Sec.~\ref{subsec:generators}.}
However, if the quintuplet represents only a fraction of the DM or was produced through a non-standard cosmology, a wider range of TeV-scale masses becomes possible.
A brief review of quintuplet DM and our conventions for it is provided in App.~\ref{app:QDM}.

The question of whether the quintuplet is the DM of our Universe can be probed through indirect detection searches for its annihilation signal, $\chi^0 \chi^0 \to \gamma + X$.
Here, $\gamma$ represents the experimentally detected hard photon with energy comparable to $M_{\chi}$, and $X$ represents additional undetected radiation.
In this context, we consider a hard photon to be one with energy near the mass of the DM, so $E_{\gamma} \sim M_{\chi}$, and these photons are of particular interest as they give a line-like signature in the photon spectrum at the multi-TeV scale, which is not expected from any plausible astrophysical backgrounds.\footnote{As discussed in Ref.~\cite{Aharonian:2012cs}, astrophysical emission processes will almost exclusively generate spectra that are broader than the energy resolution of instruments such as H.E.S.S. and CTA.
The possible exception explored in that work is inverse Compton emission deep in the Klein-Nishina regime.
In that case, where the photons acquire essentially the entire electron or positron momentum, the photon spectrum is determined solely by the initial electron and positron distribution.
Sufficiently narrow leptonic distributions to generate an apparent gamma-ray line could be produced by pulsar winds, but such a scenario is by no means generic, and the spatial distribution of such a signal would ultimately be distinct to that predicted for DM.}
There are reasons to be optimistic that the coming decade may bring with it the detection of such TeV photons from DM annihilation, primarily on account of the rapid progress in the experimental program of searching for TeV-PeV photons.
Building on the successes of DM searches with ongoing air and water Cherenkov telescopes such as HAWC~\cite{Sinnis:2004je,HAWC:2014ycj,HAWC:2017mfa,HAWC:2017udy,HAWC:2018eaa,HAWC:2019jvm}, H.E.S.S.~\cite{Hinton:2004eu,HESS:2006zwn,HESS:2007ora,HESS:2011zpk,HESS:2013rld,HESS:2014zqa,HESS:2015cda,HESS:2016mib,HESS:2016glm,Abdallah:2018qtu,HESS:2018kom,HESS:2022ygk}, VERITAS~\cite{Holder:2008ux,VERITAS:2012bez,VERITAS:2017tif,Acharyya:2023ptu}, MAGIC~\cite{Lorenz:2004ah,MAGIC:2011nta,MAGIC:2016xys,MAGIC:2017avy,MAGIC:2018tuz,MAGIC:2022acl}, and LHAASO~\cite{Bai:2019khm,LHAASO:2022yxw}, the next ten years should see both substantial advances in sensitivity at existing experiments (see {\it e.g.}~Refs.~\cite{Tak:2022vkb,Montanari:2022buj}), and new facilities such as the Cherenkov Telescope Array (CTA)~\cite{Consortium:2010bc,CTAConsortium:2012fwj,Acharya:2017ttl,CTA:2020qlo,Rinchiuso:2020skh} and Southern Wide-Field Gamma-Ray Observatory (SWGO)~\cite{swgo_instrument, swgo_science}.
For a recent review, see Ref.~\cite{Boddy:2022knd}.
The data these instruments will collect raises the exciting possibility that a signal from a heavy multi-TeV thermal DM candidate, such as the quintuplet, could be around the corner.

The central focus of the present work is to take this possibility seriously, and therefore derive a precise theoretical prediction for the photon spectrum from quintuplet annihilation.
Even though the quintuplet represents at most a single-parameter model characterized solely by its mass, there are a number of effects that must be carefully accounted for in order to produce an accurate spectrum.
Arguably the most important of these is the fact that as two $\chi^0$ approach one another, once they come within a distance $r \sim \mW^{-1}$, they can exchange virtual electroweak bosons, and thereby experience a potential that can significantly perturb their initial wavefunctions.
This effect is referred to as Sommerfeld enhancement, and can modify the expected annihilation rate by orders of magnitude~\cite{Hisano:2003ec,Hisano:2004ds,Cirelli:2007xd,ArkaniHamed:2008qn,Blum:2016nrz}.
Further, the problem involves several hierarchies of scale, which necessitates an effective field theory treatment.
The large hierarchy between the DM mass and the electroweak scale, $M_{\chi} \gg \mW$, manifests itself through large Sudakov double logarithms, $\aW \ln^2(M_{\chi}/\mW)$, which lead to a breakdown in naive perturbation theory.
Perturbative control can be restored by resumming these logarithms using the techniques of effective field theory (EFT), in particular one built using Soft-Collinear Effective Theory (SCET)~\cite{Bauer:2000ew,Bauer:2000yr,Bauer:2001ct,Bauer:2001yt}.
This solution has previously been implemented in heavy dark-matter models of neutralinos, including the wino and higgsino~\cite{Baumgart:2014vma,Bauer:2014ula,Ovanesyan:2014fwa,Baumgart:2014saa,Baumgart:2015bpa,Ovanesyan:2016vkk,Baumgart:2017nsr,Beneke:2018ssm,Baumgart:2018yed,Beneke:2019vhz,Beneke:2022eci,Beneke:2022pij}.
An additional source of large logs occurs due to our insistence on hard photons near the endpoint of the spectrum, with $E_{\gamma} \sim M_{\chi}$.
More carefully, we focus on photons with $z = E_{\gamma}/M_{\chi}$, such that $(1-z) \ll 1$.
This kinematic restriction gives rise to large terms of the form $\ln(1-z)$, which can also be resummed as shown in, for instance, Ref.~\cite{Baumgart:2017nsr}.
In summary, the effects of Sommerfeld enhancement, electroweak Sudakov logarithms, and endpoint contributions have been incorporated in a number of scenarios of neutralino DM.
For instance, Refs.~\cite{Baumgart:2017nsr,Baumgart:2018yed} developed an EFT framework for including all of these effects in the case of wino-like DM, where DM is a $\mathbf{3}$ or triplet of SU(2), allowing the spectrum to be computed to next-to-leading logarithmic (NLL) accuracy.
However, the power of EFT and factorization is that many aspects of those calculations should not depend on the specific DM representation, so that results for the wino -- an SU(2) triplet -- should lift to the quintuplet and in fact, as we will discuss, to any higher representation.
In Sec.~\ref{sec:DA} of the present work we will demonstrate this explicitly, and in particular compute the NLL quintuplet direct annihilation spectrum.

The quintuplet also represents an opportunity to extend this formalism to incorporate an additional (potentially) important source of photons: bound state formation and decay.\footnote{Sommerfeld enhancement and bound state contributions can be important whenever the DM experiences a long range interaction.
As another example where these play a role, Ref.~\cite{Biondini:2023ksj} considered a scenario where DM experiences a long range interaction in the dark sector.}
The starting point for these bound states is similar to that of the Sommerfeld enhancement to annihilation.
As there, once the initial $\chi^0$ pair comes within $r \sim \mW^{-1}$, they can exchange a $W$-boson and thereby convert into a $\chi^+ \chi^-$ pair.
This pair of charged particles can now emit a photon, capturing into a bound state, which can be thought of as the DM quintuplet analogue of positronium.
At higher DM masses, it is also possible to emit a $W$ boson and capture into a charged bound state, and further on-shell $Z$ boson emission becomes a significant contributor to the formation of neutral bound states.
All of these bound states are generally unstable and may decay either to lighter bound states in the spectrum, or directly to SM particles.
In principle the contribution of bound states could also be important for the wino, however, as shown in Ref.~\cite{Asadi:2016ybp} it is irrelevant for present-day indirect detection (albeit possibly relevant in the early Universe, see for example Ref.~\cite{Mitridate:2017izz}).
For the thermal wino, there is only one bound state present in the spectrum, and capture to it via emission of a dipole photon is forbidden by spin statistics. 
For heavier winos or wino-like particles (with non-thermal histories to prevent overclosure of the Universe), there is a non-zero rate for capture to a spectrum of bound states via dipole photon emission (and at sufficiently high masses, $W$ and $Z$ emission).
However, dimensionless numerical factors arising from the wavefunction overlap integral render these rates small compared to direct annihilation, even in the limit of very heavy DM where SU(2) can be treated as approximately unbroken.

For the quintuplet, it was already anticipated in Ref.~\cite{Asadi:2016ybp} that the higher preferred mass would allow for a more complicated spectrum of bound states, and the suppression from dimensionless factors should not be as severe.
Furthermore, previous studies of quintuplet DM have found that bound state formation plays a key role in setting the abundance of DM in the early Universe \cite{Mitridate:2017izz} and can be non-negligible in indirect searches for quintuplet annihilation in the Milky Way halo \cite{Mahbubani:2020knq}.
In the present work, we apply our SCET-based formalism to precisely compute contributions to the annihilation signal from bound state formation followed by decay.
The existence of bound states in the quintuplet spectrum roughly induces the following independent ({\it i.e.} non-interfering) contribution to the annihilation spectrum,
\bea
\frac{d\sigma}{dz} = \sum_B \sigma(\chi^0 \chi^0 \rightarrow  B+ X_{\text{us}} )\frac{1}{\Gamma_B}\frac{d\Gamma_{B\rightarrow \gamma+X}}{dz},
\label{eq:BSfact}
\eea
where the sum is over the set of bound states in the theory, $\sigma(\chi^0 \chi^0 \rightarrow  B+ X_{\textrm{us}})$ is the production cross section for the bound state which happens via the emission of an unobserved ultra-soft particle $X_{\textrm{us}}=\gamma,W,Z$, $\Gamma_B$ is the total decay width for the bound state, and $d\Gamma_{B\rightarrow \gamma+X}/dz$ is the differential decay rate to a measured photon carrying energy $zM_{\chi}$.
This result is essentially a consequence of the narrow width approximation; we will justify it in App.~\ref{app:UnstableET}.
Taking Eq.~\eqref{eq:BSfact} as given, the problem of introducing the bound state contribution is reduced to computing three ingredients.
The first of these is the capture cross section, $\sigma(\chi^0 \chi^0 \rightarrow  B+ X_{\text{us}} )$, or heuristically the probability for a given bound state to form.
We compute this using first-order perturbation theory in non-relativistic quantum mechanics, following Refs.~\cite{Asadi:2016ybp, Harz:2018csl}, with an extension to handle bound states produced by $W/Z$ emission instead of photon emission.
The second of these is determining the fully inclusive decay rate for the bound state, $\Gamma_B$.
To obtain this result we need both the decay rates for excited states to transition to lower-energy states (calculated using perturbation theory, similar to the initial capture process) and the rates for decay via annihilation into SM particles.
The final ingredient is to compute the photon spectrum of the bound state decay.
As emphasized above we are primarily interested in the spectrum of hard photons, with $E_{\gamma} \sim M_B/2$.
This implies that the photons are kinematically restricted to the endpoint region, and this will allow us to draw on the SCET-based machinery for calculating endpoint spectra developed in Refs.~\cite{Baumgart:2017nsr,Baumgart:2018yed}.
For this final step, we will also need to consider additional operators mediating the hard process that are not needed for the direct annihilation process.
The reason for this is that the decaying bound states may have different spin and angular momentum quantum numbers than those configurations that dominate the direct annihilation.

Our calculation of the aforementioned bound state effects is divided between two sections: in Sec.~\ref{sec:BSF} we compute  the formation rates, while in Sec.~\ref{sec:BSA} we determine the fate of each bound state, accounting for transitions between bound states and their ultimate decay to SM particles.
When these results are combined, we observe that the contribution of the bound state annihilation to the hard photon spectrum is only a few percent of that from direct annihilation at most masses.
The contribution from bound state annihilation is small for a number of reasons: firstly, bound-state formation from a $\chi^0 \chi^0$ initial state with $L=0$ requires capture into an excited state, which is generically suppressed by a wavefunction overlap factor, and the transition from the $L=0$ state to the lowest-lying $L=1$ states also has an accidentally small numerical prefactor in the cross section---for the wino, this coefficient is zero in the limit of unbroken SU(2).
If we instead consider $L>0$ initial states, these contributions are velocity suppressed due to the small non-relativistic speed of DM in our Milky Way Galaxy.
In addition, odd-$L$ initial states must have $S=1$ to ensure the asymmetry of the $\chi^0\chi^0$ wavefunction.
They thus give rise to $S=1$ bound states (ignoring spin-flip transitions, which are suppressed); we find that such bound states have power-suppressed contributions to the endpoint photon spectrum.
All these effects are discussed in App.~\ref{app:analytic}, where we also show that in the limit of high DM mass and large representation size, the ratio of bound-state formation to direct annihilation is expected to decrease further for larger representations (in the context of indirect detection of hard gamma rays).

In Sec.~\ref{sec:Numerics} we combine the pieces to obtain the full quintuplet endpoint spectrum and annihilation cross section as a function of DM mass.
As for the wino, the annihilation cross section exhibits a rich structure and rapid variation associated with near-zero energy bound states, characteristic of the Sommerfeld enhancement.
Unlike the wino, however, we also see a strong variation in the shape of the spectrum itself: the energy distributions of photons resulting from a quintuplet annihilation can depend sensitively on its exact mass.
Using these results, we estimate the sensitivity of existing H.E.S.S. data to the thermal quintuplet, finding that for commonly adopted DM profiles in the Milky Way, the signal should already be either observable or in tension.
If we adopt a more conservative DM profile, however, our estimate is that the final word on the quintuplet will await the data that will soon be collected by CTA.
Finally, our conclusions are presented in Sec.~\ref{sec:conclusions}, with several extended discussions and details relegated to appendices.

\section{Direct Annihilation}
\label{sec:DA}

\begin{figure*}[!t]
\centering
\includegraphics[width=0.49\textwidth]{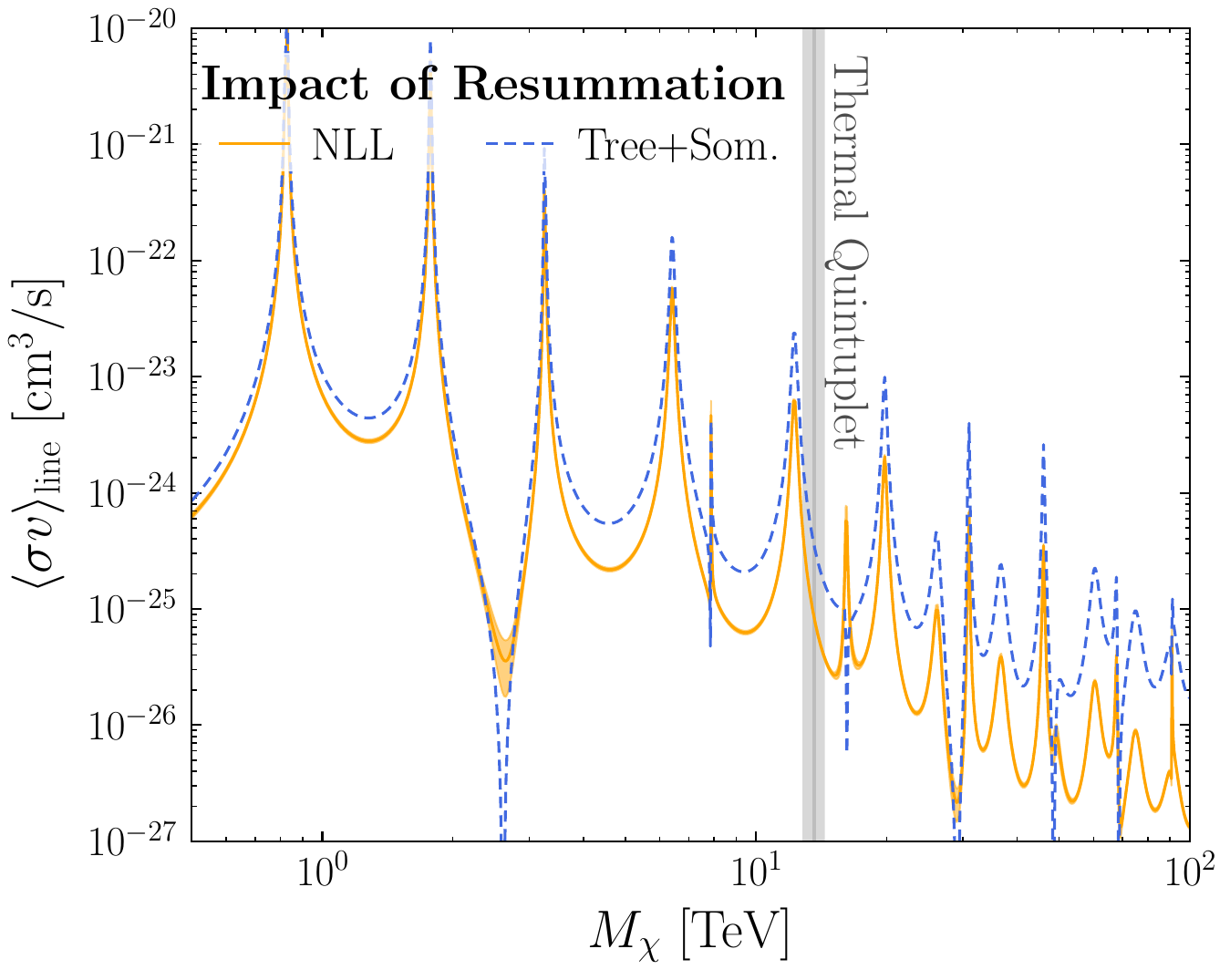}\hspace{0.1cm}
\includegraphics[width=0.49\textwidth]{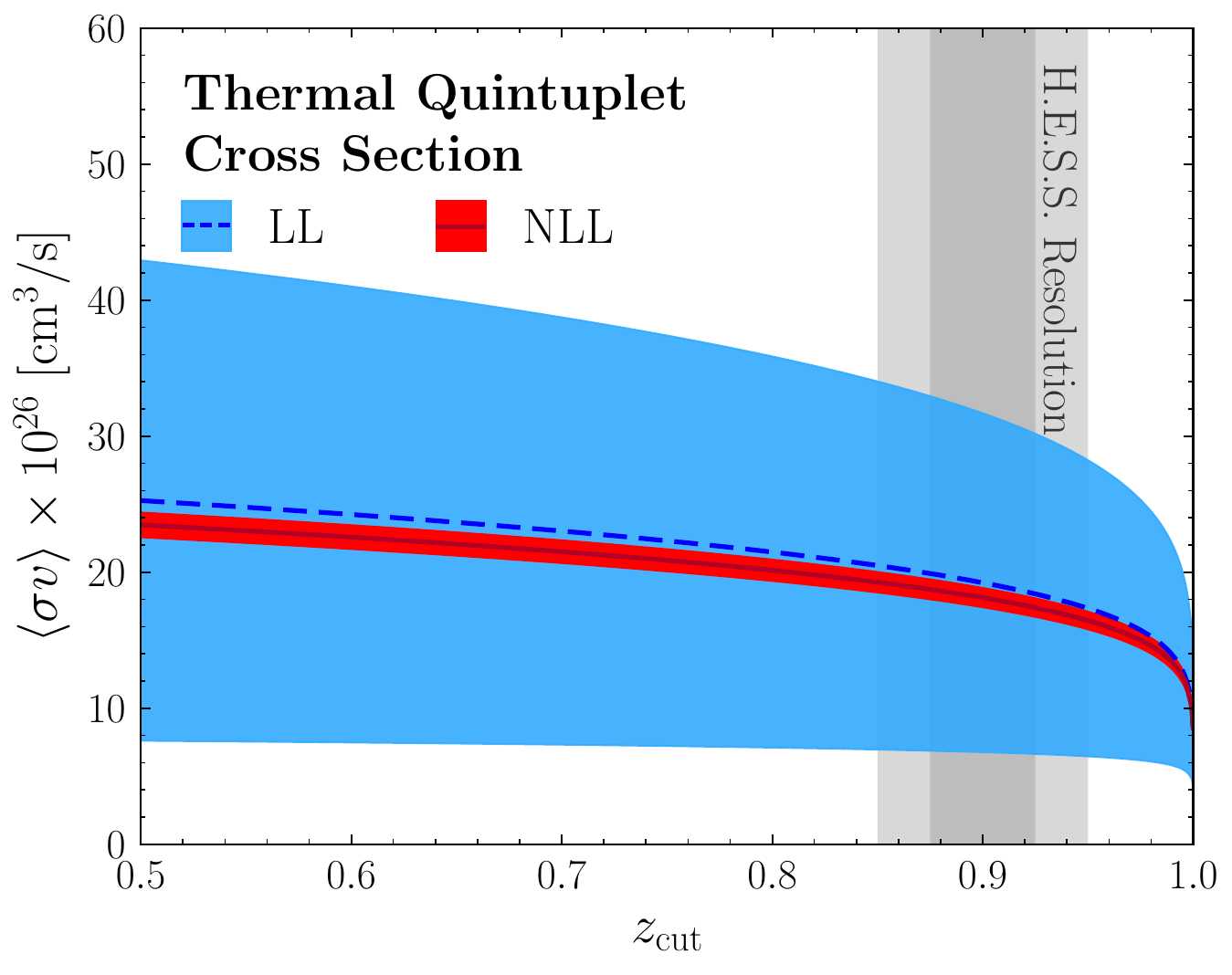} 
\caption{A demonstration of the importance of performing the NLL resummation of the direct quintuplet annnihilation cross-section.
(Left) A comparison of the line cross-section between the NLL resummed result and that obtained using only the tree-level calculation combined with the Sommerfeld enhancement.
At the thermal mass, $M_{\chi} = 13.6$ TeV, the resummed cross-section is more than a factor of four smaller.
(Right) The cross section for the thermal quintuplet integrated over $z \in [z_{\rm cut},1]$.
A significant reduction in the theoretical uncertainty is obtained for the NLL result.
Further, we see an increase in the cross section in the range appropriate for the H.E.S.S. energy resolution, which arises due to the presence of endpoint photons.
(The right figure can be directly contrasted to that of the wino, presented in Ref.~\cite{Baumgart:2018yed}.)
Both figures consider solely the contribution from direct annihilation; incorporating the bound state contributions will be a major focus of this work.
}
\label{fig:Thermal-Spectrum}
\end{figure*}

While bound states represent a novel addition to the spectrum for the case of the quintuplet, direct annihilation via $\chi^0 \chi^0 \to \gamma + X$ remains the dominant contribution to the hard photon spectrum for most masses, and we will compute it in this section.
To do so we will draw on the EFT formalism developed to compute the leading log (LL) spectrum in Ref.~\cite{Baumgart:2017nsr}, and then extended to NLL in Ref.~\cite{Baumgart:2018yed}.
The formalism there was applied to the wino, yet as emphasized in those references the approach can be readily extended to other DM candidates, especially to cases where the DM is simply charged under SU(2).
This section represents an explicit demonstration of that claim.
We begin in Sec.~\ref{eq:revHDMA} by briefly reviewing the framework developed in Refs.~\cite{Baumgart:2017nsr,Baumgart:2018yed}.
In doing so, we will focus on the aspects of the formalism that we will generalize to make it clear how to extend the calculation to additional SU(2) representations of DM, and we defer to those references for a complete discussion of all the relevant ingredients.
Having done this, in Sec.~\ref{eq:extquint} we will then demonstrate explicitly how the calculation can be extended to the quintuplet, and provide results for the LL and NLL spectrum.

Before moving into the details, however, let us already demonstrate the importance of performing the NLL resummation, and the precision obtained by doing so.
In the left of Fig.~\ref{fig:Thermal-Spectrum} we show a comparison of the line cross section computed to NLL accuracy, with the associated uncertainties, compared to the result if we had only performed a tree level computation of the rates, augmented with the Sommerfeld enhancement determined from the NLO potentials of Ref.~\cite{Beneke:2020vff} (discussed further in Sec.~\ref{subsec:generators}).
Note, the line cross section is the annihilation rate to a two-photon final state, specifically the rate for $\chi^0 \chi^0 \to \gamma \gamma$ + half the rate for $\chi^0\chi^0 \rightarrow \gamma Z$, with the photons having an energy $E_{\gamma} = M_{\chi}$.
We see that at larger masses the difference between the two methods can be significant.
The NLL result is already a factor of four smaller than the tree level approximation at the thermal mass, and by 100 TeV the difference is more than an order of magnitude.
The EFT formalism also incorporates endpoint photons (described in detail below) that have $E_{\gamma} \lesssim M_{\chi}$, which, given the finite energy resolution of IACTs like H.E.S.S., are indistinguishable from the line.
This is shown on the right of Fig.~\ref{fig:Thermal-Spectrum}, where we plot the integrated cumulative cross section down to a given $z_{\rm cut}$, which is almost a factor of two larger within the H.E.S.S. resolution as compared to the value for $z_{\rm cut} = 1$.
In addition, this figure demonstrates the considerable reduction in theoretical uncertainty of the cross section obtained at NLL.

\subsection{Review: An EFT for the endpoint spectrum from DM annihilation}
\label{eq:revHDMA}

As described in the introduction, a precise prediction of the hard photon spectrum arising from the annihilation of heavy DM charged under SU(2) mandates an accounting of a number of physical effects.
The benefit of approaching this problem using the EFT formalism reviewed in this section is that the various effects will factorize; heuristically, we will be able to separate the physics associated with different scales into objects that can be calculated independently.

To reiterate, the calculation we are interested in is the spectrum, $d \sigma/dz$, of hard photons resulting from DM annihilations, where ``hard'' means that we are interested in photons carrying away large energy fractions $z = E_{\gamma}/M_{\chi} \sim 1$.
The starting point is two incoming neutral DM particles, $\chi^0$, which are asymptotically momentum eigenstates described by plane waves.
These states, initially with momenta $p_{1,2} \sim M_{\chi} v \sim 10^{-3} M_{\chi}$ (in the Milky Way's halo) will eventually be within a distance $r \sim \mW^{-1}$, at which point they will experience an interaction potential that perturbs their wave-functions away from plane waves.
The perturbed wave-functions can yield a significantly enhanced probability for the particles to have a separation $r \sim M_{\chi}^{-1}$, where the hard annihilation occurs, and thereby provide a large boost to the cross section.
Restricting our attention to the case where $M_{\chi} \gg \mW$, the Sommerfeld enhancement occurs on a parametrically larger distance scale than the annihilation, and so we can factorize it out from the cross section as follows,\footnote{For a more detailed discussion of this factorization, we refer to Refs.~\cite{Bauer:2014ula,Baumgart:2017nsr,Beneke:2020vff}.}
\begin{equation}
\frac{d\sigma}{dz} = \sum_{a'b'ab} F^{a'b'ab}\frac{d\hat{\sigma}^{a'b'ab}}{dz}.
\label{eq:winosep}
\end{equation}
Here $a'b'ab$ are adjoint SU(2) indices, and for the wino we can write
\begin{equation}
F^{a'b'ab} = \Big\langle (\chi^0 \chi^0)_S \Big\vert (\chi_v^{a'T} i \sigma_2 \chi_v^{b'})^{\dagger} \Big\vert 0 \Big\rangle \Big\langle 0 \Big\vert (\chi^{aT}_v i \sigma_2 \chi_v^b) \Big\vert (\chi^0 \chi^0)_S \Big\rangle.
\label{eq:SommerfeldWino}
\end{equation}
Here $\chi_v$ is the field describing the non-relativistic DM, and the label $v$ appears just as for a heavy quark in Heavy Quark Effective Theory (HQET).
The non-relativistic DM effective theory that governs the dynamics of the field is reviewed in Ref.~\cite{Baumgart:2017nsr}, and as shown there the above expressions for $F^{a'b'ab}$ can be directly related to conventional Sommerfeld enhancement factors, as we now review.
In the broken phase, the triplet wino is described by a neutral Majorana fermion $\chi^0$, and a heavier charged Dirac fermion $\chi^{\pm}$.
DM annihilations proceed through the neutral states, and to ensure the antisymmetry of the state, at lowest order in the DM velocity ($s$-wave), the initial state must be a spin singlet, which we represent through the notation $(\chi^0 \chi^0)_S$.
From this initial state, Eq.~\eqref{eq:SommerfeldWino} describes the fact that through the exchange of electroweak bosons, not only will the incident wave-functions be perturbed, but further there are two final states that the system could evolve into by the time the hard process is initiated, $\chi^0 \chi^0$ or $\chi^+ \chi^-$.
In Ref.~\cite{Baumgart:2017nsr} these matrix elements were determined as,
\begin{equation}\begin{aligned}
\Big\langle 0 \Big\vert (\chi^{0T}_v i \sigma_2 \chi_v^0) \Big\vert (\chi^0 \chi^0)_S \Big\rangle
=\, &4 \sqrt{2} M_{\chi} s_{00}, \\
\Big\langle 0 \Big\vert (\chi^{+T}_v i \sigma_2 \chi_v^-) \Big\vert (\chi^0 \chi^0)_S \Big\rangle
=\, &4 M_{\chi} s_{0\pm}.
\label{eq:s00s0pm-wino}
\end{aligned}\end{equation}
Here $s_{00}$ and $s_{0\pm}$ are the Sommerfeld factors that need to be computed, and note that if the Sommerfeld effect were neglected, they would take the values $s_{00}=1$ and $s_{0\pm}=0$.
We emphasize again that the above expressions only hold for the wino; for the quintuplet, we would also need to account for the presence of the doubly-charged states.

To NLL accuracy, with the Sommerfeld enhancement stripped at the first stage of the matching, the differential cross section with factorized dynamics in SCET is given in terms of the hard function $H$, the jet functions $J^{'}_{\bar n}$, $J_{\gamma}$, and the soft function $S'$~\cite{Baumgart:2017nsr,Baumgart:2018yed}
\bea
\left(\frac{d\hat \sigma}{dz}\right)^{\text{NLL}}= H_{ij}(M_{\chi})J_{\gamma}(\mW)J^{'}_{\bar n}(M_{\chi},1-z, \mW) \otimes S^{'}_{ij}(1-z,\mW)
\label{eq:stage1}
\eea
 The cross section can then be refactorized into a combination of the following seven factors~\cite{Baumgart:2017nsr,Baumgart:2018yed}
\begin{equation}\begin{aligned}
\left(\frac{d \hat \sigma}{d z}\right)^{\rm NLL} = 
\,&H(M_{\chi},\mu)\, J_{\gamma}(\mW,\mu,\nu)\, J_{\bar n}(\mW,\mu,\nu)\, S(\mW,\mu,\nu) \\
&\times H_{J_{\bar n}}(M_{\chi},1-z,\mu)\otimes H_S(M_{\chi},1-z,\mu)\otimes C_S(M_{\chi},1-z,\mW,\mu,\nu),
\label{eq:Fact}
\end{aligned}\end{equation}
where we have suppressed the color indices on the hard scattering cross section in Eq.~\eqref{eq:winosep}.
Let us briefly provide some intuition for this expression.
The DM annihilation will occur an astrophysical distance from our telescope, and therefore no matter how complex the final state is, we should only expect to see a single photon from the decay.
This implies we are sensitive to the annihilation $\chi^0 \chi^0 \to \gamma + X$, where we must be inclusive over the unobserved $X$.
Although unobserved, $X$ cannot in fact be completely arbitrary.
Our choice to search for photons with energy $E_{\gamma} = z M_{\chi}$, with $(1-z) \ll 1$, implies that the invariant mass of the recoiling states is constrained to be small: $m_{\scriptscriptstyle X} = 2 M_{\chi} \sqrt{1-z} \ll M_{\chi}$.
This implies that the spray of radiation the photon recoils against must be a jet.
With this picture in mind, we can apply a conventional SCET factorization to our problem, breaking it into a function for the hard scattering ($H$), the collinear radiation in the direction of the photon ($J_{\gamma}$) and the recoiling jet ($J_{\bar{n}}'$), and finally the soft wide angle radiation ($S'$).
As shown in Ref.~\cite{Baumgart:2017nsr}, this factorization can be achieved, but it is insufficient to fully separate the scales that appear when accounting for the finite masses of the electroweak bosons.
For this, one must further factorize $J_{\bar{n}}'$ into the two functions $J_{\bar{n}}$ and $H_{J_{\bar{n}}}$, and $S'$ into the three functions $S$, $H_S$, and $C_S$.
The full details of this argument, together with the field theoretic definition of each function, is provided in Ref.~\cite{Baumgart:2017nsr}, with Ref.~\cite{Baumgart:2018yed} demonstrating that the factorization remains valid even when computing to NLL accuracy.
The central utility to Eq.~\eqref{eq:Fact} is that each of the functions can be computed separately -- and independently of the DM representation -- and then brought to a common scale using renormalization group evolution.
This facilitates a full result which resums logarithms of $\mW/M_{\chi}$, but also of $(1-z)$, which can be large given that we are searching for photons near the endpoint with $z \sim 1$.

The above factorization is perfectly sufficient for the wino.  However, the fact that the DM matrix elements in Eq.~\eqref{eq:SommerfeldWino} index the DM with an adjoint SU(2) label demonstrates that this form can only be appropriate for the wino, and further implies that $d\hat{\sigma}/dz$ is not representation independent.
We will now generalize this.
To do so, we must revisit the matching of the full theory to our EFT, as this is where the DM representation enters the calculation.
Doing so, it is straightforward to show that the tree-level matching can be achieved by way of a single hard scattering operator, given by
\begin{equation}
{\cal O} = \left( \chi_v^T i \sigma_2 \left\{ T_{\chi}^d,\,T_{\chi}^c \right\} \chi_v \right) \left( {\cal B}_{\perp n}^{ic} {\cal B}_{\perp \bar{n}}^{jd} \right) i \epsilon^{ijk} (n-\bar{n})^k, 
\label{eq:hardscatteringoperator}
\end{equation}
with Wilson coefficient,
\begin{equation}
C(\mu) = - \pi \frac{\aW(\mu)}{2M_{\chi}}.
\end{equation}
In the above result $T_{\chi}$ are the generators of SU(2), and therefore $c,d$ correspond to adjoint indices.
However, $T_{\chi}$ is written in whatever representation is appropriate for DM; a review of the relevant form for the generators in the triplet and quintuplet representations is given in App.~\ref{app:QDM}.

Equation~\eqref{eq:hardscatteringoperator} provides the hard operator before a BPS (Bauer-Pirjol-Stewart) field redefinition~\cite{Bauer:2001yt}.
This transformation must be performed in order to factorize the interactions of the heavy DM from the ultrasoft radiation.
Accordingly, we now perform a field redefinition,
\begin{equation}
\chi_v \to S_v \chi_v,\hspace{0.5cm} {\cal B}_{\perp n}^a \to Y_n^{aa'} {\cal B}^{a'}_{\perp n},
\end{equation}
where $S_v$ and $Y_n^{aa'}$ are both ultrasoft Wilson lines, but the former is in the $v$ direction and the same representation as DM, whereas the latter is in the $n$ direction and adjoint representation.
The operator then transforms as
\begin{equation}\begin{aligned}
{\cal O} \to &\left( \chi_v^T i \sigma_2 S_v^{\dagger} \left\{ T_{\chi}^{d'},\,T_{\chi}^{c'} \right\} S_v \chi_v \right) \left( Y_n^{c'c} {\cal B}_{\perp n}^{ic} Y_{\bar{n}}^{d'd} {\cal B}_{\perp \bar{n}}^{jd} \right) i \epsilon^{ijk} (n-\bar{n})^k \\
= &\left( \chi_v^T i \sigma_2 \left\{ T_{\chi}^a,\,T_{\chi}^b \right\} \chi_v \right) \left( Y^{abcd} {\cal B}_{\perp n}^{ic} {\cal B}_{\perp \bar{n}}^{jd} \right) i \epsilon^{ijk} (n-\bar{n})^k,
\label{eq:1hardop}
\end{aligned}\end{equation}
where in the final line we have defined,\footnote{We note that in order to reproduce the operator definitions in Ref.~\cite{Ovanesyan:2014fwa} and the works that followed it, Eq.~\eqref{eq:Ydef} would read $Y^{abcd} = (Y_v^{ae} Y_n^{ce}) (Y_v^{bf} Y_{\bar{n}}^{df})$, where all indices have been transposed.
We believe the index ordering in that work simply had a typo, and note that to the order all wino calculations have been performed so far, flipping these indices would not impact the results.}
\begin{equation}
Y^{abcd} = (Y_v^{ea} Y_n^{ec}) (Y_v^{fb} Y_{\bar{n}}^{fd}).
\label{eq:Ydef}
\end{equation}
To arrive at this result, we made use of the identity $S_v^{\dagger} T_{\chi}^a S_v = Y_v^{aa'} T_{\chi}^{a'}$, which we demonstrate in App.~\ref{app:WilsonLineProof}.
To be explicit, the identity has allowed us to replace a pair of $S_v$ Wilson lines, which are in the DM representation, with a single adjoint $Y_v$ Wilson line.

For the wino, the anticommutator can be readily evaluated,
\begin{equation}
\chi_v^T i \sigma_2 \left\{ T_{\chi}^a,\,T_{\chi}^b \right\} \chi_v = 
\chi_v^{a'T} i \sigma_2 \left(
2 \delta^{a'b'} \delta^{ab} - \delta^{a'a} \delta^{b'b} - \delta^{a'b} \delta^{b'a} \right) \chi_v^{b'}.
\end{equation}
Substituting this into Eq.~\eqref{eq:1hardop}, and using $Y_v^{ea} Y_v^{fa} = Y_v^{ea} (Y_v^{\dagger})^{af} = \delta^{ef}$, we obtain two separate operators,
\begin{equation}\begin{aligned}
{\cal O}_1 &= \left( \chi_v^{aT} i \sigma_2 \chi_v^{b} \right) \left( \left[ \delta^{ab} Y_n^{ec}Y_{\bar{n}}^{ed} \right] {\cal B}_{\perp n}^{ic} {\cal B}_{\perp \bar{n}}^{jd} \right) i \epsilon^{ijk} (n-\bar{n})^k, \\
{\cal O}_2 &= \left( \chi_v^{aT} i \sigma_2 \chi_v^{b} \right) \left( Y^{abcd} {\cal B}_{\perp n}^{ic} {\cal B}_{\perp \bar{n}}^{jd} \right) i \epsilon^{ijk} (n-\bar{n})^k,
\label{eq:O1O2oldbasis}
\end{aligned}\end{equation}
with Wilson coefficients
\begin{equation}
C_1(\mu) = - C_2(\mu) = - \pi \frac{\aW(\mu)}{M_{\chi}},
\label{eq:C1C2oldbasis}
\end{equation}
exactly matching those determined in Ref.~\cite{Ovanesyan:2014fwa}.
Nevertheless, in order to fully factorize the DM representation, we should keep the anticommutator unexpanded.
In particular, by so doing only the combination $\chi_v^T i \sigma_2 \left\{ T_{\chi}^a,\,T_{\chi}^b \right\} \chi_v$ retains any knowledge of the DM representation.
We can then introduce a modified definition of Eqs.~\eqref{eq:winosep} and \eqref{eq:SommerfeldWino} which achieves a complete separation of the DM representation from the factorized expressions in Eq.~\eqref{eq:Fact}.
In detail, we write
\begin{equation}\begin{aligned}
\frac{d\sigma}{dz} &= \sum_{a'b'ab}F_{\chi}^{a'b'ab}\frac{d\hat{\sigma}^{a'b'ab}}{dz}, \\
F_{\chi}^{a'b'ab} &= \Big\langle (\chi^0 \chi^0)_S \Big\vert \left(\chi_v^{T} i \sigma_2 \left\{ T_{\chi}^{a'},\,T_{\chi}^{b'} \right\} \chi_v\right)^{\dagger} \Big\vert 0 \Big\rangle \Big\langle 0 \Big\vert \left(\chi^{T}_v i \sigma_2 \left\{ T_{\chi}^a,\,T_{\chi}^b \right\} \chi_v \right) \Big\vert (\chi^0 \chi^0)_S \Big\rangle.
\label{eq:Som}
\end{aligned}\end{equation}
With this factorization we move the DM dependence into $F_{\chi}$, and the remaining factors in Eq.~\eqref{eq:1hardop} determine the hard matching onto the SCET calculation of $d\hat{\sigma}/dz$.
Importantly, when the DM factors are stripped, what remains in ${\cal O}$ is $\big( Y^{abcd} {\cal B}_{\perp n}^{ic} {\cal B}_{\perp \bar{n}}^{jd} \big) i \epsilon^{ijk} (n-\bar{n})^k$, which exactly matches the DM stripped contribution in ${\cal O}_2$.
This implies that if we intend to compute the cross section for the wino using Eq.~\eqref{eq:Som}, we need to make two minor changes to the approach used to compute it with Eqs.~\eqref{eq:winosep} and \eqref{eq:SommerfeldWino}.
Firstly, we must determine the new DM matrix elements $F_{\chi}^{a'b'ab}$ in terms of the Sommerfeld factors $s_{00}$ and $s_{0\pm}$, specified in Eq.~\eqref{eq:s00s0pm-wino}, and then evaluate the contraction of $F_{\chi}$ into $d\hat{\sigma}/dz$.
Secondly, in the SCET calculation, we match onto the hard function with $C_1(\mu) = 0$ and $C_2(\mu) = - \pi \aW(\mu)/2M_{\chi}$ as opposed to the values in Eq.~\eqref{eq:C1C2oldbasis}.
Of course, there was nothing fortuitous in the fact that the DM representation can be factored out so simply, this is simply a manifestation of the power of the EFT approach developed in Refs.~\cite{Baumgart:2017nsr,Baumgart:2018yed}.
The factorization in Eq.~\eqref{eq:Fact} is determined by the relevant degrees of freedom in the theory at scales below $M_{\chi}$, and this is not altered by changing the DM representation, so the factorization remains unaffected.

To make this point explicit, let us demonstrate that at LL these two approaches yield the same result for the wino; a straightforward generalization of the argument below confirms this conclusion persists at NLL.
As outlined above, the calculation in Ref.~\cite{Baumgart:2017nsr} is modified in two ways: a new set of DM matrix elements is computed in $F_{\chi}$, and then in the SCET calculation an alternative matching is provided onto the hard function, $H$.
After the BPS field redefinition, the DM representation contracts into the ultrasoft Wilson lines, and therefore in the SCET calculation is contracted into the soft function, $\tilde{S}$.
(Here, $\tilde{S} = C_S S$ is the combination of the collinear-soft and soft functions from Eq.~\eqref{eq:Fact}.)
In full, what we will need to recompute is the combination $H H_S \tilde{S}$, as the first and last of these factors is modified, and although $H_S$ remains unchanged, it connects these two objects.

Let us begin by reviewing the relevant part of the calculation as it appeared in Ref.~\cite{Baumgart:2017nsr}.
Accounting for the running of the hard function between $2M_{\chi}$ and $\mW$, we have
\begin{equation}
H_i(\mW) = U_H H_i^{\rm tree}, \hspace{0.5cm}
U_H = \exp \left( - 8 C_A \TaW \ln^2 \left( \frac{\mW}{2M_{\chi}} \right) \right)\!,
\end{equation}
where the renormalization group evolution is encoded in $U_H$, which is defined in terms of $\TaW = \aW/4\pi$ and $C_A=2$ the SU(2) adjoint Casimir.
The tree level matching coefficients $H_i^{\rm tree}$, are determined by the operator Wilson coefficients $C_1$ and $C_2$, in particular $H_1^{\rm tree} = C_1^* C_1$, $H_2^{\rm tree} = C_2^* C_2$, and $H_3^{\rm tree} = C_1^* C_2 = C_2^* C_1$.
Given the two different structures of the ultrasoft Wilson lines in Eq.~\eqref{eq:O1O2oldbasis}, there are four soft functions at the amplitude square level that evaluate to,
\begin{align}
\begin{array}{ll}
{\tilde S}_1^{aba'b'} = \delta^{ab} \delta^{a'b'},\quad\quad
&
{\tilde S}_2^{aba'b'} = \delta^{a3} \delta^{a' 3} \delta^{bb'}, \\[5pt]
{\tilde S}_3^{aba'b'} = \delta^{a3} \delta^{b3} \delta^{a'b'} + \delta^{a'3} \delta^{b'3} \delta^{ab},\quad\quad
&
{\tilde S}_4^{aba'b'} = \delta^{aa'} \delta^{bb'}.
\end{array}
\label{eq:EFTsoft}
\end{align}
These must then be contracted into $F^{a'b'ab}$ as defined in Eq.~\eqref{eq:SommerfeldWino}, which using Eq.~\eqref{eq:s00s0pm-wino} can be evaluated in terms of $s_{00}$ and $s_{0\pm}$, with explicit expressions provided in Ref.~\cite{Baumgart:2017nsr}.
The renormalization group evolution of the soft function, and the contraction between it and the hard function, are controlled by $H_S$, which is given by,\footnote{We note that to obtain this result we used $H_{S,22}^{\rm tree}=1$, as opposed to $H_{S,22}^{\rm tree}=2$, which was stated in Ref.~\cite{Baumgart:2017nsr} that we believe was a typo.}
\begin{equation}\begin{aligned}
&H_{S,11}(\mW) = 1,\hspace{0.5cm}
H_{S,33}(\mW) = U_{H_S},\hspace{0.5cm}
H_{S,31}(\mW) = \frac{2}{3} [1 - U_{H_S}],\\
&\hspace{1.55cm}H_{S,22}(\mW) = U_{H_S},\hspace{0.5cm}
H_{S,24}(\mW) = \frac{1}{3} [1 - U_{H_S}],
\end{aligned}\end{equation}
where $U_{H_S}$ quantifies the evolution in a similar fashion to $U_H$.
Combining these results, we conclude
\begin{equation}\begin{aligned}
H_i H_{S,ij} \tilde{S}_j^{aba'b'} F^{a'b'ab} =&\, 16 \pi^2 \aW^2 U_H\left[ \left( \frac{4}{3} |s_{00}|^2 + 2 |s_{0\pm}|^2 + \frac{4\sqrt{2}}{3} {\rm Re}(s_{00}s_{0\pm}^*) \right) \right.\\
&\hspace{0.75cm}\left.+ U_{H_S} \left( -\frac{4}{3} |s_{00}|^2 + 2 |s_{0\pm}|^2 - \frac{4\sqrt{2}}{3} {\rm Re}(s_{00}s_{0\pm}^*) \right) \right] \\
\equiv&\, 16 \pi^2 \aW^2 U_H \left[ F_0 + U_{H_S} F_1 \right]\!.
\label{eq:wino-defaultbasis}
\end{aligned}\end{equation}
The functions $F_0$ and $F_1$ encode the two combinations of Sommerfeld factors that appeared in parentheses, and appear in the wino LL result.

The above is a direct repetition of the calculation performed in Ref.~\cite{Baumgart:2017nsr}, we now demonstrate that the same result is achieved in our modified approach.
Firstly, in this approach, the hard matching coefficients are modified, with only $H_2^{\rm tree}$ non-zero now, because $C_1 = 0$ (recall, we have a single operator here with the same structure as ${\cal O}_2$).
The only other modifications are the contractions of $\tilde{S}$ in Eq.~\eqref{eq:EFTsoft} into $F_{\chi}$ rather than $F$.
These can be determined straightforwardly, for instance,
\begin{align}
\tilde{S}_1^{aba'b'} F_{\chi}^{a'b'ab} 
=&\, \Big\langle (\chi^0 \chi^0)_S \Big\vert \left(\chi_v^{T} i \sigma_2 \left\{ T_{\chi}^{a'},\,T_{\chi}^{a'} \right\} \chi_v\right)^{\dagger} \Big\vert 0 \Big\rangle \Big\langle 0 \Big\vert \left(\chi^{T}_v i \sigma_2 \left\{ T_{\chi}^a,\,T_{\chi}^a \right\} \chi_v \right) \Big\vert (\chi^0 \chi^0)_S \Big\rangle \nn \\
=&\,16 \left|\Big\langle 0 \Big\vert \left(\chi^{T a}_v i \sigma_2 \chi_v^a \right) \Big\vert (\chi^0 \chi^0)_S \Big\rangle \right|^2 \nn \\
=&\,16 \left|\Big\langle 0 \Big\vert \left(\chi^{T 0}_v i \sigma_2 \chi_v^0 \right) \Big\vert (\chi^0 \chi^0)_S \Big\rangle
+ 2 \Big\langle 0 \Big\vert \left(\chi^{T +}_v i \sigma_2 \chi_v^- \right) \Big\vert (\chi^0 \chi^0)_S \Big\rangle\right|^2 \nn\\
=&\,256 M_{\chi}^2 \left| \sqrt{2} s_{00} + 2 s_{0\pm} \right|^2\!.
\label{eq:winocontract1}
\end{align}
The remaining combinations are given by,
\begin{equation}\begin{aligned}
\tilde{S}_2^{aba'b'} F_{\chi}^{a'b'ab} &= 256 M_{\chi}^2 |s_{0\pm}|^2, \\
\tilde{S}_3^{aba'b'} F_{\chi}^{a'b'ab} &= 256 M_{\chi}^2 \left( 4 |s_{0\pm}|^2 + 2 \sqrt{2} {\rm Re}(s_{00} s_{0\pm}^*) \right)\!, \\
\tilde{S}_4^{aba'b'} F_{\chi}^{a'b'ab} &= 128 M_{\chi}^2 \left( 2 |s_{00}|^2 + 3 |s_{0\pm}|^2 + 2 \sqrt{2} {\rm Re}(s_{00} s_{0\pm}^*) \right)\!.
\label{eq:winocontract2}
\end{aligned}\end{equation}
Using these modified results we find that $H_i H_{S,ij} \tilde{S}_j^{aba'b'}F_{\chi}^{a'b'ab}$ in the new basis exactly matches Eq.~\eqref{eq:wino-defaultbasis}, as it must.
The utility of this approach, is that having formulated the calculation in this way, if we changed the DM representation, the only part of the calculation that would need to be modified is that the appropriate generalizations of contractions in Eqs.~\eqref{eq:winocontract1} and \eqref{eq:winocontract2} would need to be computed.
We will evince this by showing that results for the quintuplet can be derived straightforwardly in the next subsection.
Again, at NLL an almost identical modification to the approach in Ref.~\cite{Baumgart:2018yed} is required, one must simply account for the more complicated forms the hard and soft functions take at that order.

\subsection{Extension to the quintuplet}
\label{eq:extquint}

In the previous subsection, we reorganized the formalism of Refs.~\cite{Baumgart:2017nsr,Baumgart:2018yed} in such a way that the dependence on the DM representation is fully encoded in $F_{\chi}$ as defined in Eq.~\eqref{eq:Som}, and explicitly demonstrated this alternative procedure produces the same result at LL.
This reorganization has the benefit that the quintuplet calculation (and that for any higher odd-$n$ representation, see {\it e.g.} Ref.~\cite{Bottaro:2021snn}) is  almost identical to that of the wino; there are unique Sommerfeld expressions to compute, and a modification for how the new $F_{\chi}$ contracts into the soft Wilson lines in $Y^{abcd}$, but in essence the computation is the same.

For the Sommerfeld factors, in the broken phase, the five degrees of freedom of the quintuplet reorganize themselves into a neutral Majorana fermion $\chi^0$, a heavier singly charged Dirac fermion $\chi^{\pm}$, and a doubly-charged Dirac fermion $\chi^{\pm\pm}$ that is even heavier.
(Again, a more complete discussion is provided in App.~\ref{app:QDM}.)
This implies that there are now three two-body states that can initiate the hard annihilation that are coupled to the initial state through the potential,\footnote{This ignores for the moment the possibility of annihilation through bound states with different quantum numbers, which we will discuss later.} and we parameterize the various matrix elements as\footnote{We emphasize that despite the repeated notation, the functions $s_{00}$ and $s_{0\pm}$ controlling the Sommerfeld enhancement have numerically different values for the wino (Eq.~\eqref{eq:s00s0pm-wino}) and quintuplet (Eq.~\eqref{eq:s00s0pms0pmpm-quintuplet}).}
\begin{equation}\begin{aligned}
\Big\langle 0 \Big\vert (\chi^{0T}_v i \sigma_2 \chi_v^0) \Big\vert (\chi^0 \chi^0)_S \Big\rangle
=\, &4 \sqrt{2} M_{\chi} s_{00}, \\
\Big\langle 0 \Big\vert (\chi^{+T}_v i \sigma_2 \chi_v^-) \Big\vert (\chi^0 \chi^0)_S \Big\rangle
=\, &4 M_{\chi} s_{0\pm}, \\
\Big\langle 0 \Big\vert (\chi^{++T}_v i \sigma_2 \chi_v^{--}) \Big\vert (\chi^0 \chi^0)_S \Big\rangle
=\, &4 M_{\chi} s_{0\pm\pm}.
\label{eq:s00s0pms0pmpm-quintuplet}
\end{aligned}\end{equation}
Note that if we performed our entire calculation at tree level or neglected the Sommerfeld effect, we would take $s_{00}=1$ and $s_{0\pm}=s_{0\pm\pm}=0$.
Using these, we can compute the full $F_{\chi}$ for an arbitrary set of indices.

From these functions we can immediately derive the relevant spectra.
At LL, all that is required is to derive the analogue of $F_0$ and $F_1$ as they appeared in Eq.~\eqref{eq:wino-defaultbasis}.
The main calculation is to compute the equivalent contractions for Eqs.~\eqref{eq:winocontract1} and \eqref{eq:winocontract2}, which are given by
\begin{align}
\tilde{S}_1^{aba'b'} F_{\chi}^{a'b'ab} &= 2384 M_{\chi}^2 \left|2 s_{0\pm\pm} + 2 s_{0\pm} + \sqrt{2} s_{00} \right|^2\!, \nn \\
\tilde{S}_2^{aba'b'} F_{\chi}^{a'b'ab} &= 256 M_{\chi}^2 \left|4 s_{0\pm\pm} + s_{0\pm} \right|^2\!, \nn \\
\tilde{S}_3^{aba'b'} F_{\chi}^{a'b'ab} &= 1536 M_{\chi}^2 \left[ 8 |s_{0\pm\pm}|^2 + 2 |s_{0\pm}|^2 + 10 {\rm Re}(s_{0\pm} s_{0\pm\pm}^*) + 4 \sqrt{2} {\rm Re}(s_{00} s_{0\pm\pm}^*) + \sqrt{2} {\rm Re}(s_{00} s_{0\pm}^*) \right]\!, \nn \\
\tilde{S}_4^{aba'b'} F_{\chi}^{a'b'ab} &= 128 M_{\chi}^2 \left| 2 s_{0\pm\pm} + 5 s_{0\pm} + 3 \sqrt{2} s_{00} \right|^2 + 256 M_{\chi}^2 \left|4 s_{0\pm\pm} + s_{0\pm} \right|^2\!.
\end{align}
Using these, we can evaluate,
\begin{align}
H_i H_{S,ij} \tilde{S}_j^{aba'b'} F^{a'b'ab} = &16 \pi^2 \aW^2 U_H \left[ \left( \frac{4}{3} | 4 s_{0\pm\pm} + s_{0\pm} |^2 + \frac{2}{3} |2 s_{0\pm\pm} + 5 s_{0\pm} + 3 \sqrt{2} s_{00}|^2 \right) \right. \nn\\
&\left.+\, U_{H_S} \left( \frac{8}{3} | 4 s_{0\pm\pm} + s_{0\pm} |^2 - \frac{2}{3} |2 s_{0\pm\pm} + 5 s_{0\pm} + 3 \sqrt{2} s_{00}|^2 \right) \right]\!,
\end{align}
from which we conclude
\begin{equation}\begin{aligned}
F_0 &= \frac{4}{3} | 4 s_{0\pm\pm} + s_{0\pm} |^2 + \frac{2}{3} |2 s_{0\pm\pm} + 5 s_{0\pm} + 3 \sqrt{2} s_{00}|^2, \\
F_1 &= \frac{8}{3} | 4 s_{0\pm\pm} + s_{0\pm} |^2 - \frac{2}{3} |2 s_{0\pm\pm} + 5 s_{0\pm} + 3 \sqrt{2} s_{00}|^2.
\label{eq:F0F1-Quintuplet}
\end{aligned}\end{equation}

With these modified forms for $F_0$ and $F_1$, the LL result is then identical to that derived for the wino in Ref.~\cite{Baumgart:2017nsr}.
For completeness, we restate it below.
\begin{align}
\left( \frac{d\sigma}{dz} \right)^{\rm LL} = \,&(F_0 + F_1) \sigma^{\rm tree} e^{-8 C_A \TaW L_{\chi}^2} \delta(1-z) \nn\\
+\, & 4 \sigma^{\rm tree} e^{-8 C_A \TaW L_\chi^2} \bigg\{ C_A \TaW F_1 \Big( 3 {\cal L}_1^S(z) - 2 {\cal L}_1^J(z) \Big) e^{8 C_A \TaW \left( \Theta_J L_J^2(z) - \tfrac{3}{4} \Theta_S L_S^2(z) \right)} \nn\\
&\hspace{3cm}- 2 C_A \TaW F_0 {\cal L}_1^J(z) e^{8 C_A \TaW L_J^2(z)} \bigg\},
\label{eq:QLL}
\end{align}
where again $\TaW = \aW/4\pi$ and $C_A=2$.
The first line in this result describes the two photon final state, arising from $\chi \chi \to \gamma \gamma$, and is given in terms of a cross section that parameterizes the tree-level rate and a massive Sudakov logarithm,
\begin{equation}
\sigma^{\rm tree} = \frac{\pi \aW^2 \sW^2}{2M_{\chi}^2 v},\hspace{1.0cm}
L_{\chi} = \ln \left( \frac{\mW}{2M_{\chi}} \right)\!,
\label{eq:LL1}
\end{equation}
where $\sW = \sin \thetaW$, $\cW = \cos \thetaW$, and $v = |{\bf v}_1-{\bf v}_2|$ is the relative velocity between the incident DM particles---notations we will use throughout.
Substituting the Sommerfeld factors from Eq.~\eqref{eq:F0F1-Quintuplet} into Eq.~\eqref{eq:QLL}, we see that the line cross-section is proportional to $F_0 + F_1 = 4 |4 s_{0\pm\pm} + s_{0\pm}|^2$.
This shows that the purely doubly-charged contribution to the line emission is a factor of sixteen larger than the purely singly charged, and also that the two contributions interfere.
The equivalent result for the wino is $4 |s_{0\pm}^{\rm wino}|^2$ (again $F_0 + F_1$ using the wino equivalent values in Eq.~\eqref{eq:wino-defaultbasis}), which has the same form as the quintuplet when the doubly-charged contribution is turned off, although we caution that in the full result taking $s_{0\pm\pm} \to 0$ does not reproduce the wino cross-section.

The second and third lines of Eq.~\eqref{eq:QLL} correspond to the endpoint, arising from $\chi \chi \to \gamma + X$, where the invariant mass of $X$ is constrained to be near the lightcone.
This contribution depends on an additional pair of logarithms and thresholds, associated with the jet ($J$) and soft ($S$) scales in the problem.
In detail,
\begin{equation}\begin{aligned}
L_J = \ln \left( \frac{\mW}{2M_{\chi}\sqrt{1-z}} \right)\!,\hspace{0.5cm}
\Theta_J = \Theta \left( 1 - \frac{\mW^2}{4M_{\chi}^2} - z \right)\!,\hspace{0.5cm}
{\cal L}_1^J = \frac{L_J}{1-z} \Theta_J, \\
L_S = \ln \left( \frac{\mW}{2M_{\chi}(1-z)} \right)\!,\hspace{0.5cm}
\Theta_S = \Theta \left( 1 - \frac{\mW}{2M_{\chi}} - z \right)\!,\hspace{0.5cm}
{\cal L}_1^S = \frac{L_J}{1-z} \Theta_S.
\label{eq:LL2}
\end{aligned}\end{equation}

The extension to NLL proceeds identically.
The expressions are more involved, but will be schematically identical to Eq.~\eqref{eq:QLL}: all EFT functions will be identical between the wino and the quintuplet, with only the Sommerfeld contributions varying.
To begin with, the differential NLL quintuplet cross-section can be written as\footnote{In this result we have set all EFT functions to their canonical scales.
For instance, the weak coupling in the prefactor is evaluated at the hard matching scale $\mu^0_H = 2 M_{\chi}$.
A common technique for estimating the size of the theoretical uncertainty arising from neglecting higher order contributions is to vary these scales by a factor of $\sim$2.
For this, the result with the scales unfixed is required, and can be obtained by extending the result in Eq.~\eqref{eq:QNLL} to an arbitrary scale, exactly as done in Ref.~\cite{Baumgart:2018yed}.}
\begin{equation}\begin{aligned}
\left( \frac{d\sigma}{dz} \right)^{\rm NLL}
&= \frac{\pi \aW^2(2M_{\chi}) \sW^2(\mW)}{9 M_{\chi}^2 v} U_H \left[ \left. \left( {\cal F}_0 + {\cal F}_1 \right) \right|_{\Lambda \to 1} \right] \delta(1-z) \\
& +\frac{\pi \aW^2(2M_{\chi}) \sW^2(\mW)}{9 M_{\chi}^2 v (1-z)} U_H \left( (V_J - 1) \Theta_J +1 \right) \left[ {\cal F}_0\, \frac{e^{\gamma_E \omega_J}}{\Gamma(-\omega_J)} \vphantom{\frac{e^{\gamma_E (\omega_J+2\omega_S)}}{\Gamma(-\omega_J-2\omega_S)}} \right. \\
&\hspace{3.95cm}\left.+ \left( (V_S - 1) \Theta_S +1 \right) {\cal F}_1\, \frac{e^{\gamma_E (\omega_J+2\omega_S)}}{\Gamma(-\omega_J-2\omega_S)} \right]\!.
\label{eq:QNLL}
\end{aligned}\end{equation}
The result is written in a similar form to the LL result of Eq.~\eqref{eq:QLL}, with the exclusive two-photon final state on the first line, and the endpoint contribution on the last two.
We note that $\Gamma(x)$ is the Euler gamma function, not to be confused with the cusp anomalous dimensions introduced below.
All functions in the result are identical to the NLL wino expression given in Ref.~\cite{Baumgart:2018yed}, except for ${\cal F}_0$ and ${\cal F}_1$, which account for the various Sommerfeld channels.
For completeness, let us first restate the elements common to the wino, again referring to Ref.~\cite{Baumgart:2018yed} for additional details.
Firstly, the evolution of the hard function is encapsulated in
\begin{equation}\begin{aligned}
U_H = r_H^2 \exp \left\{ - \frac{2\Gamma_0}{\beta_0^2} \right. &\left[ \frac{1}{\TaW(2M_{\chi})} \left( \ln r_H + \frac{1}{r_H} - 1 \right) \right. \\
&\left.\left.+ \left( \frac{\Gamma_0}{\Gamma_1} - \frac{\beta_1}{\beta_0} \right) (r_H - 1 - \ln r_H) - \frac{\beta_1}{2\beta_0} \ln^2 r_H \right] \right\}\!.
\end{aligned}\end{equation}
This result is written in terms of the first two perturbative orders of the $\beta$ function and cusp anomalous dimension,
\begin{equation}
\beta_0 = \frac{19}{6}, \hspace{0.5cm}
\beta_1 = - \frac{35}{6}, \hspace{0.5cm}
\Gamma_0 = 4,\hspace{0.5cm}
\Gamma_1 = \frac{8}{9} ( 35-\pi^2),
\end{equation}
as well as the ratio of the coupling between scales $r_H = \aW(\mW)/\aW(2M_{\chi})$.
The evolution of the jet and soft functions is contained in
\begin{align}
V_J &= \exp \left\{ \frac{2\Gamma_0}{\beta_0^2} \left[ \frac{1}{\TaW(\mu^0_J)} \left( \ln r_J + \frac{1}{r_J} - 1 \right) + \left( \frac{\Gamma_1}{\Gamma_0} - \frac{\beta_1}{\beta_0} \right) (r_J - 1 - \ln r_J) - \frac{\beta_1}{2\beta_0} \ln^2 r_J \right] - \ln r_J \right\}\!, \nn\\
V_S &= \exp \left\{ - \frac{3\Gamma_0}{2\beta_0^2} \left[ \frac{1}{\TaW(\mu_S^0)} \left( \ln r_S + \frac{1}{r_S} - 1 \right) + \left( \frac{\Gamma_1}{\Gamma_0} - \frac{\beta_1}{\beta_0} \right) (r_S - 1 - \ln r_S) - \frac{\beta_1}{2\beta_0} \ln^2 r_S \right] \right\}\!, \nn\\
\omega_J &= - \frac{2\Gamma_0}{\beta_0} \left[ \ln r_J + \TaW(\mu_J^0) \left( \frac{\Gamma_1}{\Gamma_0} - \frac{\beta_1}{\beta_0} \right) (r_J-1) \right] \Theta_J, \\
\omega_S &= \frac{3\Gamma_0}{2\beta_0} \left[ \ln r_S + \TaW(\mu_S^0) \left( \frac{\Gamma_1}{\Gamma_0} - \frac{\beta_1}{\beta_0} \right) (r_S - 1) \right] \Theta_S. \nn
\end{align}
Here the ratio of scales are given by $r_J = \aW(\mW)/\aW(\mu^0_J)$ and $r_S = \aW(\mW)/\aW(\mu^0_S)$, written in terms of the canonical scales $\mu^0_J = 2M_{\chi}\sqrt{1-z}$ and $\mu^0_S = 2M_{\chi}(1-z)$.
Further, $\Theta_J$ and $\Theta_S$ are as defined in Eq.~\eqref{eq:LL2}.

The last terms to be defined are those unique for the quintuplet.
For those, we have
\begin{equation}\begin{aligned}
{\cal F}_0 &= \left[ 36 \Lambda^d + 18 r_{HS}^{12/\beta_0} \Lambda^c \right] |s_{00}|^2
+ \left[ 72 \Lambda^d + 9 r_{HS}^{12/\beta_0} \Lambda^c \right] |s_{0\pm}|^2 \\
&+\left[ 72 \Lambda^d + 36 r_{HS}^{12/\beta_0} \Lambda^c \right] |s_{0\pm\pm}|^2
+ \sqrt{2} \left[ 72 \Lambda^d + 18 r_{HS}^{12/\beta_0} \Lambda^c \right] \textrm{Re} (s_{00} s_{0\pm}^*) \\
&+\sqrt{2} \left[ 72 \Lambda^d - 36 r_{HS}^{12/\beta_0} \Lambda^c \right] \textrm{Re} (s_{00} s_{0\pm\pm}^*)
+ \left[ 144 \Lambda^d - 36 r_{HS}^{12/\beta_0} \Lambda^c \right] \textrm{Re} (s_{0\pm} s_{0\pm\pm}^*),
\end{aligned}\end{equation}
and
\begin{align}
{\cal F}_1 &= r_H^{6/\beta_0} \left(
  \left[ 18 r_{HS}^{6/\beta_0} \Lambda^a - 72 c_H \Lambda^b \right] |s_{00}|^2
+ \left[ 9 r_{HS}^{6/\beta_0} \Lambda^a - 72 c_H \Lambda^b \right] |s_{0\pm}|^2 \right. \\
&+\left[ 36 r_{HS}^{6/\beta_0} \Lambda^a + 144 c_H \Lambda^b \right] |s_{0\pm\pm}|^2
+ \sqrt{2} \left[ 18 r_{HS}^{6/\beta_0} \Lambda^a - 108 c_H \Lambda^b \right] \textrm{Re}(s_{00}s_{0\pm}^*) \nn\\
&+\sqrt{2} \left[ -36 r_{HS}^{6/\beta_0} \Lambda^a \right] \textrm{Re}(s_{00} s_{0\pm\pm}^*)
+ \left[ -36 r_{HS}^{6/\beta_0} \Lambda^a + 72 c_H \Lambda^b \right] \textrm{Re}(s_{0\pm}s_{0\pm\pm}^*) \nn\\
&\left.+\sqrt{2} \left[ -36 s_H \Lambda^b \right] \textrm{Im}(s_{00} s_{0\pm}^*)
+ \sqrt{2}\left[ -144 s_H \Lambda^b \right] \textrm{Im}(s_{00}s_{0\pm\pm}^*)
+ \left[ -216 s_H \Lambda^b \right] \textrm{Im}(s_{0\pm}s_{0\pm\pm}^*) \right)\!. \nn
\end{align}
These expressions introduce $r_{HS} = r_H/r_S$, as well as
\begin{equation}
c_H = \cos \left( \frac{6\pi}{\beta_0} \ln r_H \right)\!, \hspace{1cm}
s_H = \sin \left( \frac{6\pi}{\beta_0} \ln r_H \right)\!,
\end{equation}
and further four functions $\Lambda^{a-d}$.
These functions are as follows, and are the same form as appears for the wino,
\begin{equation}\begin{aligned}
\Lambda^a &= 1 + \TaW (\mu^0_J) \left[ \Gamma_0 \Delta^{(2)}_{JSJ}  + \beta_0 \Delta^{(1)}_{JS} \right] \Theta_J - 12 \TaW(\mu_S^0) \left[ \Delta^{(2)}_{JSS} - \Delta^{(1)}_{JS} \right] \Theta_S, \\
\Lambda^b &= 1 + \TaW (\mu^0_J) \left[ \Gamma_0 \Delta^{(2)}_{JSJ}  + \beta_0 \Delta^{(1)}_{JS} \right] \Theta_J - 12 \TaW(\mu_S^0) \Delta^{(2)}_{JSS} \Theta_S, \\
\Lambda^c &= 1 + \TaW (\mu^0_J) \left[ \Gamma_0 \Delta^{(2)}_{J}  + \beta_0 \Delta^{(1)}_{J} \right] \Theta_J + 24 \TaW(\mu_S^0) \Delta^{(1)}_{J} \Theta_S, \\
\Lambda^c &= 1 + \TaW (\mu^0_J) \left[ \Gamma_0 \Delta^{(2)}_{J}  + \beta_0 \Delta^{(1)}_{J} \right] \Theta_J,
\label{eq:NLLLambda}
\end{aligned}\end{equation}
with the functions $\Delta$ written in terms of the polygamma function of order $m$, $\psi^{(m)}$, as follows,
\begin{equation}\begin{aligned}
\Delta_J^{(1)} &= \gamma_E + \psi^{(0)}(-\omega_J), \\
\Delta_J^{(2)} &= \left( \gamma_E + \psi^{(0)}(-\omega_J) \right)^2 - \psi^{(1)}(-\omega_J), \\
\Delta_{JS}^{(1)} &= \gamma_E + \psi^{(0)}(-\omega_J-2\omega_S), \\
\Delta_{JSJ}^{(2)} &= \Delta_{JSS}^{(2)} = \left( \gamma_E + \psi^{(0)}(-\omega_J-2\omega_S) \right)^2 - \psi^{(1)}(-\omega_J-2\omega_S).
\end{aligned}\end{equation}
Finally, note that in the exclusive contribution to Eq.~\eqref{eq:QNLL}, we use the notation $\Lambda \to 1$ to imply all of the functions in Eq.~\eqref{eq:NLLLambda} are set to unity.

Whilst the NLL expression is more involved than the LL result, the associated theoretical uncertainties are significantly reduced.
This is demonstrated in the right of Fig.~\ref{fig:Thermal-Spectrum}, where we show the cumulative, or integrated $d\sigma/dz$, taken from a given $z_{\rm cut}$ to 1.
The fact H.E.S.S. (or indeed any real imaging air Cherenkov telescope) does not have perfect energy resolution is represented by the fact the physically appropriate $z_{\rm cut}$ is away from unity.

We will further explore these results in Sec.~\ref{sec:Numerics}.
Before doing so, however, we turn to the additional contribution to the spectrum that can result from bound states.

\section{Bound State Formation}
\label{sec:BSF}

In this section we work out the rate of formation of relevant bound states, before considering the application of the SCET formalism to their annihilation in Sec.~\ref{sec:BSA}.
The general formalism we employ is based on the methods of Ref.~\cite{Harz:2018csl} for general non-Abelian gauge groups (see also Refs.~\cite{Asadi:2016ybp, Mitridate:2017izz}).
Throughout this work, we consider only single-vector-boson emission in the dipole approximation.
We first review the key equations and define our notation, then work out the form of the generators and potential with our basis conventions (Sec.~\ref{subsec:generators}), and use these results to evaluate the cross sections for bound-state formation via emission of photon (Sec.~\ref{subsec:photoncapture}) and weak gauge bosons (Sec.~\ref{subsec:Wzcapture}).
We note already that when SU(2) is broken, the velocity dependence of bound-state formation differs from that of Sommerfeld-enhanced direct annihilation; we will present a discussion of this issue and the uncertainties associated with the velocity distribution of the DM halo when we turn to our numerical results in Sec.~\ref{subsec:vdep}.

\begin{figure*}[!t]
\centering
\includegraphics[width=0.4\textwidth]{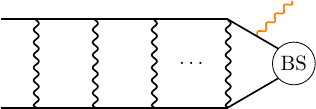}
\hspace{2.cm}
\includegraphics[width=0.4\textwidth]{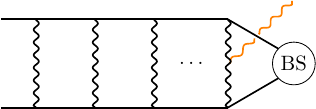}
\caption{Due to the non-Abelian nature of the gauge bosons generating the potential, the emission necessary for bound state (BS) formation can occur either from the external fermion line (left), or from the potential itself (right).}
\label{fig:BS-Diags}
\end{figure*}

If we label the different states in the multiplet containing DM as $\chi_i$ where $i$ runs from 1 to 5, then for capture of a $\chi_i \chi_j$ initial state into a $\chi_{i^\prime} \chi_{j^\prime}$ bound state with quantum numbers $(nlm)$, where all particles have equal masses $M_\chi$ and the emitted particle has color index $a$ and can be approximated as massless, the amplitude for radiative capture into a bound state (stripping off the polarization vector for the outgoing gauge boson) is given by Ref.~\cite{Harz:2018csl}:
\begin{equation}
\mathcal{M}^a_{ii^\prime\!,\, jj^\prime} = 
-\sqrt{2^8 \pi \alpha_\text{rad} M_\chi}
\left\{ -i f^{abc} (T_1^b)_{i^\prime i} (T_2^c)_{j^\prime j} \mathcal{Y} + \frac{1}{2} \left[ (T_1^a)_{i^\prime i} \delta_{j^\prime j} - (T_2^a)_{j^\prime j} \delta_{i^\prime i}\right] \mathcal{J} \right\}\!. 
\label{eq:matrixelement} 
\end{equation}
Let us define the various notations introduced in this amplitude.
Firstly, $\alpha_\text{rad}$ specifies the coupling associated with the radiated gauge boson: in our case, $\alpha_\text{rad} = \alpha$ for a radiated photon,\footnote{We denote the electromagnetic fine structure constant by $\alpha = \sW^2 \aW$.} $\alpha_\text{rad}= \cW^2 \aW$ for a radiated $Z$ boson, and $\alpha_\text{rad}= \aW$ for a radiated $W$ boson.
$T_1$ and $T_2$ denote the generators of the representation associated with the $\chi_i$ and $\chi_j$ particles respectively, whilst $f^{abc}$ are the structure constants.
(We emphasize that for the moment we are discussing the more general result where in principle $\chi_i$ and $\chi_j$ could be in different representations.
Shortly, we will specialize to the case where $T_1 = T_2 = T_{\mathbf{5}}$ appropriate for the quintuplet.) 
Finally, we have
\begin{equation}\begin{aligned}
\mathcal{Y} & = 4 \pi M_\chi \, \aNA \int \frac{d^3 \mathbf{p}}{(2\pi)^3} \frac{d^3 \mathbf{q}}{(2\pi)^3} \frac{\mathbf{q} - \mathbf{p}}{(\mathbf{q} - \mathbf{p})^4} \tilde{\psi}_{nlm}^*(\mathbf{p}) \tilde{\phi}(\mathbf{q}), \\
\mathcal{J} & \simeq \int \frac{d^3 \mathbf{p}}{(2\pi)^3} \, \mathbf{p} \, \tilde{\psi}^*_{nlm}(\mathbf{p}) \tilde{\phi}(\mathbf{p}). 
\label{eq:yjequations} 
\end{aligned}\end{equation}
In the expression for $\mathcal{J}$ we have made the approximation of neglecting the momentum of the outgoing gauge boson.
$\aNA$ is the coupling between the fermions and the $t$-channel gauge boson exchanged between them to support the potential, for the diagram where the bound state formation occurs through the emission of a gauge boson from the potential line.
The two possible emission channels are depicted in Fig.~\ref{fig:BS-Diags}.
For example, when the bound-state formation occurs through emission of a photon or $Z$ boson, the exchanged gauge boson will be a $W$ boson, and so we will have $\aNA = \aW$.
The $\tilde{\psi}_{nlm}$ and $\tilde{\phi}$ wavefunctions are the momentum-space wavefunctions of the final and initial states respectively (the corresponding real-space wavefunctions are labeled $\psi_{nlm}$ and $\phi$).

The $\mathcal{Y}$ and $\mathcal{J}$ coefficients can be rewritten in position space as,\footnote{We follow the Fourier transformation conventions of Ref.~\cite{Harz:2018csl}.}
\begin{equation}\begin{aligned}
\mathcal{Y}
& = - \frac{i}{2} M_\chi \aNA \int d^3 \mathbf{r} \, \hat{\mathbf{r}} \, \psi_{nlm}^*(\mathbf{r}) \phi(\mathbf{r}),\\
\mathcal{J} & \simeq - i  \int d^3 \mathbf{r} \, \psi^*_{nlm}(\mathbf{r}) \nabla \phi(\mathbf{r}).  
\end{aligned}\end{equation}
In subsequent equations we will suppress the $\mathbf{r}$ dependence of the position-space wavefunctions for notational convenience.
We emphasize that these expressions tacitly assume the two particles are distinguishable; we will follow the conventions of Ref.~\cite{Asadi:2016ybp} for the normalization of two-particle states, which can introduce a factor of $\sqrt{2}$ into terms involving transitions between states of identical and non-identical particles.\footnote{We emphasize that there is a subtlety here associated with the ordering of particles in the two-particle states, which can induce a sign flip that must be treated carefully.
We discuss this in App.~\ref{app:signs}.}
Breaking SU(2) can also introduce an additional multiplicative factor inside the integral for $\mathcal{Y}$, arising from the propagators in the potential line from which the particle is emitted; for the case of photon or $Z$ emission, these are $W$ propagators, and the additional factor takes the form $e^{-m^{}_{\scaleto{W}{3.5pt}} r}$ \cite{Asadi:2016ybp}. 
We will work out the correct replacement in the case of $W$ emission later in this section.

Throughout this section, when solving for the wavefunctions for the initial and final states given a specific potential, we will adopt the normalization conventions and numerical approach of Ref.~\cite{Asadi:2016ybp}.
(We note that there was a minus sign error in the equation for bound-state formation in the original published version of Ref.~\cite{Asadi:2016ybp}, which has since been corrected in an erratum.)

\subsection{Generators and the potential}
\label{subsec:generators}

In general there is a degree of freedom to choose the basis for our generators, but since we are interested in transitions between two-body states whose constituents are mass eigenstates distinguished by their charges, it is convenient to use the basis discussed in App.~\ref{app:QDM}, where the $\chi_1$, $\chi_2$, $\chi_3$, $\chi_4$, $\chi_5$ states correspond to states with electric charges $+2$, $+1$, 0, $-1$ and $-2$, as given in Eq.~\eqref{eq:Qstvector}.
It is important that the basis used to compute the bound-state formation rate and the basis used to compute the potential are identical; we will require the potential when solving for the initial- and final-state wavefunctions, which are relevant both for bound-state formation and for the Sommerfeld enhancement to direct annihilation.
In this basis, we obtain for the generators:
\begin{equation}\begin{aligned}
T^1_{\mathbf{5}} & = \frac{1}{\sqrt{2}} \begin{pmatrix} 0& \sqrt{2} & 0 & 0 & 0 \\ \sqrt{2} & 0 & \sqrt{3} & 0 & 0 \\ 0 & \sqrt{3} & 0 & \sqrt{3} & 0 \\ 0 & 0 & \sqrt{3} & 0 & \sqrt{2} \\ 0 & 0 & 0 & \sqrt{2} & 0\end{pmatrix}\!, \quad
T^2_{\mathbf{5}} = \frac{i}{\sqrt{2}} \begin{pmatrix} 0& - \sqrt{2} & 0 & 0 & 0 \\  \sqrt{2} & 0 & - \sqrt{3} & 0 & 0 \\ 0 &  \sqrt{3} & 0 & - \sqrt{3} & 0 \\ 0 & 0 &  \sqrt{3} & 0 & - \sqrt{2} \\ 0 & 0 & 0 &  \sqrt{2} & 0\end{pmatrix}\!, \\
T^3_{\mathbf{5}} & = \textrm{diag}(2,1,0,-1,-2).
\label{eq:generators}
\end{aligned}\end{equation}

The potential, up to terms corresponding to mass splittings between the two-body states (whose contribution is spelled out in App.~\ref{app:QDM}), can be written in the following form \cite{Cirelli:2007xd, Beneke:2014gja}
\begin{equation}
V_{i j;i^\prime j^\prime} = N_{i j} N_{i^\prime j^\prime} \sum_{AB} K_{AB} \left(T^A_{i i^\prime} T^B_{j j^\prime} + (-1)^{L+S} T^A_{i j^\prime} T^B_{j i^\prime} \right) \frac{e^{-m^{}_{\scaleto{A}{3.5pt}} r}}{4\pi r}, \quad 
K = \begin{pmatrix} 1 & 0 & 0 & 0 \\ 0 & 1 & 0 & 0 \\ 0 & 0 & 0 & 1 \\ 0 & 0 & 1 & 0\end{pmatrix}\!.
\end{equation}
Here $N_{ij} =1$ if $i\ne j$ and $1/\sqrt{2}$ if $i=j$ (this corresponds to the aforementioned change in normalization for two-body states composed of identical vs distinguishable particles), and the indices $A$, $B$ run over $\{\gamma, Z, W^+, W^-\}$.
The gauge couplings are included in the generators in this notation, following the conventions of Ref.~\cite{Cirelli:2007xd}: explicitly, $T^\gamma = \gW \sin \thetaW T^3$, $T^Z = \gW \cos \thetaW T^3$, $T^{W^+} = \gW T^+$, $T^{W^-} = \gW T^-$, with $T^{\pm} = (T^1 \pm i T^2)/\sqrt{2}$.
Note that the $(-1)^{L+S}$ factor arises from treating the $ij$ and $ji$ states as representatives of a single two-body state, using the conventions employed in method-2 of Ref.~\cite{Beneke:2014gja} and also discussed in App.~\ref{app:signs}.
(Here $L$ denotes orbital angular momentum and $S$ denotes spin; see Ref.~\cite{Beneke:2014gja} for a detailed discussion.)
This sign was also discussed in the context of the potential for two-body states with net charge $Q=\pm 1$ in Ref.~\cite{Mitridate:2017izz}.

Thus, with the basis above, we obtain the following potential for the case with $L+S$ even, where the 1st row/column corresponds to the $\chi^{++} \chi^{--}$ two-body state ($\chi_1 \chi_5$), the 2nd row/column corresponds to the $\chi^+\chi^-$ state ($\chi_2 \chi_4$), and the 3rd row/column corresponds to the $\chi^0 \chi^0$ state ($\chi_3 \chi_3$):
\begin{equation} 
V(r) = \aW \begin{pmatrix} -4 \left(\cW^2 \frac{e^{-m^{}_{\scaleto{Z}{3.pt}} r}}{r} + \frac{s^2_{\scaleto{W}{3.pt}}}{r}\right) & 2 \frac{e^{-m^{}_{\scaleto{W}{3.pt}} r}}{r} & 0 \\ 2 \frac{e^{-m^{}_{\scaleto{W}{3.pt}} r}}{r} & - \left(\cW^2 \frac{e^{-m^{}_{\scaleto{Z}{3.pt}} r}}{r} + \frac{s^2_{\scaleto{W}{3.pt}}}{r}\right) & 3 \sqrt{2} \frac{e^{-m^{}_{\scaleto{W}{3.pt}} r}}{r} \\ 0 & 3 \sqrt{2} \frac{e^{-m^{}_{\scaleto{W}{3.pt}} r}}{r} & 0 \end{pmatrix}\!. 
\label{eq:quintpot} 
\end{equation}
Note that the signs of the off-diagonal terms are opposite to the potential matrix given in Ref.~\cite{Cirelli:2007xd} (and the analogue for the wino employed in Ref.~\cite{Asadi:2016ybp}); this is a basis-dependent choice and either option is correct provided it is used self-consistently throughout the calculation.
The effect of changing the basis in a way that modifies the signs in the off-diagonal terms of the potential is to flip the sign of one or more components of the resulting solution for the wavefunction; this compensates the changes in sign in generator elements in the new basis, when computing the bound-state wavefunctions.

It is also possible to go beyond the tree-level potential of Eq.~\eqref{eq:quintpot} and include NLO corrections.
Especially in proximity to resonances, the resulting modifications to the Sommerfeld enhancement and bound-state formation rate can be substantial~\cite{Beneke:2019qaa}.
We employ the analytic fitting functions for the NLO potential calculated by Refs.~\cite{Beneke:2019qaa, Urban:2021cdu}.
Specifically, we make the following replacements in Eq.~\eqref{eq:quintpot}, where $L\equiv \ln(\mW r)$:
\begin{align}
     e^{-m^{}_{\scaleto{W}{3.pt}} r} & \rightarrow e^{-m^{}_{\scaleto{W}{3.pt}} r} + \frac{2595}{\pi} \aW 
     \begin{cases} (-1) \text{exp}\left[-\frac{79 \left(L - \frac{787}{12}\right)\left(L - \frac{736}{373}\right) \left(L - \frac{116}{65}\right)\left(L^2 - \frac{286 L}{59} + \frac{533}{77}\right) }{34 \left(L - \frac{512}{19}\right) \left(L - \frac{339}{176}\right) \left(L - \frac{501}{281}\right) \left(L^2 -  \frac{268 L}{61} + \frac{38}{7}\right)} \right]\!, & \mW r < 555/94 \\
     \text{exp}\left[-\frac{13267 \left(L - \frac{76}{43}\right)\left(L - \frac{28}{17}\right) \left(L + \frac{37}{30}\right)\left(L^2 - \frac{389 L}{88} + \frac{676}{129}\right) }{5 \left(L - \frac{191}{108}\right) \left(L - \frac{256}{153}\right) \left(L + \frac{8412}{13}\right) \left(L^2 -  \frac{457 L}{103} + \frac{773}{146}\right)} \right] & \mW r > 555/94\end{cases} 
     \nn \\
     \cW^2 e^{-m^{}_{\scaleto{Z}{3.pt}} r} & \rightarrow \cW^2 e^{-m^{}_{\scaleto{Z}{3.pt}} r} + \aW \left[ -\frac{80}{9} \frac{\sW^4 \left(\ln(\mZ r) + \gamma_E\right)}{2\pi (1 + (32/11) (\mW r)^{-22/9})}   + \frac{\left( \frac{19}{6}  \ln(\mZ r) - \frac{1960}{433} \right)}{2\pi(1 + (7/59) (\mW r)^{61/29})} \right. \nn \\
     & \left. - \frac{\sW^2 \left(-\frac{1}{30} + \frac{4}{135} \ln(\mW r) \right)}{1 + (58/79) (\mW r)^{-17/15} + (1/30) (\mW r)^{119/120} + (8/177) (\mW r)^{17/8}}\right]\!.
\label{eq:vnlo}     
\end{align}
We will use the NLO potential for all calculations performed in this work.
In particular, in addition to using it to compute the capture cross-sections in this section, we will also use it to compute the Sommerfeld enhancement appropriate for the direct annihilation discussed in Sec.~\ref{sec:DA}.
We note, however, that the relic abundance calculations in Refs.~\cite{Mitridate:2017izz,Bottaro:2021snn}, which we rely on for our value of the thermal mass $M_{\chi} = 13.6 \pm 0.8~{\rm TeV}$, did not use the NLO potential.\footnote{Ref.~\cite{Bottaro:2023wjv} emphasized the possible importance of NLO corrections to the potential in a U(1) model coming from hard loops that are not captured by Eq.~\ref{eq:vnlo}.
Understanding the size of this effect for both relic abundance and indirect detection is worthy of future investigation.}
As we will see in Sec.~\ref{sec:Numerics}, our findings as to the quintuplet will vary considerably across the thermal mass range, and so updating the calculation to NLO is an important improvement left to be done.

\subsection{Bound state formation through emission of a photon}
\label{subsec:photoncapture}

For the quintuplet it will often be possible to form a bound state through emission of $W$ or $Z$ gauge bosons, but let us first consider the case where the radiated particle is a photon, as this channel is available for all DM masses and provides the closest analogy to previous studies of the wino ({\it e.g.}~Ref.~\cite{Asadi:2016ybp}).
In this case we have $\alpha_\text{rad}=\alpha$, $\aNA = \aW$, and $a=3$ (since the photon obtains its SU(2) couplings through the $W^3$ component).
Let us also assume both incoming particles are in the same representation (as appropriate for Majorana fermions), and $T$ denotes the generators of that representation, so we can write:
\begin{align}
\mathcal{M}^3_{ii^\prime, jj^\prime} & = i \sqrt{2^6 \pi \alpha M_\chi} \left\{ -i f^{123} \left[  (T^1)_{i^\prime i} (T^2)_{j^\prime j} - (T^2)_{i^\prime i} (T^1)_{j^\prime j} \right] M_\chi \aW \int d^3 \mathbf{r} \, \hat{\mathbf{r}}\, e^{-m^{}_{\scaleto{W}{3.5pt}} r} \psi_{nlm}^* \phi \right. \nn \\
& + \left.  \left[ (T^3)_{i^\prime i} \delta_{j^\prime j} - (T^3)_{j^\prime j} \delta_{i^\prime i}\right] \int d^3 \mathbf{r}\, \psi^*_{nlm} \nabla \phi\right\}\!.
\end{align}

From here, substituting in the generators above, we find the following non-zero matrix elements for bound-state formation
\begin{align}
\mathcal{M}^3_{32,34} 
& = i \sqrt{2^6 \pi \alpha M_\chi} \left\{ 3 M_\chi \aW \int d^3 \mathbf{r}\, e^{-m^{}_{\scaleto{W}{3.5pt}} r}\, \hat{\mathbf{r}}\,  \psi_{nlm}^* \phi \right\}\!, \nonumber \\
\mathcal{M}^3_{22,44} 
&= i \sqrt{2^6 \pi \alpha M_\chi}  \left\{2 \int d^3 \mathbf{r}\, \psi^*_{nlm} \nabla \phi\right\}\!, 
\label{eq:quintcapture1}\\
\mathcal{M}^3_{12,54} 
& = i \sqrt{2^6 \pi \alpha M_\chi}  \left\{ -2 M_\chi  \aW \int d^3 \mathbf{r}\, e^{-m^{}_{\scaleto{W}{3.5pt}} r} \,\hat{\mathbf{r}}\, \psi_{nlm}^* \phi\right\}\!, \nonumber
\end{align}
which correspond to capture into the $\chi^+ \chi^-$ bound state component from the $\chi^0 \chi^0$, $\chi^+ \chi^-$, and $\chi^{++}\chi^{--}$ initial state components respectively.
The equivalent matrix elements for capture into the $\chi^{++} \chi^{--}$ bound-state component are given by
\begin{equation}\begin{aligned}
\mathcal{M}^3_{11,55} 
& = i \sqrt{2^6 \pi \alpha M_\chi} \left\{ 4 \int d^3 \mathbf{r}\, \psi^*_{nlm} \nabla \phi\right\}\!, \\
\mathcal{M}^3_{21,45}
& = i \sqrt{2^6 \pi \alpha M_\chi}\left\{ 2 M_\chi  \aW \int d^3 \mathbf{r}\, e^{-m^{}_{\scaleto{W}{3.5pt}} r} \,\hat{\mathbf{r}}\,  \psi_{nlm}^* \phi \right\}\!.
\label{eq:quintcapture2}
\end{aligned}\end{equation}

Combining these matrix elements, and including a factor of $\sqrt{2}$ for the capture from the $\chi^0 \chi^0$ to $\chi^+ \chi^-$ state \cite{Asadi:2016ybp} to account for the differing normalization of states built from identical and distinguishable particles, we can write the cross section for bound-state formation as \cite{Harz:2018csl}:
\begin{equation}\begin{aligned} 
\sigma v & = \int d\Omega_k \frac{k}{2^7\pi^2 M_\chi^3} | {\boldsymbol \epsilon}(\hat{\mathbf{k}}) \cdot \mathcal{M}|^2 \\
& =  \frac{2 \alpha k}{\pi M_\chi^2} \int d\Omega_k \left| \int d^3 \mathbf{r}\, {\boldsymbol \epsilon}(\hat{\mathbf{k}}) \cdot \left[  \left(2  \psi^*_{CC} \nabla \phi_{CC} +  \psi_{C}^* \nabla \phi_C \right) \right. \right. \\
& \left. \left. + \frac{1}{2} M_\chi \aW \hat{\mathbf{r}}\, e^{-m^{}_{\scaleto{W}{3.5pt}} r}  \left(2 \psi_{CC}^* \phi_C  - 2 \psi_{C}^* \phi_{CC} + 3 \sqrt{2}  \psi_C^* \phi_N \right) \right] \right|^2\!,
\label{eq:BSxsec} 
\end{aligned}\end{equation}
where $k$ and ${\boldsymbol \epsilon}(\hat{\mathbf{k}})$ respectively denote the momentum and polarization of the outgoing photon; a $CC$ subscript indicates the $\chi^{++} \chi^{--}$ component, $C$ indicates $\chi^+ \chi^-$, and $N$ indicates $\chi^0 \chi^0$.
As previously, $\psi$ and $\phi$ indicate the final bound-state and initial wavefunctions, respectively.

Note that the potential of Eq.~\eqref{eq:quintpot} is only accurate as written for two-body states with angular momentum quantum numbers $L+S$ summing to an even value.
If $L+S$ is odd, the state is symmetric under particle exchange and cannot support a pair of identical fermions, and consequently the rows and columns corresponding to the $\chi^0\chi^0$ state must be zeroed out.
We are primarily interested in the behavior of quintuplet DM in the Milky Way halo, where on-shell charginos are not likely to be kinematically allowed (exciting the $\chi^+ \chi^-$ state requires 164 MeV of kinetic energy per particle, which for a Milky Way escape velocity of $\sim$500 km/s requires $M_\chi \gtrsim 120$ TeV).
Consequently we will always assume the initial two-body state is $\chi^0\chi^0$ (at large separation) and so has $L+S$ even; this means the bound state formed by the leading-order vector-boson emission will have $L+S$ odd (the dipole selection rule is $\Delta L=\pm 1$, $\Delta S=0$).
The appropriate potentials are used to compute the wavefunctions for the scattering state (Eq.~\eqref{eq:quintpot} as written) and the bound state (Eq.~\eqref{eq:quintpot} with the third row/column removed).

\subsection{Bound state formation through emission of $W$ and $Z$ bosons}
\label{subsec:Wzcapture}

Since the SU(2) couplings of the $Z$ boson are controlled by its $W^3$ component, we can re-use the expression for capture via photon emission in the case of the $Z$ boson, with the replacement $\alpha \rightarrow \cW^2 \aW$ in the prefactor, and with the momentum $k$ now depending on the mass of the $Z$ boson,
\begin{equation} 
k = \sqrt{ \left( -E_n + \frac{M_\chi v^2}{4} \right)^2 - \mZ^2}. 
\label{eq:Zphasespace} 
\end{equation}
Here the energy of the outgoing bound state is $2M_{\chi} + E_n$, where $E_n$ denotes the (negative) binding energy.\footnote{The convention here sets $E_n = 0$ as the rest mass energy of a pair of $\chi^0$s.
For bound states that do not have a $\chi^0\chi^0$ component, putting their constituents at infinite separation still leaves finite positive energy because of the charged/neutral mass-splitting, $\delta_0$.
Thus, in a capture process, the bound state carries off energy $A\delta_0 - (|E_n| + A\delta_0) = -|E_n|$, where $A$ is an integer that depends on the number of charged particles in the lightest component of the bound state.
The second term, $-(|E_n| + A\delta_0)$ corresponds to the ``ionization energy'' necessary to separate the bound state into its constituents.
We have neglected the subleading bound-state recoil kinetic energy.
In this way we see that the boson emitted in the capture process has an energy independent of $\delta_0$.}  
In particular, this process is forbidden when $\mZ$ exceeds the available energy ({\it i.e.} the kinetic energy of the incoming particles and the (absolute value of the) binding energy of the final state).

The emission of $W^\pm$ bosons is more complicated as it involves a different set of matrix elements and Feynman diagrams, corresponding to formation of bound states with unit charge.
In particular, when a $W$ boson is emitted from the potential, the $t$-channel propagator must be a mixed $W-Z$ or $W-\gamma$ propagator, which modifies the structure of the matrix element.
Performing the Fourier transform of the mixed propagator, we find that the appropriate replacement (compared to the photon-emission case where the propagator involves only $W$ bosons) is:
\begin{equation} 
e^{-m^{}_{\scaleto{W}{3.5pt}} r} \rightarrow \frac{2}{r^2 (\mW^2 - m_0^2)} \left[e^{-m^{}_{\scaleto{0}{3.5pt}} r}(1 + m_0 r) - e^{-m^{}_{\scaleto{W}{3.5pt}} r} (1 + \mW r) \right]\!,
\end{equation}
where $m_0 = \mZ$ or $0$ for the mixed $W-Z$ and mixed $W-\gamma$ propagator, respectively.
Since the diagrams with these two propagator structures are identical except for the propagators and the coupling of the $\gamma$ or $Z$ to the fermion line, the sum of their contributions can be captured by inserting the propagator factor:
\begin{equation}\begin{aligned}
\zeta(r) \equiv \frac{2}{r^2} &\left[ \frac{\cW^2}{\mZ^2 - \mW^2} \left(e^{-m^{}_{\scaleto{W}{3.5pt}} r} (1 + \mW r) - e^{-m^{}_{\scaleto{Z}{3.5pt}} r}(1 + \mZ r) \right) \right.\\
&\hspace{0.2cm}\left.+ \frac{\sW^2}{\mW^2} \left( 1- e^{-m^{}_{\scaleto{W}{3.5pt}} r} (1 + \mW r)\right) \right] \!.
\label{eq:propagatorfactor}
\end{aligned}\end{equation}

Now repeating the calculation from the photon case for the case where the emitted gauge boson is $W^1$ or $W^2$ instead of $W^3$, and inserting the $\zeta(r)$ factors in the terms corresponding to $W$ emission from the potential, we obtain the cross section for the $Q=1$ case (the $Q=-1$ case is identical):
\begin{align} 
\sigma v & = \frac{2 \alpha k}{\pi M_\chi^2} \int d\Omega_k \, \Bigg| \boldsymbol{\epsilon}(\hat{\mathbf{k}}) \cdot \int d^3 \mathbf{r}  \left[  \sqrt{\frac{3}{2}} \psi^*_{+0} \nabla \phi_{N} -  \frac{\sqrt{3}}{2} \psi^*_{+0} \nabla \phi_C \right. \nn \\
& \left. \left. + \frac{1}{\sqrt{2}} \psi^*_{++-} \nabla \phi_{C} - \frac{1}{\sqrt{2}}  \psi^*_{++-} \nabla \phi_{CC}  \right. \right. \label{eq:Q1capture}  \\
& \left. + \frac{1}{2} \hat{\mathbf{r}} \zeta(r) M_\chi \aW \left( \sqrt{3}  \psi_{+0}^* \phi_C + \sqrt{2} \psi_{++-}^* \phi_C + 2  \sqrt{2} \psi_{++-}^* \phi_{CC} \right) \right] \Bigg|^2. \nn
\end{align}
Here the $++-$ subscript denotes the component in the $\chi^{++} \chi^-$ state, and the $+0$ subscript denotes the component in the $\chi^+ \chi^0$ state, for the final $Q=1$ bound state.

Note that the phase-space factor $k$ for the outgoing $W$ boson must be modified to:
\begin{equation} 
k = \sqrt{\left( -E_n + \frac{M_\chi v^2}{4} \right)^2 - \mW^2}. 
\label{eq:Wphasespace} 
\end{equation}
As for $Z$ emission, this process is forbidden when $\mW$ exceeds the kinetic energy of the incoming particles + the binding energy of the final state.

The potential for the $Q=\pm1$ sector, needed to derive the wavefunctions for the bound states, is similarly given by:
\begin{equation} 
V(r)  = \alpha_W \begin{pmatrix} - 2 \left(\frac{s^2_{\scaleto{W}{3.pt}}}{r} + \frac{c^2_{\scaleto{W}{3.pt}} e^{-m^{}_{\scaleto{Z}{3.pt}} r}}{r} \right) & \sqrt{6}\, \frac{e^{-m^{}_{\scaleto{W}{3.pt}} r}}{r}\\  \sqrt{6}\, \frac{e^{-m^{}_{\scaleto{W}{3.pt}} r}}{r} & (-1)^{L+S} \,3\, \frac{e^{-m^{}_{\scaleto{W}{3.pt}} r}}{r} \end{pmatrix}\!,
\end{equation}
where the first row/column corresponds to the $\chi^{++} \chi^-$ state ($Q=+1$) or $\chi^{--}\chi^+$ state ($Q=-1$), and the second row/column corresponds to the $\chi^+ \chi^0$ state ($Q=+1$) or $\chi^- \chi^0$ state ($Q=-1$).
Again note that the off-diagonal terms disagree with Ref.~\cite{Cirelli:2007xd} by a sign; this is due to our choice of basis.

In principle there may also be $Q=\pm 2, 3, 4$ bound states in the spectrum, which can be accessed by a series of transitions involving emission of $W$ bosons. 
However, for the $Q=4$ case, the only available state is $\chi^{++} \chi^{++}$ (or $\chi^{--} \chi^{--}$ in the $Q=-4$ case), and the potential is a repulsive Coulomb potential mediated by $\gamma$ and $Z$ exchange, which does not support bound states.
For $Q= \pm3$, the only available two-particle states are $\chi^{++} \chi^+$ ($\chi^{--} \chi^-$), so again the potential is a scalar, and its value can be computed as
\begin{equation}
V(r) = \aW \left[ 2\, (-1)^{L+S}\, \frac{e^{-m^{}_{\scaleto{W}{3.5pt}} r}}{r} + 2 \left( \frac{\cW^2e^{-m^{}_{\scaleto{Z}{3.5pt}} r}}{r} + \frac{\sW^2}{r}\right) \right]\!.
\end{equation}
We observe that for $L+S$ even, this potential is always repulsive; for $L+S$ odd, the potential vanishes in the unbroken limit and in the broken regime a residual repulsive potential remains.
In either case, we do not expect bound states (this analysis also accords with the discussion in Ref.~\cite{Mitridate:2017izz}).

The $Q=\pm 2$ case is more interesting.
There are two relevant states: for $Q=2$ they are $\chi^+\chi^+$ and $\chi^{++}\chi^0$, with the former only being allowed for even $L+S$.
The potential then reads as follows,
\begin{equation}
V(r) = \aW \begin{pmatrix} \frac{s^2_{\scaleto{W}{3.pt}}}{r} + \frac{c^2_{\scaleto{W}{3.pt}} e^{-m^{}_{\scaleto{Z}{3.pt}} r}}{r} & 2\sqrt{3}\, \frac{e^{-m^{}_{\scaleto{W}{3.pt}} r}}{r} \\  2\sqrt{3}\, \frac{e^{-m^{}_{\scaleto{W}{3.pt}} r}}{r} & 0 \end{pmatrix}\!,
\end{equation}
for even $L+S$, where the 1st row/column corresponds to the $\chi^+\chi^+$ state and the 2nd row/column corresponds to the $\chi^{++}\chi^0$ state.
This potential has an attractive eigenvalue that can support bound states, asymptoting to $-3/r$ in the unbroken limit.
For odd $L+S$ only the $\chi^{++}\chi^0$ state exists, which experiences no potential.

Thus in addition to the $Q=0\leftrightarrow Q=\pm 1$ transitions through $W$ emission already considered, the only bound-bound transitions we need to compute involving higher-charge states are $Q=\pm 1 \leftrightarrow Q=\pm 2$ (proceeding via $W$ emission) and $Q=\pm 1 \leftrightarrow Q=\pm 1$,  $Q=\pm 2 \leftrightarrow Q=\pm 2$ transitions via photon or $Z$ emission.

\subsection{Capture rate results}

\begin{figure*}[!t]
\centering
\includegraphics[width=0.49\textwidth]{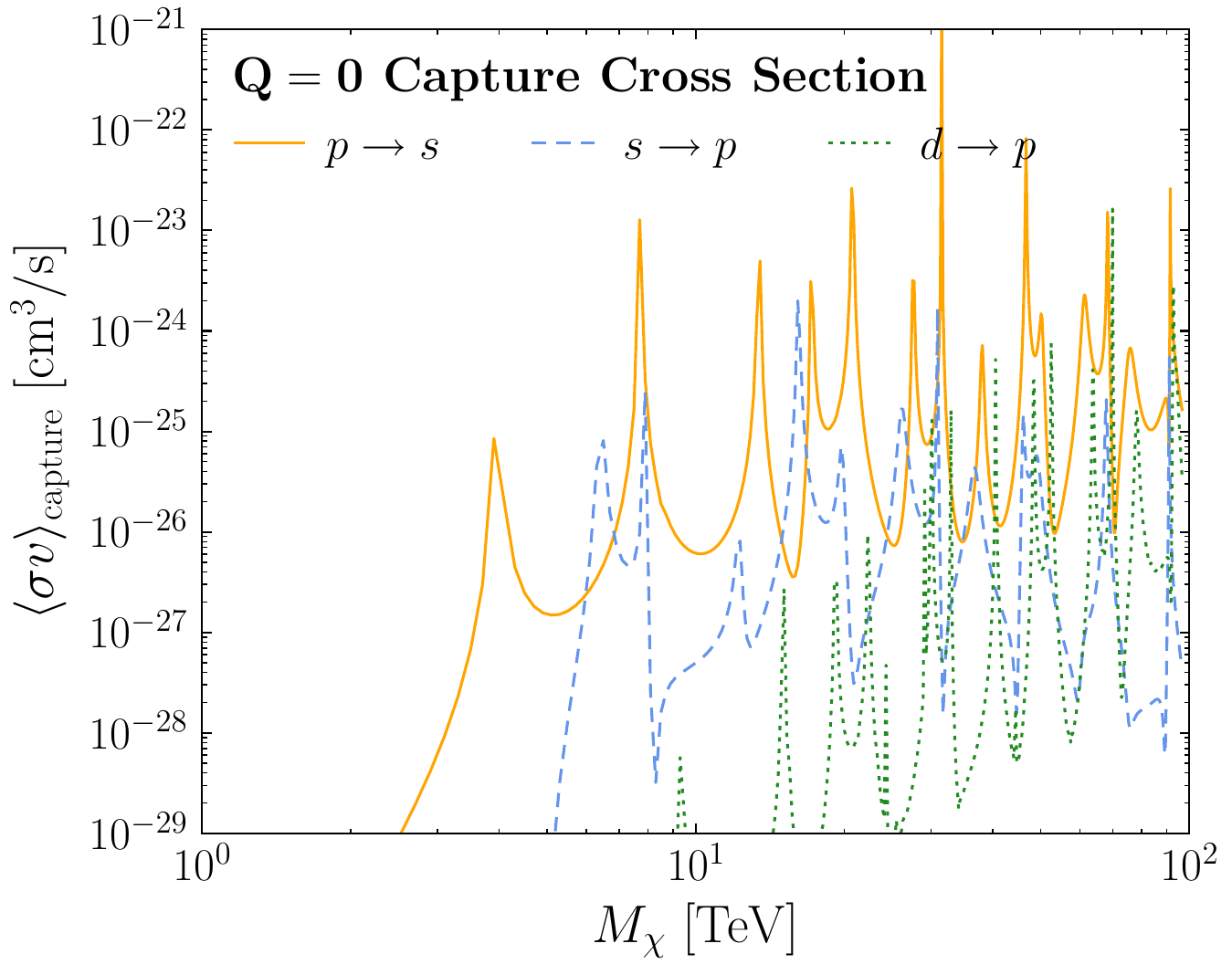}\hspace{0.1cm}
\includegraphics[width=0.49\textwidth]{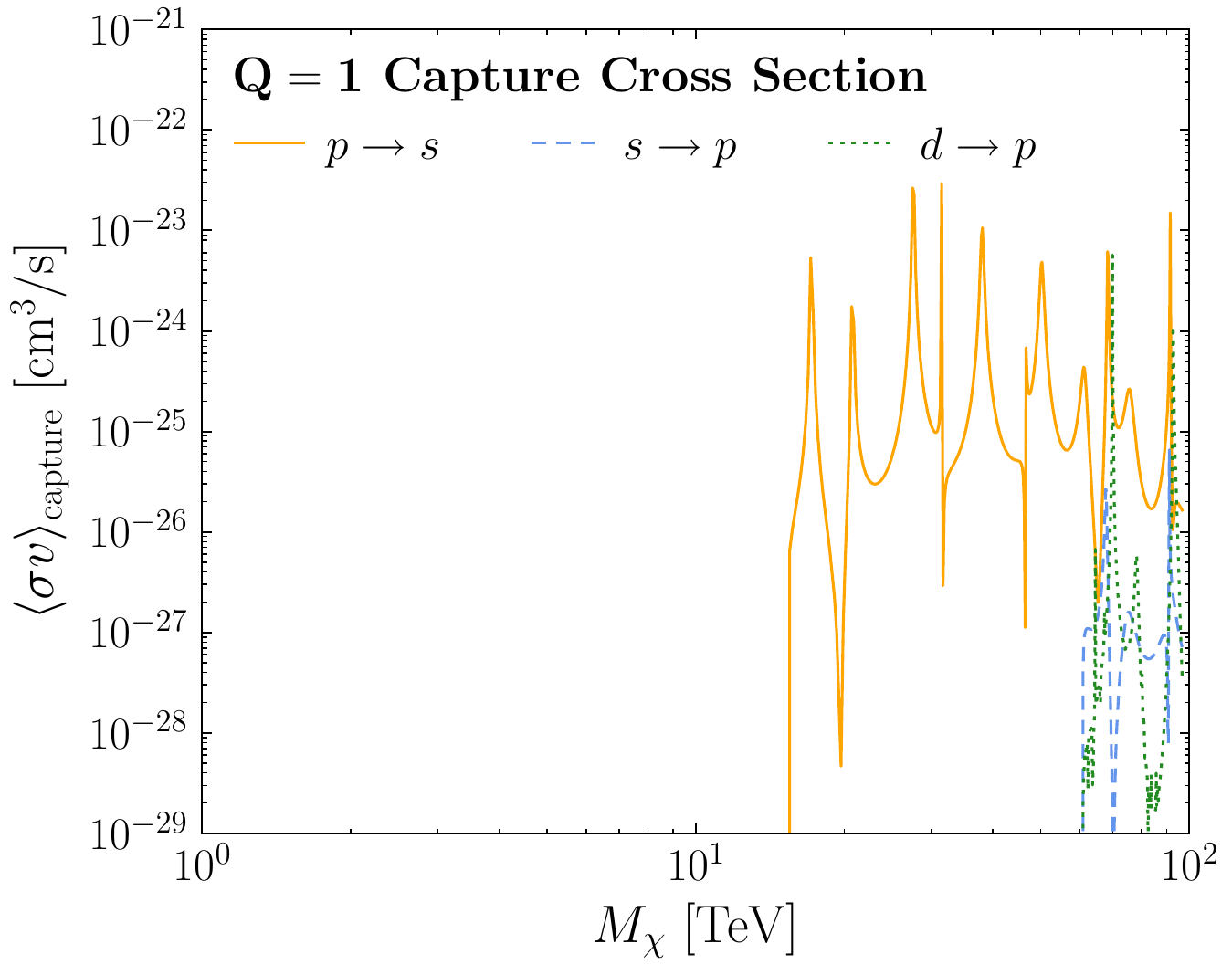} 
\caption{The capture cross section into bound states with $Q=0$ (left) and $Q=1$ (right), for three different transition: $p \to s$, $s \to p$, and $d \to p$.
These cross sections describe the total capture rate into all available states with these quantum numbers, not simply the most tightly bound.}
\label{fig:Capture-Xsec}
\end{figure*}

In Fig.~\ref{fig:Capture-Xsec}, we show examples of the formation cross-section for bound states with different quantum numbers, corresponding to capture from various initial partial waves.
Formation rates for $Q=1$ bound states become non-zero at masses high enough that the binding energies (plus the kinetic energy of the collision) exceed the $W$ boson mass.
We observe that at most mass points, the dominant capture rate is to $s$-wave bound states, corresponding to the $p$-wave ($L=1$) component of the initial state.

The overall size of the formation rate and its scaling with mass can be estimated analytically, as discussed in detail in App.~\ref{app:analytic}. To summarize, in the limit of high DM mass we expect the leading rates for bound-state formation and direct annihilation to take the form:
\begin{equation}
(\sigma v)_{\text{bsf}}^{n=1,L=0} \simeq \frac{700\pi^2\aW^3}{M_\chi^2 v},\hspace{0.5cm}
(\sigma v)_{\text{ann}} \simeq \frac{720 \pi^2 \aW^3}{M_\chi^2 v}.
\label{eq:analyticleadingxsecs}
\end{equation}
For $M_\chi v \lesssim \mW$, we expect the $p\rightarrow s$ capture cross section to experience a velocity suppression due to the $p$-wave initial state, which is parametrically of order $(M_\chi v/\mW)^2$.

\begin{figure*}[!t]
\centering
\includegraphics[width=0.49\textwidth]{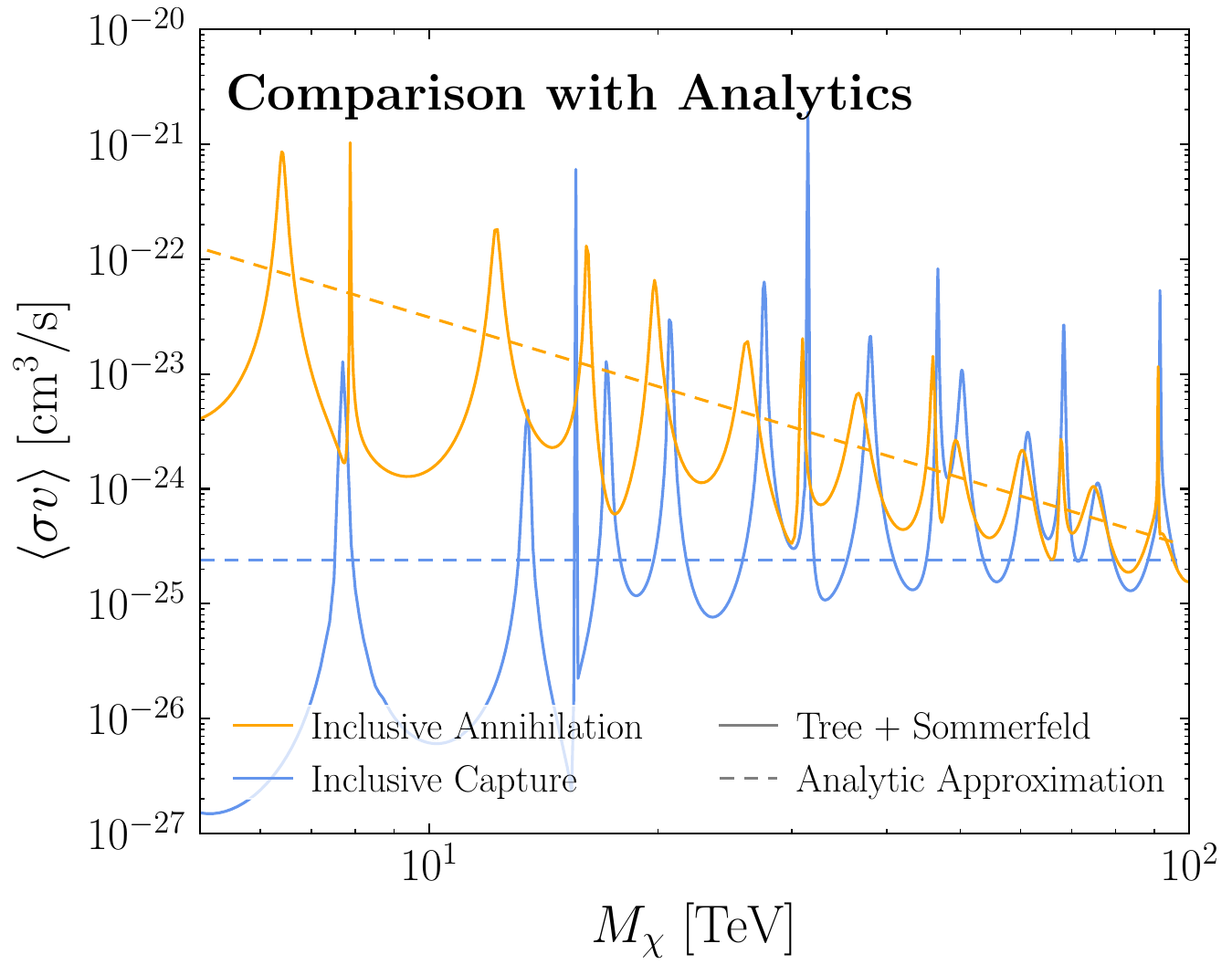}
\caption{Bound-state formation rate for the spin-triplet ground state (blue) compared to the inclusive annihilation cross section (orange).
Dashed lines indicate analytic estimates for the corresponding rates; see text for details.
}
\label{fig:analytic-comparison}
\end{figure*}

In Fig.~\ref{fig:analytic-comparison}, we compare the dominant $p\rightarrow s$ bound-state formation rate for capture into the ground state with the inclusive direct annihilation rate; the latter is computed including the Sommerfeld enhancement but without any SCET corrections.
We overplot the analytic estimates given in Eq.~\eqref{eq:analyticleadingxsecs}, with a $p$-wave correction factor of $(M_\chi v/\mW)^2$ for the estimate corresponding to the bound-state formation rate.
We observe that at the thermal quintuplet mass (13.6 TeV), we expect the direct annihilation to dominate due to the $p$-wave suppression of the leading bound-state formation channel, but this suppression is lifted at high masses; furthermore, even at lower masses the bound-state capture rate may exceed the rate for direct annihilation at specific mass values (such as at the peak near 13.5 TeV).
However, recall that these are inclusive rates; to understand the relative contributions to the line and endpoint spectrum, we must now understand how the bound states eventually annihilate to SM particles.

\section{Bound State Annihilation}
\label{sec:BSA}

Having computed all the relevant bound-state formation rates, the second part of our calculation involves determining the differential branching ratios for bound states to decay, producing hard photons.
To compute these we need the differential decay rate of the bound state to a final state including a photon, as well as its total decay rate to all SM particles.
For the differential decay rate, we can recycle our EFT developed for direct annihilation as described in Sec.~\ref{sec:DA}.
The factorized form of the differential cross section remains identical to Eq.~\eqref{eq:Som} which we reproduce here for convenience,
\begin{equation}
\frac{d\sigma}{dz} = \sum_{a'b'ab} F_{\chi}^{a'b'ab}\frac{d\hat{\sigma}^{a'b'ab}}{dz}.
\label{eq:quinsep}
\end{equation}

To apply this expression to bound states, we will need to update the initial state wavefunctions encoded in $F_{\chi}$.
For direct annihilation, $F_{\chi}$ as given in the second line of Eq.~\eqref{eq:Som}, described an initial state of two free DM particles in the $s$-wave spin-singlet configuration.
Here, however, our initial state is described by the two-body bound state wavefunctions computed in Sec.~\ref{sec:BSF}.
The bound states can be classified according to their value of total orbital angular momentum $L$, total spin $S$, and charge $Q$, and we will need to track all bound states at a given mass.
Beyond this, however, the differential cross section, $d\hat{\sigma}/dz$, is again given by Eq.~\eqref{eq:Fact}, and each of the associated objects such as the jet and soft functions are identical to those used in the direct annihilation computation of Sec.~\ref{sec:DA}; this is the advantage of the EFT approach, the infrared (IR) physics is identical for direct and bound-state annihilation.
To compute the decay rate, one needs to simply alter the details of the initial state, such as overall kinematic factors and a modified form of $F_{\chi}$.
Nevertheless, as our interest is in the branching ratio of bound states to various decay rates, we will be computing ratios and will find the kinematic differences cancel (see Sec.~\ref{sec:inc}), further increasing the similarity to the direct annihilation computation.
We note, however, that to fully describe the possible end state of all bound states in the quintuplet spectrum, we would need to include additional operators in hard matching beyond the single operator we used for the direct annihilation given in Eq.~\eqref{eq:hardscatteringoperator}.
That operator described the annihilation of an $L=S=Q=0$ initial state, so for the annihilation from states with $L > 0$, $Q > 0$, or $S=1$, a new set of operators is required.
Nevertheless, we will show that the contributions of the $L > 0$ and $S=1$ states to the endpoint spectrum are suppressed, and the contributions from $Q > 0$ states can be captured within our existing framework, so that in fact the form of $d\hat{\sigma}/dz$ we have already computed is sufficient.
Formalizing the logic above, the decay cross section into the hard photon can be written as 
\begin{equation}
\frac{d\sigma}{dz}\Big|_\text{bound} = \sum_B \sigma(\chi^0 \chi^0 \rightarrow  B+ X_{\text{us}} ) \frac{1}{\Gamma_{B}} \frac{d\Gamma_{B\rightarrow \gamma+X}}{dz},
\label{eq:Br}
\end{equation}
where $\sigma(\chi^0 \chi^0 \rightarrow  B+ X_{\text{us}})$ is the total production cross section for the $L=0$, $S=0$ state including any decays from shallower bound states and $\Gamma_B$ is the decay rate into all possible SM particles.

\begin{figure}
\centering
\hspace*{2.5cm}\includegraphics[width=0.9\linewidth]{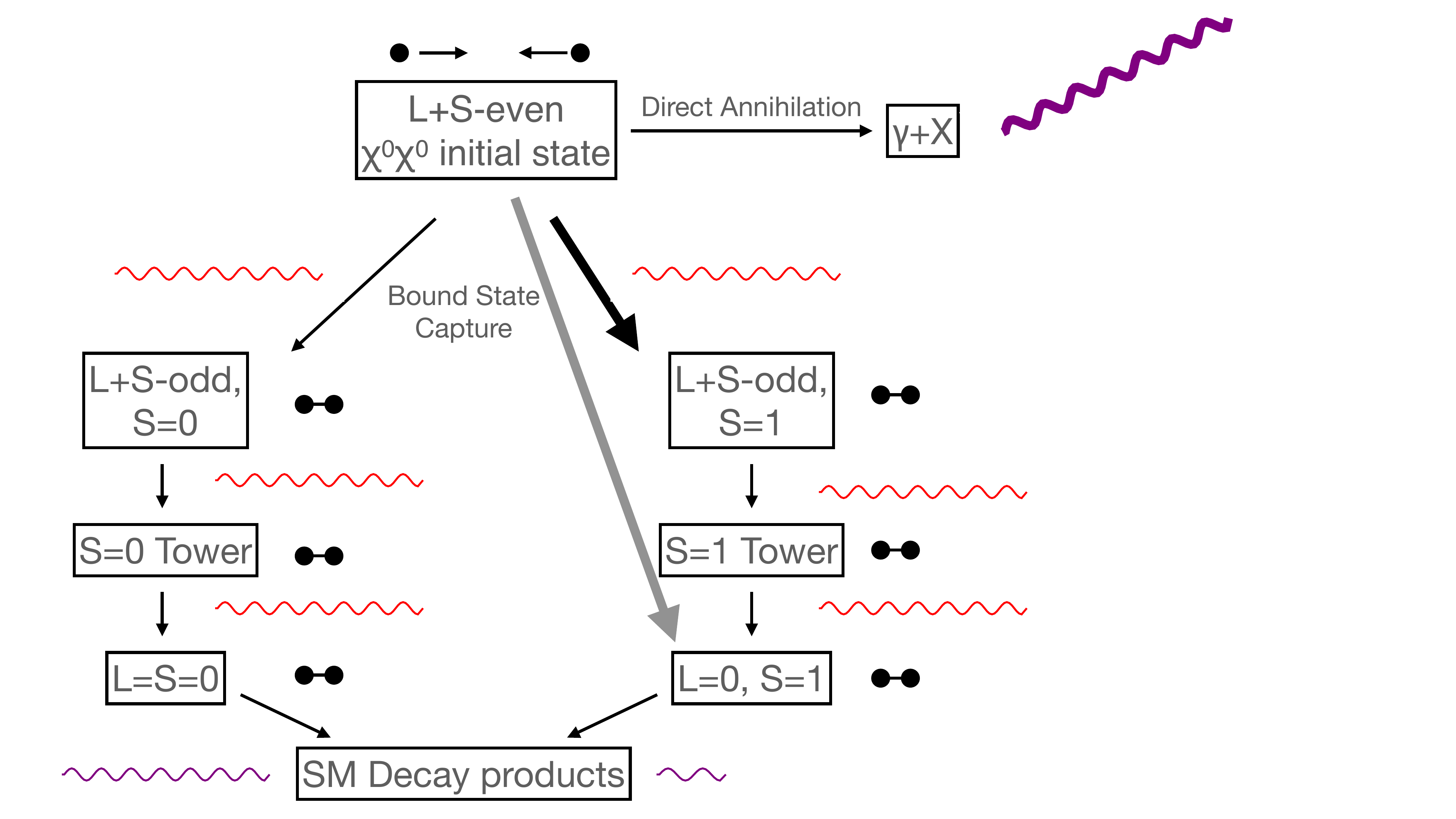}
\caption{Schematic diagram that shows how two initial quintuplet particles evolve into various final states, as we work out in detail in this section.
Endpoint photons are violet, whereas the soft photons that arise in the dipole transitions involving bound states are red.
Capturing to quintuplet bound-states can give an additional source of endpoint photons.
In this work, we include those that result from the ``$S\!=\!0$ Tower.''
In general, their contribution is suppressed compared to those from direct annihilation.
As we discuss in Sec.~\ref{sec:slpop}, there is an enhanced capture rate to the ``$S=1$ Tower,'' including direct capture to the lowest-lying $L\!=\!0,\,S\!=\!1$ state, indicated by the thicker arrows.
However, the decay to endpoint photons from this tower is power suppressed compared to $S\!=\!0$ (hence the smaller endpoint photon on the right).
This combination turns out to balance such that both towers give parametrically similar contributions.
However, since this is a subleading overall component, only  ``$S\!=\!0$ Tower'' endpoint photons are included in our analysis.}
\label{fig:cascadeschematic}
\end{figure}

We sketch the structure of the contributions to the endpoint spectrum from bound-state formation in Fig.~\ref{fig:cascadeschematic}, and work out the required ingredients in the rest of this section.
We begin in Sec.~\ref{sec:bsdecay} by understanding the general structure of the decay cascade that follows capture into an excited state, arguing that excited states will typically decay (possibly via multiple steps) to an $L=0$ state before annihilating to SM particles.
In Sec.~\ref{sec:OP} we study the operators through which $L=0$ bound states can annihilate to SM particles, and show that the contribution to the endpoint spectrum from $S=1$ bound states is power-suppressed; thus our endpoint calculation focuses on annihilation from $L=S=0$ states.
In Sec.~\ref{sec:WaveFn} we discuss how to compute the wavefunction factors needed to obtain the photon endpoint spectrum from decay (to SM particles) of a given $L=S=0$ bound state. In Sec.~\ref{sec:inc} we compute the inclusive rate for decay via annihilation into SM particles for $L=S=0$ states, while in Sec.~\ref{sec:BSD} we describe how to calculate the rates for decay into lower-lying bound states, for bound states of arbitrary $L$, $S$.
Key points of the calculation and several results are presented in Sec.~\ref{sec:bsdecaysummary}.
Ultimately, we employ these rates to compute the overall endpoint annihilation signal from decay of all $L=S=0$ states to SM particles, taking into account the possibility to populate these states by decay from all shallower $L=0,1,2$ states, as encapsulated in Eq.~\eqref{eq:Br}.

\subsection{The decay cascade}
\label{sec:bsdecay}

Starting with an initial DM pair with a specific mass, we can expect capture into a number of metastable bound states characterized by their total orbital angular momentum $L$, spin $S$, and charge $Q$.
These bound states have the option of either decaying into more tightly bound states with the same total spin but with $|\Delta L|=1$ (in a single decay step), or annihilating directly into SM particles.\footnote{Transitions where there is a change of spin or $L$ by more than one unit are allowed, but suppressed, and we ignore them in this work; see {\it e.g.} Ref.~\cite{Johnson:2016sjs}.}
To compute the final annihilation spectrum into photons, we therefore need to know the branching ratios for various annihilation and decay channels.
The decay of a shallow state into a low-lying ``stable'' state (by which we mean stable against decay to other bound states) may happen via several intermediate decay steps with their own branching ratios.
We are therefore required to implement this cascade of decays to obtain the effective production cross section for a specific ``stable'' bound state, so we can then compute the signal from its subsequent annihilation into SM particles producing a hard photon.

Determination of the full decay cascade requires three ingredients for the spectrum from bound states at a given DM mass: 1. the direct capture cross-section into all bound states; 2. the decay rate from each initial state to all more deeply-bound states; and 3. the rate for direct annihilation into SM particles.
The first of these ingredients proceeds as discussed in Sec.~\ref{sec:BSF}.
What remains to be computed is then the competition between the decay of one bound state to a deeper one, versus direct annihilation into the SM.\footnote{Our discussion of this competition follows similar arguments in the literature, {\it e.g.} Refs.~\cite{An:2016gad, Asadi:2016ybp}.}
We will determine that in this section.

Before doing so, we can already provide an analytic estimate.
The bare cross section for free electroweak DM particles to annihilate to the SM scales parametrically as $\sigma v \propto v^{2 L} \aW^2/M_\chi^2$, where $L$ is the orbital angular momentum of the two-body initial state.
The equivalent decay rate of a bound state to SM particles is related to this expression by replacing an incoming plane-wave wavefunction with the bound-state wavefunction.
We can parametrically estimate this with two steps.
Firstly, we replace $v \rightarrow \aW$ as the characteristic momentum associated with the potential is $p\sim \aW M_\chi$ and $v=p/M_\chi$ in the non-relativistic limit.
Secondly, we must account for a multiplicative factor of $(\aW M_\chi)^3$, which arises from the square of the bound-state wavefunction.
(In more detail, as the wavefunctions are normalized by $\int d^3\mathbf{r}\, |\psi|^2 = 1$ and have support over the Bohr radius, $a_0$, their characteristic value is $|\psi|^2 \sim 1/a_0^3 = (\aW M_\chi)^3$.)
Thus we expect the decay rate of a bound state with orbital angular momentum $L$ to SM particles to scale approximately as $\Gamma \propto (\aW^2/M_\chi^2) \aW^{2 L} (\aW M_\chi)^3 = \aW^{5+ 2L} M_\chi$. In contrast, a dipole-mediated decay to a lower-lying bound state scales as $\alpha\, \aW^4 M_\chi$ independent of $L$; consequently, if such decays are allowed, they will generally dominate over annihilation to SM particles for $L > 0$.
This argument suggests that unless dipole transitions to lower-lying states are forbidden, states with $L > 0$ will preferentially decay to $L=0$ states before annihilating to the SM.  

One might then ask whether the spectrum contains $L>0$ states that have no allowed dipole transitions to more deeply bound states. For such states, the branching ratio for decay to SM particles via annihilation might indeed be important.
However, we argue in App.~\ref{app:analytic} that this can only occur for very high-$L$ states (beyond the range we consider in this work) for which the formation rate is likely to be negligible.
We will thus assume that these states can be neglected, and restrict to $L=0$ states when computing the endpoint photon spectrum from bound state decay.
Similarly, in principle there can be stable $Q=\pm 1,2$, $L=S=0$ states in the spectrum.
However, we find that the branching ratio to these states is very small (sub-percent) compared to the $L=S=Q=0$ states, and so we neglect them in computing the endpoint photon spectrum.
(In fact, at lower masses the charged bound state contribution will be exactly zero when either there is no charged bound state in the spectrum, or when those available cannot be accessed due to insufficient energy to produce an on-shell $W$.)

\subsection{Operators for bound state decay}
\label{sec:OP}

\subsubsection{Leading power operators}

In the direct-annihilation case, we expect $s$-wave annihilation ($L=0$) to dominate.
However, in order to support the annihilation of bound states with higher angular momentum we need operators that are suppressed by powers of the DM velocity.
To see which will contribute, we consider the various structures that arise from a tree-level matching calculation.
The tree-level amplitude for annihilation to a final state $\gamma+X$ has contributions from $s$-, $t$- and $u$-channel diagrams, which give the following leading-order operator when expanded to $\mathcal{O}(v)$
\begin{equation}
{\cal O} = \left( \chi_v^T i \sigma_2 \left\{ T_{\chi}^d,\,T_{\chi}^c \right\} \chi_v \right) \left( {\cal B}_{\perp n}^{ic} {\cal B}_{\perp \bar{n}}^{jd} \right) i \epsilon^{ijk} (n-\bar{n})^k.
\label{eq:boundgoperator}
\end{equation}
As already mentioned, this operator is identical to that used for direct annihilation in Eq.~\eqref{eq:hardscatteringoperator} and supports a bound state with $L=S=0$ (and both are given before a BPS field redefinition).

As we will see, this is the primary operator required for computing the  dominant contribution to the end-point spectrum from bound state annihilation.
There is no operator at this order that supports $S=1,\,L=0$ bound state annihilation to gauge bosons at tree level.
Such a state can annihilate to fermions and Higgs final states at tree level, but its contribution to end-point photons via bremsstrahlung is power suppressed in our EFT.
These bound states however, can contribute substantially to the soft photon spectrum, as we will consider in Sec.~\ref{sec:Numerics}.

For higher-$L$ bound states with $S=0$, there is a competition between decay to a state with lower $L$ as compared with direct annihilation to SM particles. 
However, as discussed in Sec.~\ref{sec:bsdecay}, the decay to lower-$L$ bound states always wins out for $L > 0$, so that only the decays of $L=S=0$ states to SM particles remain relevant and the only operator we need is given in Eq.~\eqref{eq:boundgoperator}.
Nevertheless, for completeness we provide the subleading operators in App.~\ref{app:HigherL}.
At the same time, there is no interference between the direct and bound state channels so that we may treat these cross sections separately.
This is discussed in detail in App.~\ref{app:UnstableET} from an EFT perspective.
If the widths of the bound states are parametrically much smaller than the separation in their energy, then we may also safely neglect any interference between the various bound state channels.
This is essentially the narrow-width approximation.
Accordingly, to determine the total spectrum it will suffice to sum over the cross sections for the direct channels and the allowed bound state channels individually.

\subsubsection{Sub-leading power operators}
\label{sec:slpop}

As we saw in the previous section, there is no operator at leading power which supports an $S=1,\,L=0$ bound state annihilation into a hard photon.
The only operators that support an $S=1$ bound state are those which describe annihilation to fermions or scalars via an $s$-channel process.
We can conceive of a hard photon emission from the final state SM Higgs or fermions; however, this is power suppressed by our SCET power counting parameter $\lambda^{1/2}$ (where $\lambda = 1-z$) at the amplitude level.
We can see this explicitly by looking at the emission amplitude of a hard (collinear) photon off a collinear fermion in the final state as in the diagram below 
\begin{center}
\includegraphics[width=0.3\linewidth]{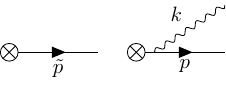}
\end{center}
The matrix element is then,
\begin{equation}\begin{aligned}
\mathcal{M} &= \bar{u}(\tilde p)+ ie\bar{u}(p)\gamma^{\mu}\epsilon^\mu(k)\frac{i(\slashed{p}+\slashed{k})}{(p+k)^2+i\epsilon}\\
&\simeq \bar{u}(\tilde p)-e \bar{u}(p) \gamma^{\mu}_{\perp}\epsilon^{\mu}(k)\frac{\frac{\slashed{n}}{2}}{n\cdot p} =\langle \tilde p|f_n|0\rangle -e \langle k, p|A_{n,\mu \perp}\frac{1}{n\cdot \mathcal{P}}\bar{f}_s \gamma^{\mu}_{\perp}\frac{\slashed{n}}{2}|0 \rangle.
\end{aligned}\end{equation}
where the first term is the tree level diagram and the second term comes from a single photon emission.
By the power counting of SCET, we see that the collinear fields scale as $A_{n,\mu \perp} , f_n\sim \lambda$, where $\lambda$ is the expansion parameter of our EFT.
The soft fermion field scales as $f_s  \sim \lambda^{3/2}$ while the soft momentum $p$ scales as $n \cdot p \sim \lambda$.
This means that compared to the tree level, the hard photon emission is suppressed by a power $\lambda^{1/2}$. If the fermion is ultra-soft, the suppression is enhanced to $\lambda$.
Further discussion of this operator is provided in App.~\ref{app:HigherL}.

Including this operator in our analysis would be justified only if the production cross section for this channel compensates for the $\lambda$ suppression (at the amplitude squared level), in order to be comparable to the $S=0$ channel. Indeed this turns out to be true based on numerical calculations (see also App.~\ref{app:analytic} for an analytic estimate and discussion)  and therefore, in principle we also need to include this sub-leading operator, in order to accurately compute the bound state contribution to the endpoint spectrum.
However, a  numerical analysis tells us that the leading $S=0$ bound state channel, which will be the focus of the next subsection, is only a few percent of the direct annihilation cross section in terms of the contribution to the endpoint.
The $S=1$ bound-state contribution {\it is} power-suppressed relative to direct annihilation; it is only appreciable compared to the leading-power term from $S=0$ bound states (whose formation rate is suppressed, by a different mechanism).
So given the relative overall unimportance of the bound state contribution to the endpoint, we will not include the contribution from the $S=1$ bound states here; a more precise calculation would need to account for this channel.

\subsection{Wavefunction factors for bound state annihilation}
\label{sec:WaveFn} 

In this section we compute the wavefunction factors $F^{aba'b'}$ relevant for the $L=S=0$ bound  states, which are needed to obtain the endpoint spectrum from the bound states' annihilation to SM particles. The  definition of these factors remains identical to the case of direct annihilation given in Eq.~\eqref{eq:Som}, but now the operator is sandwiched between the $L=S=0$ bound states.
\begin{equation}
F_{\chi}^{a'b'ab} = \Big\langle B \Big\vert \left(\chi_v^{T} i \sigma_2 \left\{ T_{\chi}^{a'},\,T_{\chi}^{b'} \right\} \chi_v\right)^{\dagger} \Big\vert 0 \Big\rangle \Big\langle 0 \Big\vert \left(\chi^{T}_v i \sigma_2 \left\{ T_{\chi}^a,\,T_{\chi}^b \right\} \chi_v \right) \Big\vert B \Big\rangle.
\end{equation}
As for the case of direct annihilation,	the wavefunction factors will be evaluated at the IR scale of our EFT, which is the electroweak scale.
At this scale, electroweak gauge symmetry is broken and the bound state wavefunctions are computed in terms of the broken eigenstates as in Eq.~\eqref{eq:s00s0pms0pmpm-quintuplet}, but now for the bound state.

Once we have the bound state analogues of Eq.~\eqref{eq:s00s0pms0pmpm-quintuplet} in the broken basis, the next step is to relate these to the bound state wavefunction determined in Sec.~\ref{sec:BSF}.
We start with the momentum-space representation of the bound state.
In general for a two-particle bound state, we can express the bound state (which is an eigenfunction of the Hamiltonian) as
\begin{equation}
|\mathbf{P}\rangle  = \int \frac{d^3\mathbf{k}}{(2\pi)^3 \sqrt{M_{\chi}}} \phi_P(\mathbf{k} )c^{\dagger}\left( \frac{\mathbf{P}}{2}+\mathbf{k},\lambda\right)d^{\dagger} \left(\frac{\mathbf{P}}{2}-\mathbf{k} ,\lambda'\right)|0\rangle, 
\end{equation}
where $\mathbf{P}$ is the momentum of the bound state, while $2\mathbf{k}$ is the relative 3 momentum of the 2 particles making up the bound state.
Further, $\phi_P(\mathbf{k})$ is the momentum space bound state wavefunction whereas $c^{\dagger}$, $d^{\dagger}$ are creation operators for the two constituents of the bound state.

The operators that we have in SCET are bilinear local operators with non-trivial Dirac structures.
Based on the definition above we can evaluate the overlap of the bilinear operators with the bound state for the $s$-wave states.
After SU(2) breaking, a general matrix element of a bound state will take the form (working in four-component notation for the moment and suppressing the color structure),
\begin{equation}
\langle 0|\bar \chi \gamma^0\gamma^5 \chi |B\rangle = \int \frac{d^3\mathbf{k}}{(2\pi)^3 \sqrt{M_{\chi}}}\phi_{P}(\mathbf{k} ) \bar v_s( \mathbf{P}/2+ \mathbf{k}) \gamma^0\gamma^5 u_r(\mathbf{P}/2- \mathbf{k}).
\end{equation}
Working in the bound state rest frame, we can expand this result to leading order in velocity, 
\begin{equation}
\langle 0|\bar \chi \gamma^0\gamma^5 \chi| B\rangle =\frac{\bar v_s(M_{\chi}) \gamma^0\gamma^5  u_r( M_{\chi})}{\sqrt{M_{\chi}}}\psi^{B}(0) = - 2\sqrt{M_{\chi}}\eta_s^{\dagger}\xi_r\psi^B(0).
\end{equation}
where $\psi^B(\mathbf{x})$ is the position space analogue of $\phi_P(\mathbf{k})$, and we have introduced basis spinors $\eta_s$ and $\xi_r$ according to,
\begin{equation}
\bar v_s(M_{\chi}) \gamma^0\gamma^5  u_r( M_{\chi}) =-2M_{\chi}\eta_s^{\dagger}\xi_r.
\end{equation}
For a bound state in the spin singlet configuration, we can evaluate the basis spinors explicitly, and we are left with,
\bea
\langle 0|\bar \chi \gamma^0\gamma^5 \chi |B\rangle = 2\sqrt{2M_{\chi}}\psi^B(0).
\eea

For the computation at hand, we need to restore the color structure, which amounts to defining the bound state analogues of Eq.~\eqref{eq:s00s0pms0pmpm-quintuplet} which we used for direct annihilation.
As our bound state is neutral, again there are only three objects to define
\begin{equation}\begin{aligned}
\Big\langle 0 \Big\vert (\chi^{0T}_v i \sigma_2 \chi_v^0) \Big\vert B \Big\rangle
=\, &4\sqrt{M_{\chi}}\psi^B_0, \\
\Big\langle 0 \Big\vert (\chi^{+T}_v i \sigma_2 \chi_v^-) \Big\vert B \Big\rangle
=\, &2\sqrt{2M_{\chi}}\psi^B_{\pm}, \\
\Big\langle 0 \Big\vert (\chi^{++T}_v i \sigma_2 \chi_v^{--}) \Big\vert B \Big\rangle
=\, &2\sqrt{2M_{\chi}}\psi^B_{\pm\pm},
\label{eq:s00s0pms0pmpm-quintuplet-BS}
\end{aligned}\end{equation}
where each of the bound state wavefunctions is the appropriate expression for that transition evaluated at the origin.
We can then evaluate the wavefunction factors for the $s$-wave spin-0 bound state as follows,
\begin{equation}\begin{aligned}
F_{\chi}^{aabb} &= 144 \, (8M_{\chi}) \left|2\psi^B_{\pm \pm}+2\psi^B_{\pm}+\sqrt{2}\psi^B_0\right|^2\!, \\
F_{\chi}^{a3a3} &= 16\, (8M_{\chi}) \left| 4\psi^B_{\pm\pm}+\psi^B_{\pm} \right|^2\!,\\
F_{\chi}^{aa33} &= 48\, (8M_{\chi}) \left( 4\psi^B_{\pm\pm}+\psi^B_{\pm} \right)\left(  2 \psi^B_{\pm \pm}+2\psi^B_{\pm}+\sqrt{2}\psi^B_0\right)^*\!, \\
F_{\chi}^{abab} &= 16 \, (8M_{\chi}) \left| 4\psi^B_{\pm\pm}+\psi^B_{\pm} \right|^2 + 8\, (8M_{\chi})  \left|2 \psi^B_{\pm\pm}+5 \psi^B_{\pm}+ 3\sqrt{2}\psi^B_0 \right|^2\!.
\end{aligned}\end{equation}
Note, other than the different definition for the bound state wavefunctions versus the Sommerfeld factors, Eq.~\eqref{eq:s00s0pms0pmpm-quintuplet-BS} versus Eq.~\eqref{eq:s00s0pms0pmpm-quintuplet}, these results are identical to those for the direct annihilation used to determine Eqs.~\eqref{eq:winocontract1} and \eqref{eq:winocontract2}.

\subsection{Bound state decay rate into SM particles}
\label{sec:inc}

In this subsection we compute rates for $L=S=0$ bound states to decay into the SM.
Because these states decay to gauge bosons, we can recast our previous results for photon emission through the $L=S=0$ operator to obtain both the inclusive cross section and the differential branching ratio to photons.
Taken together, these results will allow us to compute the hard photon spectrum from bound states, as dictated by Eq.~\eqref{eq:Br}.
For the total (inclusive) decay rate for a given state, it suffices to look at the tree level cross section, since (as we explain next) there are no large logarithms induced due to loop corrections.

According to the KLN (Kinoshita-Lee-Nauenberg) theorem~\cite{Kinoshita:1962ur,Lee:1964is}, for non-abelian gauge theories, the cross section is IR finite only if we perform a sufficiently inclusive sum over both initial and final states.
Since our initial states have a specific SU(2) color, one might expect to find IR divergences in the cross section when computing electroweak corrections.
Since the electroweak symmetry is broken, these IR divergences should manifest themselves in the form of large logarithms $\ln(M_{\chi}/\mW)$.
Most of the examples which demonstrate this violation of the KLN theorem in the literature consider light-like initial state particles as opposed to heavy time-like momenta considered in this paper.
As we shall see, this is the key difference which removes the presence of large logarithms in inclusive cross sections for the case of heavy particle annihilation.
Here, by heavy, we mean that the mass of the initial particle is of the same order as the hard scale in the EFT.

To demonstrate this explicitly, consider again the case of the wino but now imagine the wino to be a much lighter particle (mass $\sim$ electroweak scale) with a TeV scale (hard scale) energy. 
When we match the full theory onto an effective operator, the operator basis obtained after a tree level matching is fairly simple and reduces to
\begin{equation}
\begin{aligned}
O_r&=\left( \chi_{n}^T i \sigma_2 \left\{ T_{\chi}^a,\,T_{\chi}^b \right\} \chi_{\bar{n}} \right) \left( Y_r^{abcd} {\cal B}_{\perp n^{'}}^{ic} {\cal B}_{\perp \bar{n}^{'}}^{jd} \right) i \epsilon^{ijk} (n^{'}-\bar{n}^{'})^k,
\end{aligned}\end{equation}
with
\begin{equation}
Y_1^{abcd} = Y_n^{ea}Y_{\bar n}^{eb} Y_{n'}^{fc}Y_{\bar n'}^{fd}, \hspace{0.5cm} 
Y_2^{abcd}= Y_n^{ea}Y_{\bar n}^{fb}Y_{n'}^{ec}Y_{\bar n'}^{fd}, \hspace{0.5cm} 
Y_3^{abcd}= Y_n^{ea}Y_{\bar n}^{fb}Y_{n'}^{fc}Y_{\bar n'}^{ed}.
\end{equation}
Note that in this case we have three soft functions in place of two since we can distinguish between the directions of the initial states.
We are  considering the inclusive case so that we sum over the colors of the final state gauge bosons. 
We can now look at the soft operators that we get at the amplitude squared level.
As a single example, we can consider the interference term between $Y_2$ and $Y_3$, which will contain $Y_n^{ea}Y_{\bar n}^{fb}Y_n^{fa}Y_{\bar n}^{eb}$.
It is then clear that the soft Wilson lines do not cancel out due to the distinction between the n and $\bar n$ directions, if the initial state colors, $a,b$, are not summed over.
On the other hand, if instead $n= \bar{n} = v$, then the Wilson lines would cancel.
This will render the soft function trivial and hence no Sudakov logs from KLN violation exist in this case.
This implies that at NLL accuracy, for computing the inclusive cross section, we only need to consider the inclusive tree level cross section.

For the $L=S=0$ operator, we need the differential branching ratio to the photon, in addition to the inclusive cross section.
The differential decay rate is given by the following factorized formula which takes the same form as that of the differential cross section in Eq.~\eqref{eq:Fact}
\begin{equation}\begin{aligned}
\frac{d\Gamma}{dz} =\, &A_0 \int \frac{d \Omega_\gamma}{4\pi}\, F_{\chi}^{a'b'ab} HJ_{\gamma} J_{\bar n}  S H_{\bar n} \otimes H_S \otimes C_S,
\label{eq:diff}
\end{aligned}\end{equation}
where $A_0$ is an overall kinematic factor.
We do not need to know its explicit form since it will cancel out in the branching ratio.

To compute the inclusive cross section, we can make use of the stage 1 EFT given in Eq.~\eqref{eq:stage1} with all the functions evaluated to tree level
\begin{equation}\begin{aligned}
\left[\frac{d\Gamma}{dz}\right]_{\text{Stage 1}} =\, &A_0 \int \frac{d \Omega_\gamma}{4\pi}\, F_{\chi}^{a'b'ab} J_{\gamma} \int \frac{dk^+}{2\pi}\, J_{\bar{n}}(k^+) \\
\times &\int\frac{dq^+}{2\pi}\, \left(\sum_{i=1}^{4}H_{ij}S_{ij}^{'\,a'b'ab}(q^+)\right)\delta(2M_{\chi}(1-z)-k^+-q^+),
\end{aligned}\end{equation}
where we have explicitly written out the convolution, and $d\Omega_{\gamma}$ is an integral over the outgoing direction of the photon.
This form is sufficient for computing the inclusive cross section, since we as explained earlier in this section, we do not have to resum any logs.

An identical approach can be used to determine the inclusive decay rate of the $L=S=0$ state, which at tree level decays purely to gauge bosons.
Only several small alterations are required between the semi-inclusive ($\gamma + X$ final state) versus inclusive cross section.
We adjust the wavefunction factors contracted into the soft function, replace $J_{\gamma}$ by $J_n$ to allow for any final state gauge boson (we are no longer requiring a photon in the final state), and we remove the restriction on the phase space (as there is not observed endpoint photons) and therefore integrate over the full phase space.
Further, we only need the various functions at tree level, and so we use
\begin{equation}
J_{\gamma} = 1,\hspace{0.5cm}
J_{\bar n}(k^+) = \delta(2M_{\chi} k^+),\hspace{0.5cm}
S^{'\,a'b'ab}(q^+)  = \delta(q^+)\delta^{aa'}\delta^{bb'}.
\label{eq:JJStree}
\end{equation}
For the inclusive case, we have a factor of 3 due to the color sum in the final state where we are now allowing for all gauge bosons instead of just a photon, although the function remains the same which is why we have retained the $J_{\gamma}$ notation.
Then we are left with
\begin{equation}	
\left.\frac{d\Gamma}{dz}\right|_{\text{inc.}} = \frac{A_0}{(2\pi)^2} \int \frac{d\Omega_\gamma}{4\pi}\, F_{\chi}^{abab}\, \delta(2M_{\chi}(1-z)).
\end{equation}
so now when we integrate over $z$ and $\Omega_{\gamma}$ we have 
\bea
\Gamma = 3A_0 \frac{F_{\chi}^{abab}}{(2\pi)^22M_{\chi}}.
\label{eq:incl}
\eea
The branching ratio, therefore can be obtained combining Eqs.~\eqref{eq:diff} and \eqref{eq:incl} where the factor $A_0$ cancels out.

\subsection{Bound state transitions}
\label{sec:BSD}

The previous subsections have established how to compute the photon spectrum from decay of $L=S=0$ bound states via annihilation to SM particles.
The remaining necessary ingredient is to determine how these states are populated through radiative capture and decays.
The initial formation rate for bound states has already been discussed in Sec.~\ref{sec:BSF}; this subsection details the computation for shallowly-bound states to decay to lower-lying bound states.

Transitions between bound states, mediated by emission of a vector boson, can be computed using very similar expressions to those discussed in Sec.~\ref{sec:BSF} for the initial bound-state formation.
The are three salient differences: 1.~the scattering-state wavefunction is now replaced with the bound-state wavefunction; 2.~we must account for cases where the initial state has odd $L+S$ and the final state has even $L+S$; and 3.~we need address cases where the initial state has net total charge $Q\ne 0$.
As discussed in Sec.~\ref{sec:BSF}, the second issue can be taken into account by modifying the potential used to compute the initial- and final-state wavefunctions.

The expression for the decay rate between states with total charge $Q=0$ due to photon emission is given by a straightforward modification of Eq.~\eqref{eq:BSxsec},
\begin{equation}\begin{aligned} 
\Gamma & = \frac{2 \alpha k}{\pi M_\chi^2} \int d\Omega_k \left| \int d^3 {\bf r}\, {\boldsymbol \epsilon}(\hat{\bf k}) \cdot \left[  \left(2  \psi^*_{CC} \nabla \phi_{CC} +  \psi_{C} \nabla \phi_C\right) \right. \right. \\
& \left. \left. + \frac{1}{2} M_\chi \aW \hat{\bf r} e^{-m^{}_{\scaleto{W}{3.5pt}} r} \left(2 \psi_{CC}^* \phi_C  - 2 \psi_{C}^* \phi_{CC} + 3 \sqrt{2}  \psi_C^* \phi_N \right) \right] \right|^2 ,\label{eq:decayrate}
\end{aligned}\end{equation}
where as previously $k$ is the momentum of the emitted photon and ${\boldsymbol \epsilon}(\hat{\bf k})$ is its polarization vector.
Now $\psi({\bf r})$ is the wavefunction for whichever of the initial and final states has $L+S$ odd, while $\phi({\bf r})$ is the wavefunction for whichever state has $L+S$ even (the dipole selection rule ensures that states connected by a dipole transition have opposite signs for $L+S$).
We can make this simplification because the absolute value of the matrix element does not depend on the direction of the transition (although only one direction will have a positive $k$ and hence non-zero available phase space).
Consequently, once we have computed the matrix element for an $L+S$-even $\rightarrow$ $L+S$-odd transition (as in Eq.~\eqref{eq:BSxsec}), we can reuse the same matrix element for a transition in the reverse direction, only modifying the phase-space factors.
The contribution from $Z$ emission can be obtained by replacing $\alpha \rightarrow \aW \cW^2$ in the prefactor and replacing the momentum factor $k$ as described in Eq.~\eqref{eq:Zphasespace}.

To see explicitly that the matrix element is invariant under time-reversal (up to conjugation), consider relabeling $i \leftrightarrow i^\prime$, $j \leftrightarrow j^\prime$ in Eq.~\eqref{eq:matrixelement}, and likewise swapping the notation for the initial- and final-state wavefunctions $\psi_{nlm} \leftrightarrow \phi$ (that is, $\phi$ remains the wavefunction for the $ij$ two-particle state, whether it is the initial or final state).
The index relabeling has the effect of transposing the generator matrices; as the generators are Hermitian, this is equivalent to taking their complex conjugates.
The relabeling of the wavefunctions applies complex conjugation to both $\mathcal{J}$ and $\mathcal{Y}$, and additionally flips the sign of $\mathcal{Y}$, as can be seen from Eq.~\eqref{eq:yjequations}. Consequently, we obtain:
\begin{align}
\mathcal{M}^a_{i^\prime i, j^\prime j} & = -\sqrt{2^8 \pi \alpha_\text{rad} M_\chi} \left\{ -i f^{abc} (T_1^b)_{i^\prime i}^* (T_2^c)_{j^\prime j}^* (- \mathcal{Y}^*) + \frac{1}{2} \mathcal{J}^* \left[ (T_1^a)_{i^\prime i}^* \delta_{j^\prime j} - (T_2^a)_{j^\prime j}^* \delta_{i^\prime i}\right]\right\} \nn \\
& = \left[-\sqrt{2^8 \pi \alpha_\text{rad} M_\chi} \left\{ - i f^{abc} (T_1^b)_{i^\prime i} (T_2^c)_{j^\prime j}  \mathcal{Y} + \frac{1}{2} \mathcal{J} \left[ (T_1^a)_{i^\prime i} \delta_{j^\prime j} - (T_2^a)_{j^\prime j} \delta_{i^\prime i}\right]\right\}  \right]^* \nn \\
& = \mathcal{M}^{a*}_{i i^\prime, j j^\prime},
\end{align}
where $\mathcal{Y}$ and $\mathcal{J}$ are the functions computed from Eq.~\eqref{eq:yjequations} for the $ij \rightarrow i^\prime j^\prime$ matrix element.

Transitions between $Q=0$ and $Q=\pm 1$ bound states are similarly related to the cross section for capture into $Q=\pm 1$ states, Eq.~\eqref{eq:Q1capture}, with a rate given by:
\begin{equation}\begin{aligned}
\Gamma & = \frac{2 \alpha k}{\pi M_\chi^2} \int d\Omega_k \, \Bigg| {\boldsymbol \epsilon}(\hat{\mathbf{k}}) \cdot \int d^3 \mathbf{r}  \left[  \sqrt{\frac{3}{2}} \psi^*_{+0} \nabla \phi_{N} -  \frac{\sqrt{3}}{2} \psi^*_{+0} \nabla \phi_C \right.\\
& \left. \left. + \frac{1}{\sqrt{2}} \psi^*_{++-} \nabla \phi_{C} - \frac{1}{\sqrt{2}}  \psi^*_{++-} \nabla \phi_{CC}  \right. \right. \\
& \left. + \frac{1}{2} \hat{\mathbf{r}} \zeta(r) M_\chi \aW \left( \sqrt{3}  \psi_{+0}^* \phi_C + \sqrt{2} \psi_{++-}^* \phi_C + 2  \sqrt{2} \psi_{++-}^* \phi_{CC} \right) \right] \Bigg|^2, 
\end{aligned}\end{equation}
with $\zeta(r)$ as defined in Eq.~\eqref{eq:propagatorfactor}, and phase space factor $k$ as defined in Eq.~\eqref{eq:Wphasespace}.
Here $\phi$ denotes the $Q=0$ wavefunction and $\psi$ denotes the $Q=1$ wavefunction; in the case where the $Q=0$ wavefunction has $L+S$ odd, the terms containing $\phi_N$ should be set to zero.
For the $Q=1$ sector there are no states containing two identical particles, so the set of allowed states is the same for $L+S$ even or odd, although the wavefunctions for the two cases will differ due to the $L+S$-dependent potential.

Next, let us consider transitions mediated by photon or $Z$ emission between two $Q=\pm 1$ states.
In this case we obtain (suppressing the position dependence of the wavefunctions)
\begin{align}
&\Gamma  =  \frac{2 \alpha k}{\pi M_\chi^2} \int \!d\Omega_k \left| \boldsymbol{\epsilon}(\hat{\mathbf{k}}) \cdot \int d^3 \mathbf{r} \left\{ \frac{3}{2} (\psi^*_{++,-})_f \nabla (\psi_{++,-})_i + \frac{1}{2} (\psi^*_{+,0})_f \nabla (\psi_{+,0})_i \right. \right. \\
& \left. \left. + \frac{1}{2} \hat{\mathbf{r}} e^{-m^{}_{\scaleto{W}{3.5pt}} r} M_\chi \aW  \!\left[3 (-1)^{(L+S)_i} (\psi^*_{+,0})_f (\psi_{+,0})_i \!-\! \sqrt{6} (\psi^*_{+,0})_f (\psi_{++,-})_i \!+\! \sqrt{6} (\psi^*_{++,-})_f (\psi_{+,0})_i \right]  \right\} \right|^2\!, \nn
\end{align}
for photon emission, and as above we obtain the $Z$-emission contribution by replacing $\alpha \rightarrow \aW \cW^2$ and modifying the phase-space factor $k$ as defined in Eq.~\eqref{eq:Zphasespace}.
In the expression above, initial and final states are distinguished by $i$ and $f$ subscripts, and this expression can be used for transitions where the initial state is either $L+S$-odd or $L+S$-even. Note that there is an explicit $(L+S)_i$-dependent factor in the second line, which is necessary to ensure the consistency of the matrix element when the initial and final states are swapped.  We discuss the details of its origin in Appendix~\ref{app:signs}.

Finally, for transitions mediated by $W$ emission between $L+S$-odd $Q=\pm 1$ states and $L+S$-even $Q=\pm 2$ states (as discussed above, we do not expect any $L+S$-odd $Q=\pm 2$ bound states), the decay rate is given by:
\begin{equation}\begin{aligned} 
\Gamma & = \frac{k \aW}{\pi M_\chi^2} \int d\Omega_k \left| \boldsymbol{\epsilon}(\hat{\mathbf{k}}) \cdot \int d^3 \mathbf{r} \left\{ \left[\psi^*_{+,0} - \sqrt{3/2} \psi^*_{++,-} \right] \nabla \phi_{++,0} -  \sqrt{3} \psi^*_{+,0} \nabla \phi_{+,+} \right. \right. \\
& \left. \left. - \hat{\mathbf{r}} M_\chi \aW \zeta(r) \left[\sqrt{6} \psi^*_{++,-} \phi_{++,0}  + \sqrt{3} \psi^*_{+,0} \phi_{+,+} \right] \right\} \right|^2\!, 
\end{aligned}\end{equation}
where here $\psi$ denotes the wavefunction for the $Q=1$ state and $\phi$ denotes the wavefunction for the $Q=2$ state, with subscripts labeling the components as above.
The cross section is identical for transitions between $Q=-1$ and $Q=-2$ states, with the appropriate modifications to the wavefunction labels.

\subsection{Key points from the bound state decay calculation}
\label{sec:bsdecaysummary}

\begin{figure*}[!t]
\centering
\includegraphics[width=0.47\textwidth]{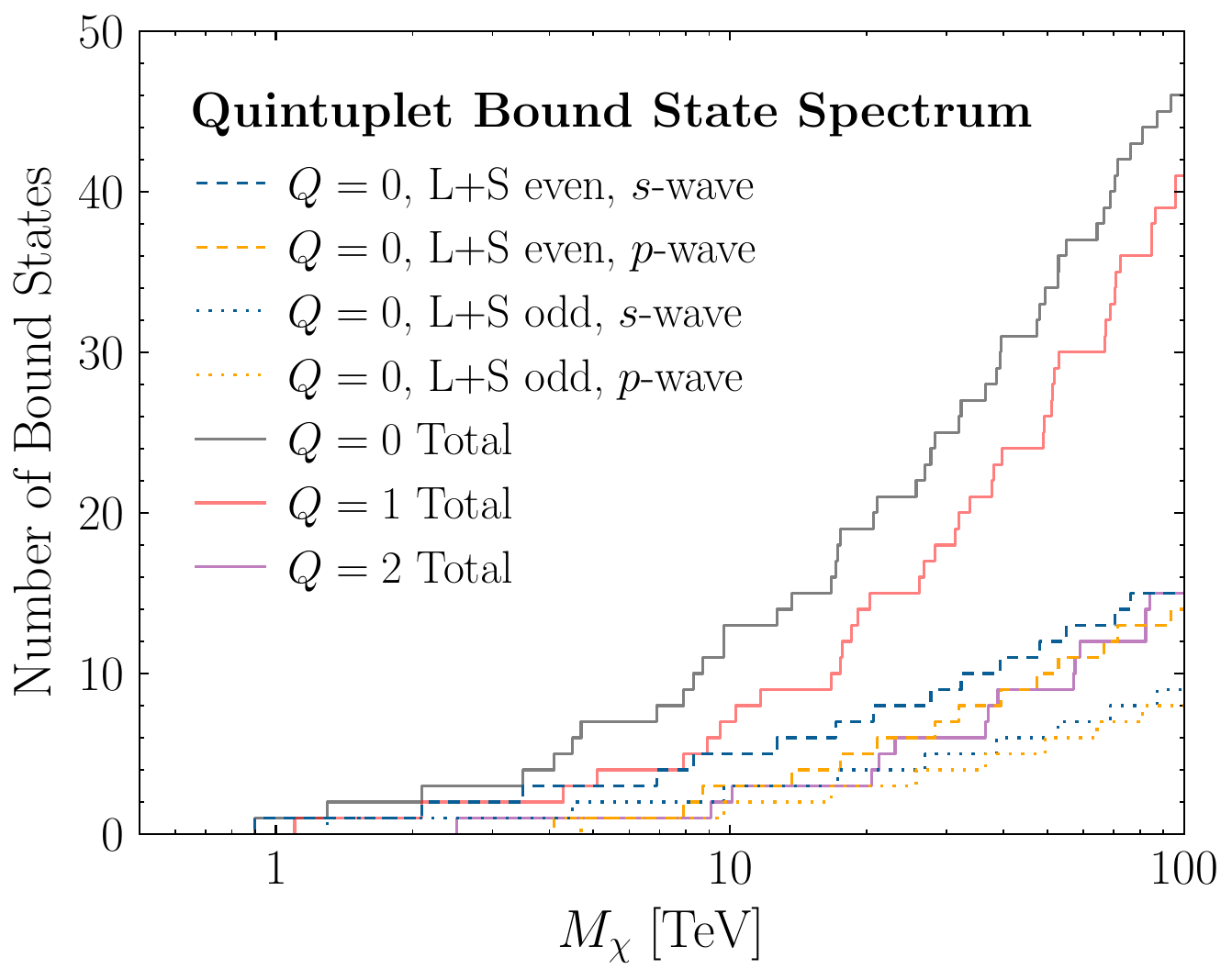} 
\hspace{0.5cm}
\includegraphics[width=0.47\textwidth]{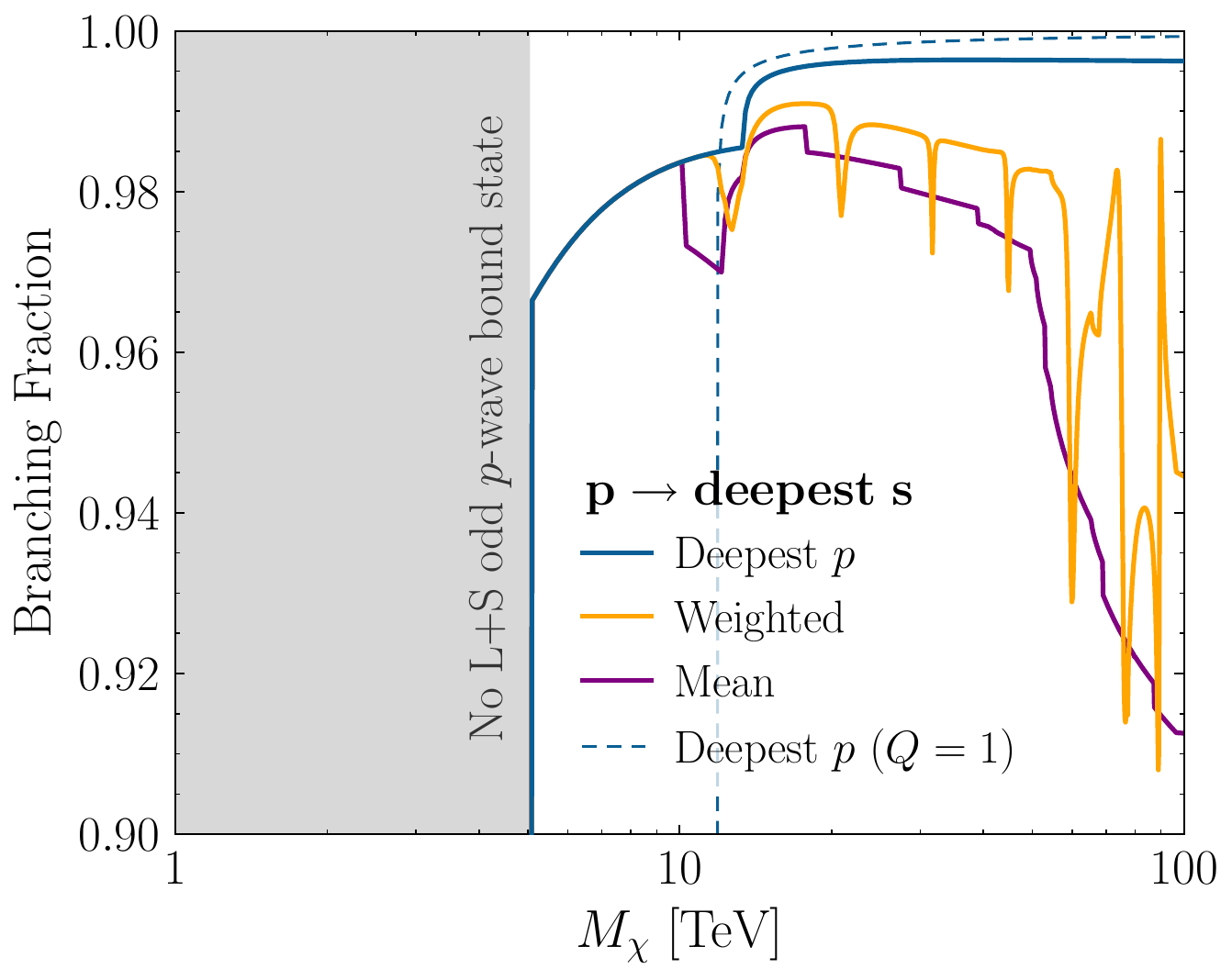}
\caption{(Left) The number of bound states of different types as a function of mass.
For $Q=0$ we show the breakdown of the bound states into various types, whereas for the charged bound states we simply show the total.
(Right) The branching fraction for $Q=0$ $L+S$ odd $p$-wave bound states to decay to the deepest $s$-wave state.
We show the branching fraction for the deepest bound state, for an average over all of the $Q=0$ $L+S$ odd $p$-wave bound states weighted by their capture cross section, and then a simple mean over all states.
Finally, we also show the branching fraction for the deepest $Q=1$ $L+S$ odd $p$-wave bound states to decay to the deepest $Q=0$ $s$-wave state.
As claimed, in all cases decay to the deepest $s$-wave state dominates.}
\label{fig:BS-spectrum-transition}
\end{figure*}

We now have all the ingredients for computing the differential decay rate of a bound state to a hard photon, so let us finally summarize our prescription (see also Fig.~\ref{fig:cascadeschematic}).
Firstly, the number of bound states at a given mass is shown on the left of Fig.~\ref{fig:BS-spectrum-transition}.
We have argued that production of a bound state with a given $S$ and $L > 0$ will generically lead to production (via one or more decays) of an $L=0$ bound state with the same $S$, so it unnecessary to compute the rate to produce SM particles directly from those $L>0$ states.
We confirm this numerically on the right of Fig.~\ref{fig:BS-spectrum-transition}: however measured, the branching fraction is always greater than 90\%.
For instance, at 100 TeV, the deepest $Q=0$ $L+S$ odd $p$-wave bound state decays to the deepest $s$-wave state with 99.6\% probability and to the second deepest $s$-wave state with a likelihood of 0.4\%.
(The next most dominant transition is to the third deepest $s$-wave bound state, and only occurs with a probability of 0.006\%.)

We also discussed decays of the $S=1,\,L=0$ bound states, noting that they generate only a power suppressed contribution to the photon endpoint spectrum.
Nevertheless, the contribution from the $S=1$ states can be comparable to that from the $S=0$ states, after accounting for both their (enhanced) production rate and (suppressed) endpoint spectrum from annihilation.
However, in practice this means that the contributions to the endpoint spectrum from both the $S=0$ and $S=1$ bound states are suppressed compared to direct annihilation (either due to power suppression or because of the small formation rate).

For the aforementioned reasons, we focus on the $L=S=0$ bound states to estimate the size of the (generally subdominant) bound-state contribution.
Doing so implies we can reuse our SCET results from the direct annihilation case, with appropriate modification of the wavefunction factors.
This choice does mean that the total bound-state contribution to the endpoint photon spectrum could increase by an $\mo(1)$ factor in an improved calculation (once the $S=1$ states are included).
Because the bound-state contribution to the endpoint spectrum is suppressed, this typically corresponds to a percent-level theoretical uncertainty in the overall endpoint spectrum, with larger uncertainties at specific mass points where the bound state formation cross section is enhanced relative to direct annihilation.
The contribution to continuum photons (not near the endpoint) from bound state formation can be markedly larger, compared to the effect on the endpoint spectrum, as there is no power suppression in the continuum contribution from annihilation of the $S=1$ bound states.
We will discuss each of these contributions to the total spectrum in the next section.

\section{The Combined Photon Spectrum and Numerical Results}
\label{sec:Numerics}

At this stage we have all ingredients required to determine the quintuplet annihilation spectrum, including both direct annihilation and the contribution of bound states.
In this section we collect our results to determine the full energy distribution of photons the quintuplet generates at the thermal mass of 13.6~TeV, but also for a wider range of masses.
We will estimate the impact of several uncertainties on our results, such as the residual theoretical uncertainty on the NLL computations, but also from astrophysical uncertainties such as the distribution of $v$ values and on the DM density in the inner Galaxy.
Finally, we will put these results together to estimate the sensitivity of existing and upcoming IACTs to quintuplet DM.

\subsection{Predictions for the spectrum and rate of photon production}

A central goal of this work is to accurately determine the distribution of photons that emerge when two SU(2) quintuplets annihilate.
This spectrum forms the signal template for telescopes searching for high energy photons, and therefore is a central theoretical input.
To achieve this, throughout we have computed differential cross sections $d \sigma / d z$, both for the direct annihilation in Eq.~\eqref{eq:QNLL}, and also for the bound state contribution by combining the results of the previous sections with Eq.~\eqref{eq:BSfact}.
For indirect detection, observables are sensitive to $d \langle \sigma v \rangle / d z$.  To begin with we will assume the DM states are incident with a fixed $v = 10^{-3}$, revisiting the validity of this approximation in the next subsection.
In order to extract the shape of the photon distribution from the differential cross section, it is common in indirect detection to introduce a photon spectrum $dN/dE$, and our convention for doing so is the following,\footnote{Further discussion of the connection between spectra used in indirect detection and the corresponding field theoretic quantities can be found in, for instance, Ref.~\cite{Bauer:2020jay}.}
\bea
\frac{d \langle \sigma v \rangle}{dE} = \langle \sigma v \rangle_{\rm line} \times \frac{dN}{dE}.
\label{eq:sigmavline-def}
\eea
This choice follows Ref.~\cite{Baumgart:2017nsr}, and implies that our spectrum is normalized with respect to the \emph{line cross section}, $\langle \sigma v \rangle_{\rm line} \equiv \langle \sigma v \rangle_{\gamma\gamma} + \tfrac{1}{2} \langle \sigma v \rangle_{\gamma Z}$, which is defined as the rate to produce two photons at exactly $E=M_{\chi}$.
By construction, $dN/dE$ will contain a contribution of exactly $2 \delta(E-M_{\chi})$ for the line, but it will also contain contributions from endpoint photons, bound state decays, and continuum photons arising primarily from the unstable particles the direct annihilations or bound state decays can produce.
For each of these latter components, however, their additions to $dN/dE$ will be weighted by a branching fraction $\langle \sigma v \rangle_i/\langle \sigma v \rangle_{\rm line}$, with $\langle \sigma v \rangle_i$ the cross section for that particular contribution.
The rationale for anchoring our calculations to the line cross section is that $\chi \chi \to \gamma \gamma$, which has a spectrum of exactly $dN/dE = 2 \delta(E-M_{\chi})$, is a common experimental target, and therefore there are a wide number of existing constraints on $\langle \sigma v \rangle_{\rm line}$ which we can then directly compare with.
Further discussion of this point can be found in Ref.~\cite{Baumgart:2017nsr}.

\begin{figure*}[!t]
\centering
\includegraphics[width=0.47\textwidth]{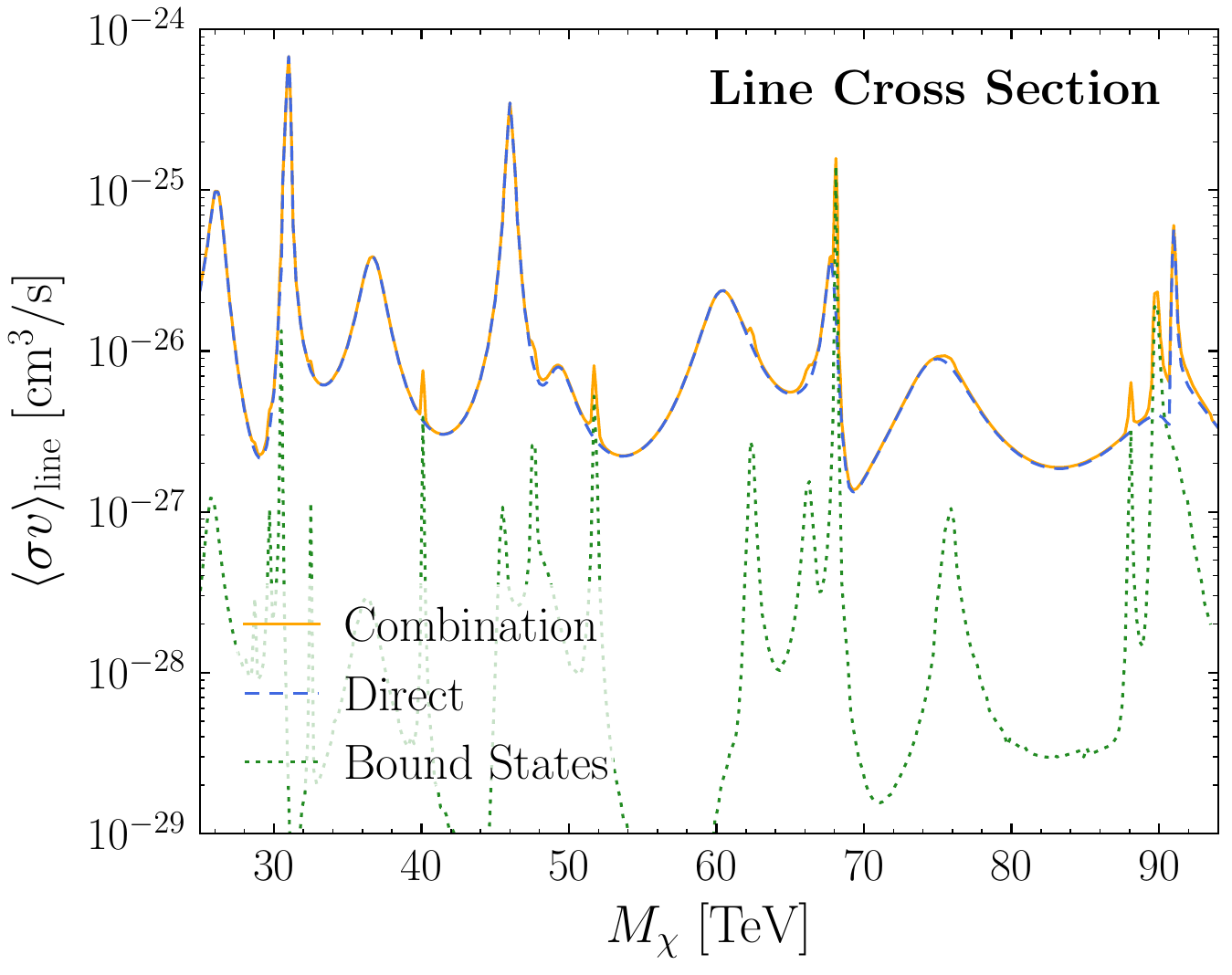} 
\caption{The cross section for line photons, breaking down the contributions from the direct annihilation and bound states.
While generically the direct annihilation dominates, for isolated masses near Sommerfeld peaks the bound state contribution can be the leading one.
For all masses lower than those shown bound states are strictly subdominant.
}
\label{fig:Line}
\end{figure*}

The spectra of line and endpoint photons produced by decay of the bound states is computed using the methods of Sec.~\ref{sec:BSA}.\footnote{In this work we do not track the much lower-energy photons radiated in bound state formation and transitions; these signals have been studied for the quintuplet in Refs.~\cite{Mitridate:2017izz, Mahbubani:2020knq}.}
For our bound state formation and transition calculations, we include only states with $L=0,1,2$. Capture into $L=3$ and higher states requires at least $L=2$ for the initial state, and we expect the contributions from components of the initial state with high $L$ to be suppressed at low velocities, by a factor that is parametrically $(M_\chi v/\mW)^{2L}$ (although for sufficiently high masses, $M_\chi \gtrsim 100$ TeV, this suppression is lifted). It is also worth noting that for essentially all the parameter space most relevant for experimental searches with H.E.S.S, at $M_\chi < 20$ TeV, we find that no $L=3+$ bound states exist in the spectrum. We independently expect capture into states with high principal quantum number $n$ (which is required for high $L$) to be suppressed, as (1) the finite range of the potential means only a limited number of states are bound at all, so unlike in unbroken gauge theories there is no infinite tower of high-$n$ states, (2) capture into weakly-bound states is suppressed by a phase space factor, and (3) analytic approximations (App.~\ref{app:analytic}) suggest that we can expect the leading contribution to the capture cross section to be exponentially suppressed for large $n$. In practice, our numerical calculation expresses the bound states as a linear combination of 30 basis states for each combination of $L$, $Q$ and $(-1)^{L+S}$, allowing us to access up to 30 distinct bound states indexed by different values of $n$ (although as we approach this upper bound we expect the spectrum to become less accurate), and we include all these states in our calculation. We have checked at sample mass points that our binding energies and cross sections for capture into lower-$n$ states are not significantly affected by doubling the number of basis states. For the reasons given above, we generally expect the error due to the omission of higher-$n$ states to be small.

Before showing the full distributions of line and endpoint photons, we can already consider one measure of the importance of bound states to the resulting photon signal: their contribution to $\langle \sigma v \rangle_\text{line}$.
This is shown in Fig.~\ref{fig:Line}, where we separate the contribution to the line from the direct annihilation to that of processes involving an intermediate bound state.
At this stage we only include bound-state contributions that produce line photons, with energy essentially at $M_{\chi}$.
The figure makes clear a point already estimated earlier: direct annihilation generally dominates the production of line photons at $E\simeq M_\chi$ by $1-2$ orders of magnitude.
However, the bound-state contribution can be significant and even dominate at isolated mass points, for instance as at $M_{\chi} = 68.1~{\rm TeV}$, and therefore a reliable prediction at arbitrary masses must include this contribution.

\begin{figure*}[!t]
\centering
\includegraphics[width=0.47\textwidth]{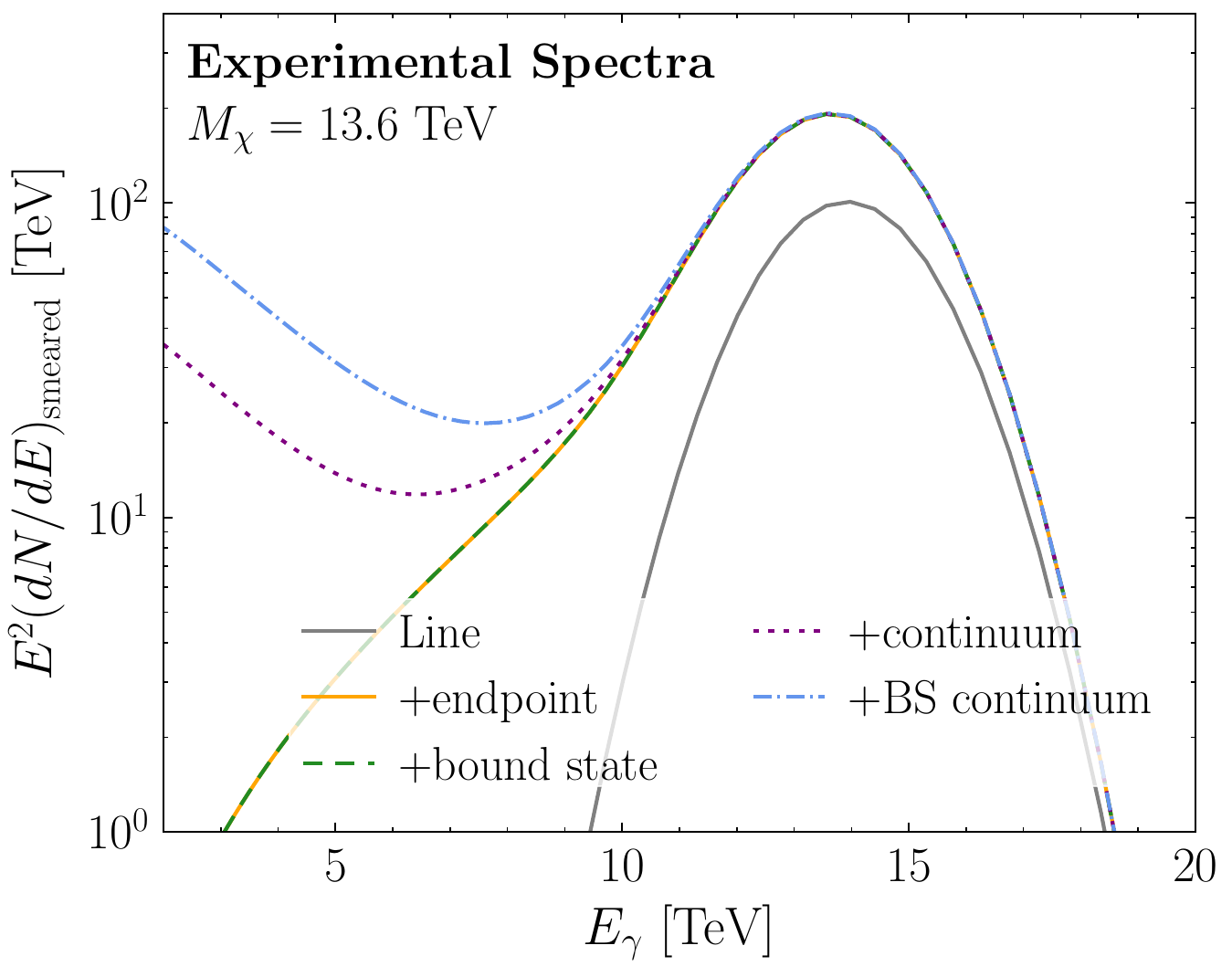} 
\hspace{0.5cm}
\includegraphics[width=0.47\textwidth]{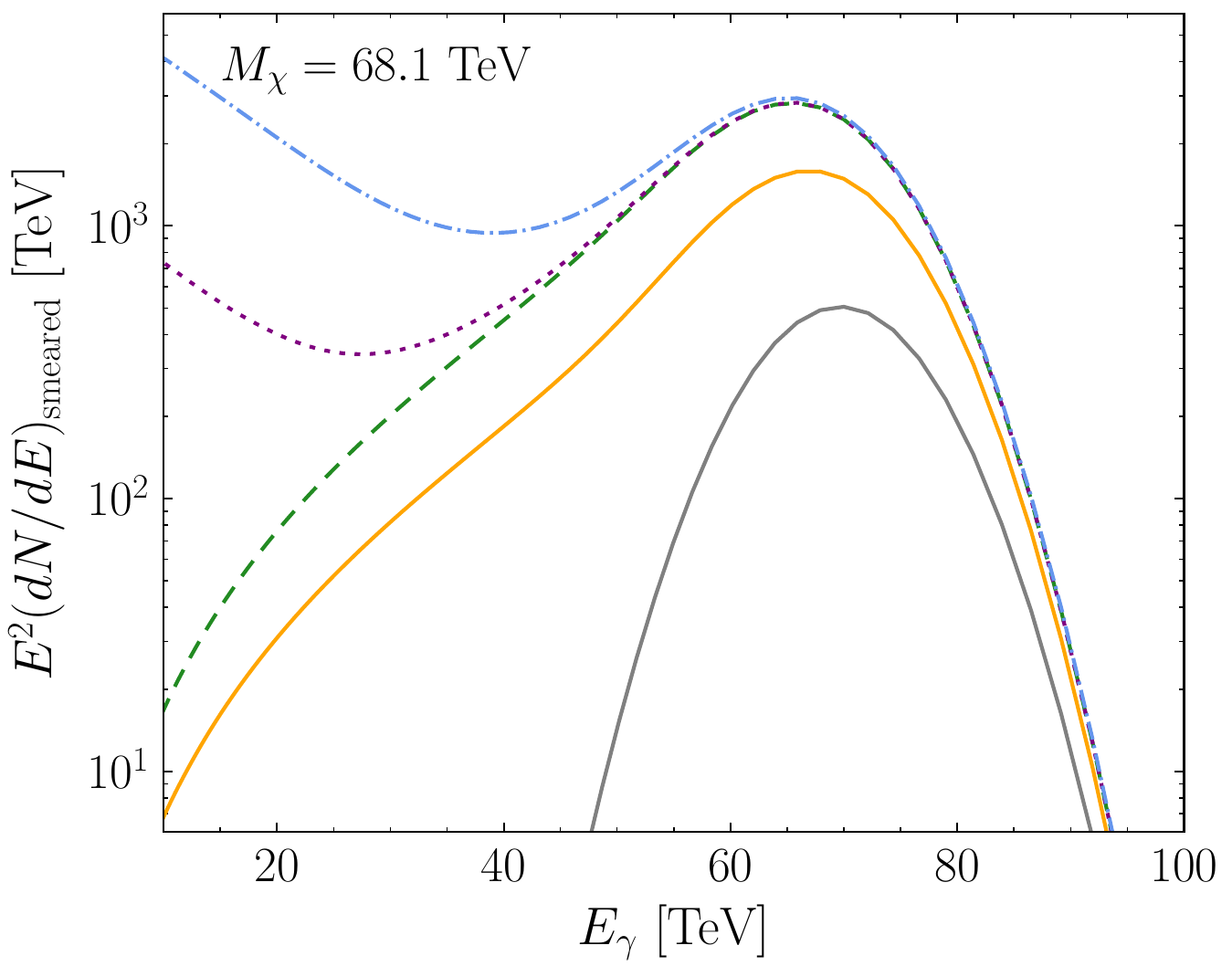}
\includegraphics[width=0.47\textwidth]{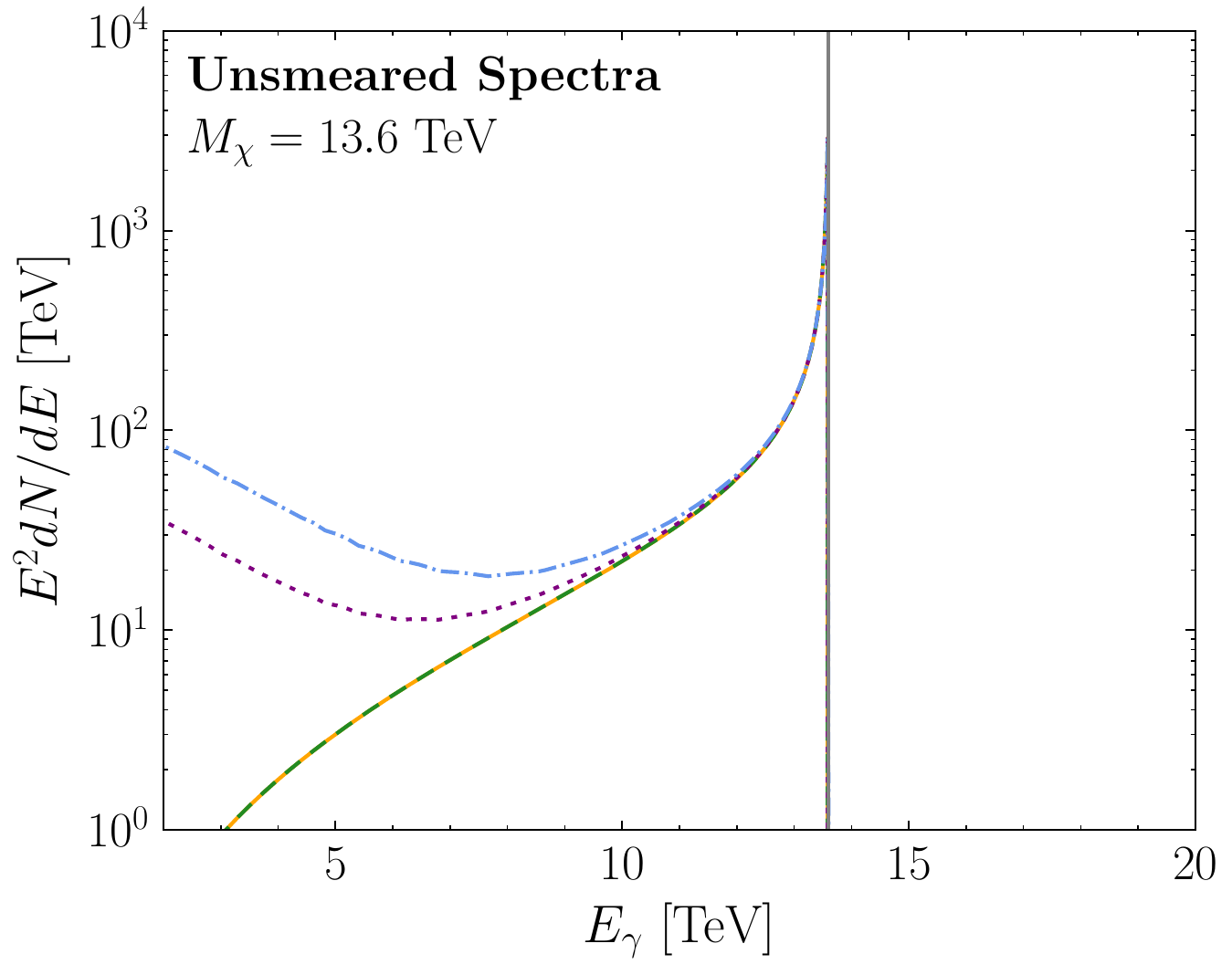} 
\hspace{0.5cm}
\includegraphics[width=0.47\textwidth]{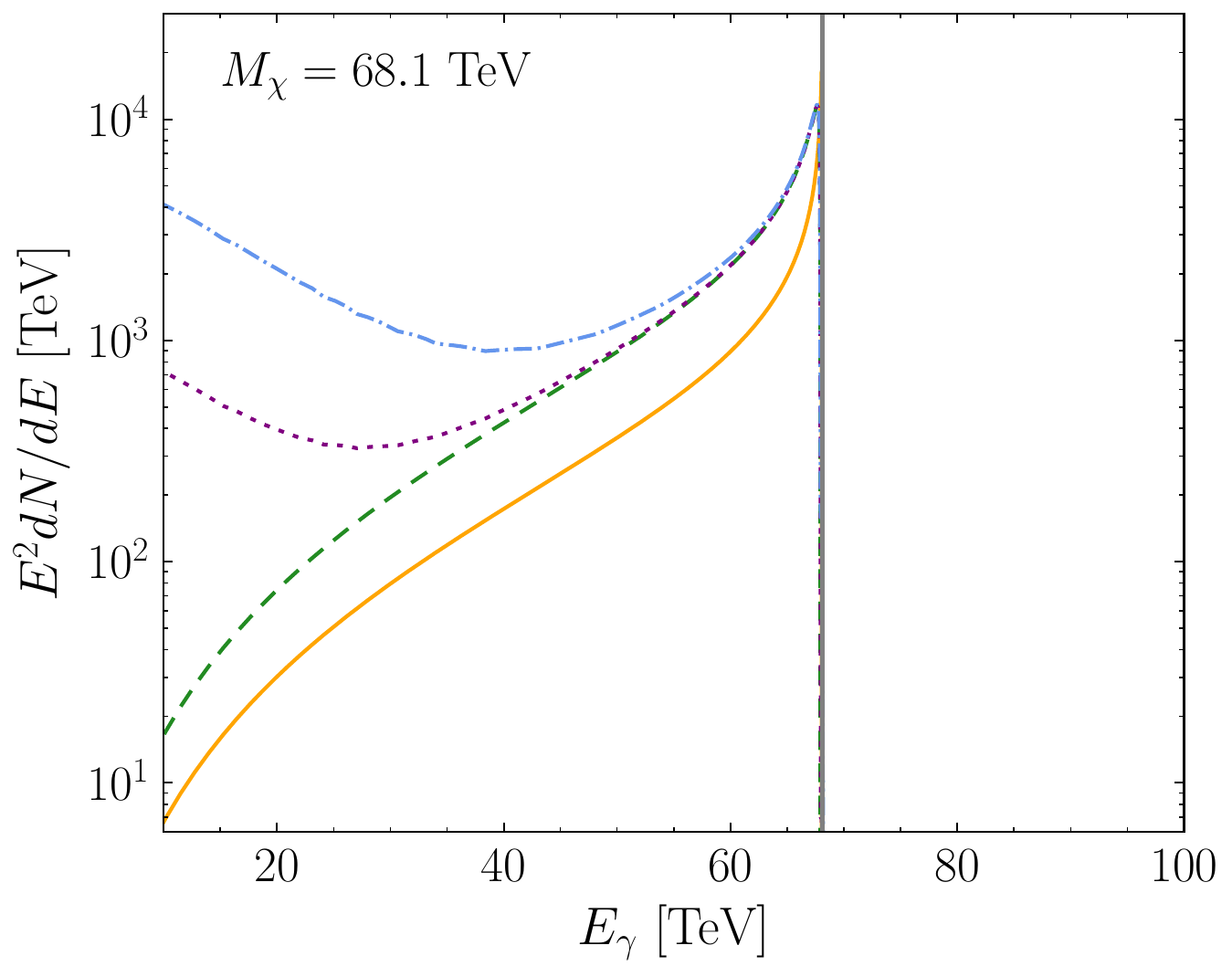}
\caption{The quintuplet annihilation spectrum for two masses, the thermal mass of 13.6 TeV (left), and a mass where bound state contributions are appreciable, 68.1 TeV (right).
For each mass we show results convolved with the H.E.S.S. energy resolution (top) and unsmeared (below).
The full spectrum is broken into five individual components: the line, endpoint, bound state line and endpoint, the direct annihilation continuum, and bound state contribution to the continuum.
Details of each are provided in the text.
}
\label{fig:Spectra}
\end{figure*}

Moving beyond the line, in Fig.~\ref{fig:Spectra} we show the full spectrum, broken down by various contributions, for two masses: the thermal mass of $M_{\chi}=13.6~{\rm TeV}$, and a mass where bound state contributions are significant, $68.1~{\rm TeV}$.
For each mass we show two versions of the spectrum.
In the lower panels, we show the unsmeared spectra, which is the distribution of photons that emerge from the annihilations.
(Note in this case the line contribution is simply $dN/dE = 2 \delta(E-M_{\chi})$ and so represented by a vertical line.)
In the upper panels, we have convolved the raw spectra with a finite experimental energy resolution in order to model what would actually be seen at a realistic instrument.
For this we take the energy resolution of the H.E.S.S. telescope, determined from Ref.~\cite{HESS:2013rld}.
In detail, we fix the relative width $\Delta E/E$ to 0.17 and 0.11 for $E = 500~{\rm GeV}$ and $E = 10~{\rm TeV}$, respectively, and then vary logarithmically between these endpoints, freezing the ratio either side.
From this we compute $(dN/dE)_{\rm smeared}$ as $dN/dE$ convolved with a Gaussian of width equal to the energy resolution.

In terms of these two notions of the spectra, Fig.~\ref{fig:Spectra} shows five contributions to the photons distributions for the two masses.
The first three of these are: 1. the direct annihilation line; 2. direct annihilation endpoint; and 3. the bound state contribution to the line and endpoint.
Again we see clearly a point noted for the wino in Refs.~\cite{Baumgart:2017nsr,Baumgart:2018yed}: the endpoint contribution makes a considerable modification to the observed number of photons with $E \sim M_{\chi}$, with the peak smeared spectra enhanced by 1.9 and 3.1 for $M_{\chi} = 13.6$ and 68.1 TeV.
The bound state contribution is more modest: it is effectively negligible at the thermal mass, and a factor of 1.7 enhancement at 68.1 TeV, which again is a mass with an anomalously large bound state contribution to the hard photon spectrum.

Beyond these three, we also show the contribution of two continuum sources, both of which can generate lower energy photons.
The first of these is the continuum emission arising from direct annihilation.
This results from tree level annihilation of the quintuplets into $W$ or $Z$ bosons.
The latter of these arises from $\gamma Z$ and $ZZ$ final states, and is a simple reweighting of the line cross section as
\be
\frac{\langle \sigma v \rangle_{ZZ} + \tfrac{1}{2} \langle \sigma v \rangle_{Z \gamma}}{\langle \sigma v \rangle_{\gamma\gamma} + \tfrac{1}{2} \langle \sigma v \rangle_{\gamma Z}} = \frac{\cW^2}{\sW^2}.
\ee
For the $W^+ W^-$ final state, the tree level annihilation rate, with the Sommerfeld effect included can be computed as,
\begin{equation}\begin{aligned}
\langle \sigma v \rangle_{\scriptscriptstyle WW} = \frac{\pi \aW^2}{M_{\chi}^2} &\Big[ 
18 |s_{00}|^2 + 25 |s_{0\pm}|^2 + 4 |s_{0\pm\pm}|^2 \\
&+ 30 \sqrt{2} {\rm Re}(s_{00} s_{0\pm}^*) + 12 \sqrt{2} {\rm Re} (s_{00} s_{0\pm\pm}^*) + 20 {\rm Re}(s_{0\pm} s_{0\pm\pm}^*) \Big].
\end{aligned}\end{equation}
As discussed for the case of the wino in Ref.~\cite{Hryczuk:2011vi}, higher order corrections to this should not be appreciable, and so we do not include them.
We can then add the $W^+ W^-$ final state to $dN/dE$ with weighting $\langle \sigma v \rangle_{\scriptscriptstyle WW}/\langle \sigma v \rangle_{\rm line}$ along with the $Z$ contribution.
In each case, to determine the spectrum of photons that result from these electroweak bosons we use PPPC4DMID~\cite{Cirelli:2010xx} with the electroweak corrections turned off, to avoid any double counting of the endpoint contributions we computed.
As seen in Fig.~\ref{fig:Spectra}, these contributions are important for $E_{\gamma} \ll M_{\chi}$.

The final contribution to the spectrum we consider is continuum photons arising from the bound state decays.
This contribution is not the main focus of this work, but to get an estimate of its size and spectrum, we assume the most common SM decay products are light quarks (equally weighted between flavors), and employ the corresponding gamma-ray spectrum from PPPC4DMID.
The motivation for this choice is that the bound states will decay through their couplings to (on-shell or off-shell) $W$ and $Z$ bosons, with the exact channel depending on their $L$ and $S$ quantum numbers, and the gauge bosons in turn decay dominantly to quarks due to their large associated degrees of freedom.
We weight the continuum spectrum by the ratio between the bound state capture cross section and $\langle \sigma v \rangle_\textrm{line}$, similar to how we weight the $Z$ and $W$ continuum components above.
At the thermal mass, this ratio is roughly 31, and the visible contribution can be seen in Fig.~\ref{fig:Spectra}.
The contribution is dominated by the $Q=0$ $p \to s$ capture cross section which sits near a Sommerfeld peak in this capture rate at 13.6 TeV ({\it cf.} Fig.~\ref{fig:analytic-comparison}).
To highlight this, at the edges of the uncertainty band on the thermal mass, 12.8 and 14.4 TeV, the equivalent ratio is reduced significantly, to 0.15 and 0.70, respectively.
At 68.1 TeV the ratio is larger still -- just over 471 -- and is dominated by the $Q=0$ and $Q=1$ $p \to s$ capture rates.

\begin{figure*}[!t]
\centering
\includegraphics[width=0.47\textwidth]{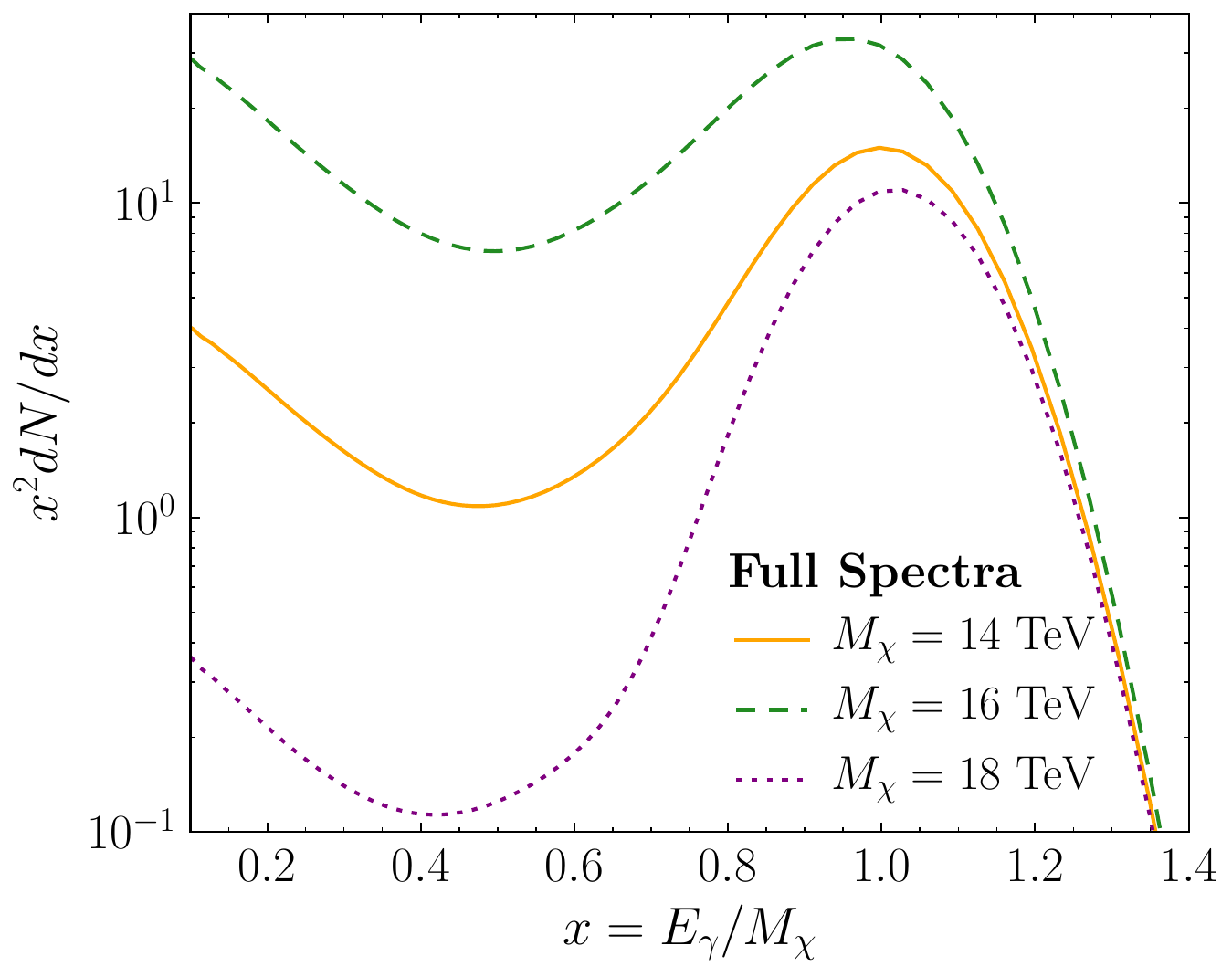}
\caption{
The full quintuplet spectrum for three different masses, $M_{\chi} = 14,\,16,$ and $18~{\rm TeV}$.
As shown, the spectrum can change significantly as a function of mass, a feature which does not arise for the wino or higgsino.
}
\label{fig:Spectra-var}
\end{figure*}

Figure~\ref{fig:Spectra} highlights the various contributions to the spectrum, but does not capture the variation of the spectrum as a function of mass.
The variation can be considerable, as shown in Fig.~\ref{fig:Spectra-var}.
From the definition, the line contribution to this spectrum is fixed at $2 \delta(E-M_{\chi})$ independent of mass.
What is not fixed, however, is the endpoint and continuum contributions, which can vary significantly even for small changes in mass.
(The bound state contributions are not significant for the masses shown.)
As shown in Ref.~\cite{Montanari:2022buj}, such rapid variations can lead to sharp features in the instrumental sensitivity to $\langle \sigma v \rangle_{\rm line}$, as the shape of the DM signal being searched for varies rapidly with mass.
These effects do not occur for the wino or higgsino, where the spectra varies relatively smoothly with mass (see Ref.~\cite{Montanari:2022buj}).

The origin of this behavior is that the various contributions to the combined spectrum -- such as the line, endpoint, and continuum -- have a different dependence on the quintuplet Sommerfeld factors, $s_{00}$, $s_{0\pm}$, and $s_{0\pm\pm}$.
The amplitude and phases of these factors can change rapidly and independently, leading to detailed interference effects in the rate to produce the different contributions to the spectrum.
The end result is that the ratio of the endpoint and continuum to the line can vary sharply and independently as a function of mass, as seen in Fig.~\ref{fig:Spectra-var}.
We discuss this behavior further in App.~\ref{app:massvariation}.

\subsection{Uncertainty associated with the velocity distribution of dark matter}
\label{subsec:vdep}

The complete initial-state wavefunctions naturally depend on the relative velocity of the incoming DM particles, which in the discussion so far we have simply set as $v = 10^{-3}$.
In this subsection we explore the systematic uncertainties associated with our modeling of the velocity.
We first discuss the detailed dependence of the cross sections on the relative DM velocity, and then explore the effects on our spectra of averaging over different plausible velocity distributions.
The effects of the long-range potential on the wavefunction saturate when $v \lesssim \mW/M_\chi$, which is true for halo velocities ($v\sim 10^{-3}$) for $M_\chi \lesssim 80~{\rm TeV}$; consequently, except near resonances or for very heavy DM, we do not expect the Sommerfeld enhancement from the weak interactions to depend sensitively on the velocity distribution.
However, the bound-state formation rate from $L > 0$ partial-wave components of the initial state {\it will} have a non-trivial velocity dependence even below this threshold.
Furthermore, for the quintuplet the thermal mass is only a factor of few below the saturation threshold, and in systems with higher velocities than the Sun's neighborhood -- such as galaxy clusters -- both the direct annihilation cross section and the bound-state formation rates are expected to depend sensitively on the velocity.

\begin{figure}
\centering
\includegraphics[width=0.49\linewidth]{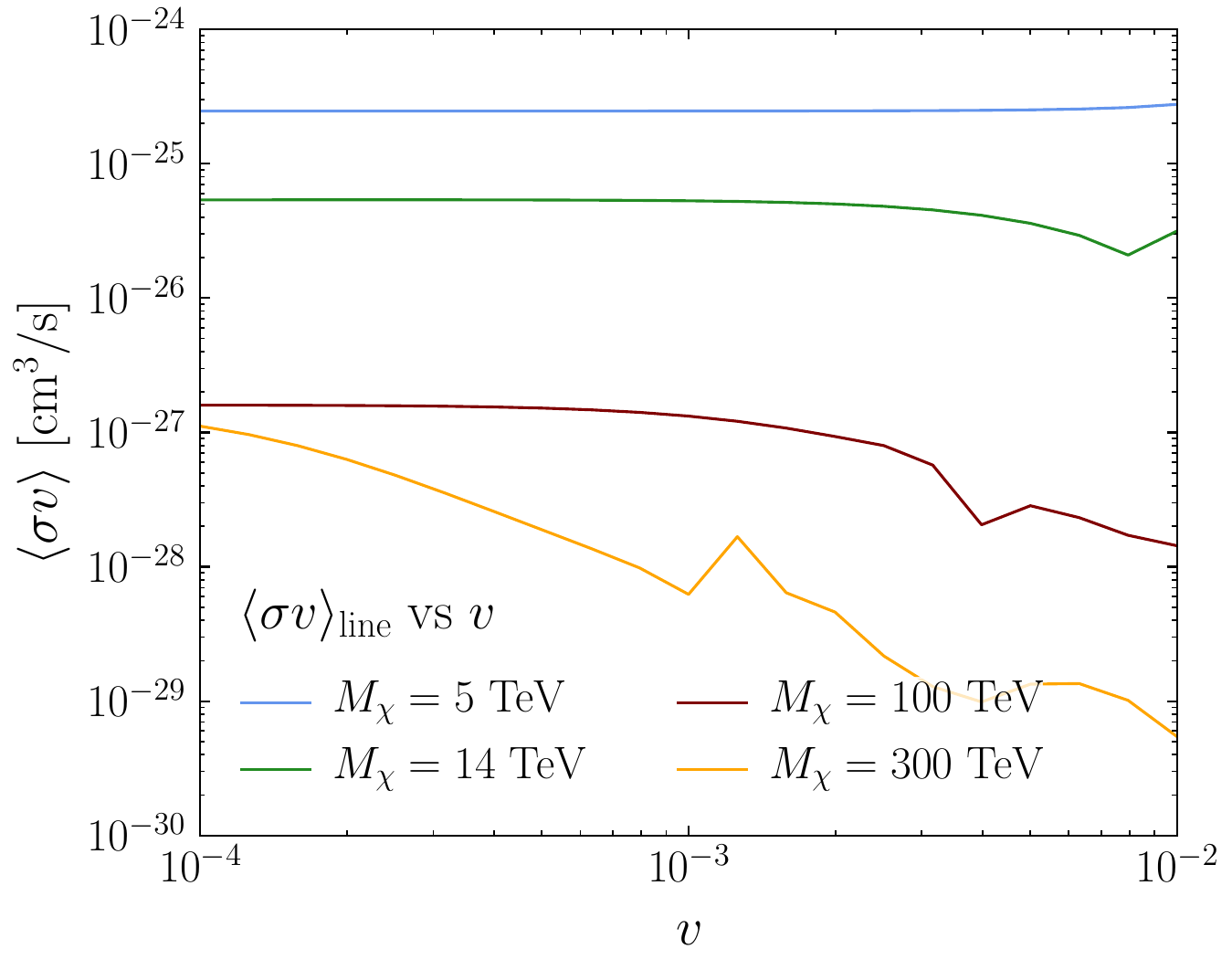}
\hspace{0.1cm}\includegraphics[width=0.49\linewidth]{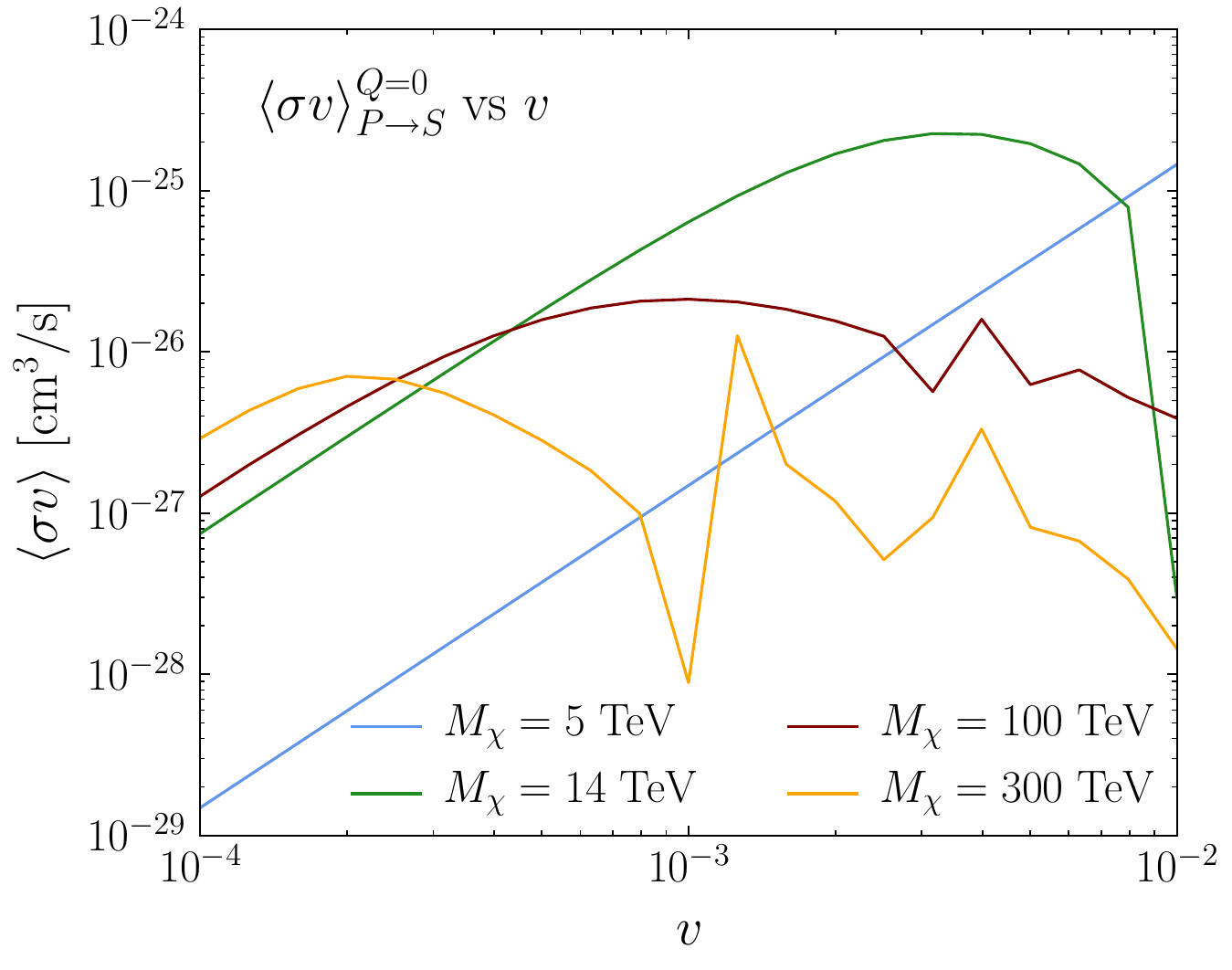}
\caption{The dependence of the cross sections on $v$ for four different fixed masses.
We show the case of direct endpoint annihilation (left, the analogue of Fig.~\ref{fig:arc}), and bound state capture (right, as in Fig.~\ref{fig:crc}).}
\label{fig:vscans}
\end{figure}

In Fig.~\ref{fig:vscans} we show how annihilation and capture vary as a function of velocity at four different masses.
We observe that for a mass of $M_\chi=14$ TeV, noticeable velocity dependence is present at $v \gtrsim 4\times10^{-3}$.
As we discuss in depth in App.~\ref{app:analytic}, the oscillatory behavior observed at high velocities can be understood in the limit of unbroken SU(2).
This behavior originates from interference between the different eigenvalues of the potential.
At low velocities, by contrast, SU(2) breaking effects are expected to suppress the oscillations.
For higher DM masses, where the velocity dependence is relevant even for $v \lesssim 10^{-3}$, our previous annihilation cross section plots should be taken as an illustrative estimate.
A full calculation at high mass would involve integrating the formulae given in this paper over the true velocity distribution in the region of interest.
The oscillatory behavior of the cross section at high velocities means that assuming a single velocity could in principle lead to large errors in this case.

We now estimate the effect of averaging over the velocity distribution.
The characteristic scale of the DM velocity dispersion should be comparable to the circular velocity of the visible matter, which in the vicinity of the Sun has been measured to be $v_\text{circ} \simeq 240$ km/s \cite{2016MNRAS.463.2623H}. 
Since the Milky Way's rotation curve is roughly flat at the Sun's location, we expect the velocity dispersion to be of a similar order over much of the Galaxy.
However, close to the Galactic Center the DM velocity is not well-known.
In DM-only simulations the velocity dispersion falls as one approaches the Galactic Center ({\it e.g.}~Ref.~\cite{Navarro:2008kc}) but simulations including baryons have demonstrated the opposite behavior ({\it e.g.}~Refs.~\cite{2010MNRAS.406..922T, Board:2021bwj, 2022MNRAS.513...55M}).
Even at the Sun's location, the full DM velocity distribution is not well-understood: the distribution is often treated as Maxwellian up to some escape velocity, although this is only a crude approximation ({\it e.g.} Ref.~\cite{Herzog-Arbeitman:2017fte}).
The escape velocity is determined to be $\sim$500 km/s at the location of the solar system~\cite{Necib:2021vxr,kh2021,mfcEtAl2018}.\footnote{Ref.~\cite{mfcEtAl2018} finds that the escape velocity increases slightly toward the Galactic Center.  However, they only present results in to a radius of around 5 kpc, where it is closer to 650 km/s.
The precise value of this cutoff is numerically unimportant for this work though, due to the exponential suppression in the distribution ({\it cf.}~Eq.~\eqref{eq:vpdf}).
In practice, we truncate the particle velocity at 500 km/s, but the numerical difference between this and 2400 km/s is at most a part per million in the annihilation rate.}
Within the Maxwellian approximation, the distribution is specified by that escape velocity and the velocity dispersion, with the latter having a greater effect on the annihilation rate.

For the Milky Way, we use the velocity dispersion values obtained in Ref.~\cite{Boddy:2018ike} for a variety of NFW-profiles.
In particular, we take the slowest and fastest velocities for locations interior to the solar system.
This gives a range $v_{\rm disp} \in [130, 330]$ km/s.
As a function of $v_{\rm disp}$, the magnitude of the relative WIMP velocity is drawn from the following 1D probability distribution,
\begin{equation}
f(v) = \sqrt{\frac{27}{4\pi}}\, \frac{v^2}{v_{\rm disp}^3} e^{-3 v^2/4 v_{\rm disp}^2}. 
\label{eq:vpdf}
\end{equation}
Here $v_{\rm disp}$ is the RMS velocity for a single DM particle, which is equal to the three-dimensional velocity dispersion $\sigma_{v,3d}$ defined in Ref.~\cite{Boddy:2018ike} by
\begin{equation}
v_{\rm disp}^2 = \sigma_{v,3d}^2 = \frac{\int dv\,v^4 f(r,v)}{\int dv\, v^2 f(r,v)}.
\end{equation}
Here $f(r,v)$ is the speed distribution for a single DM particle at a distance $r$ from the Galactic Center.

\begin{figure}
\centering
\includegraphics[width=0.49\linewidth]{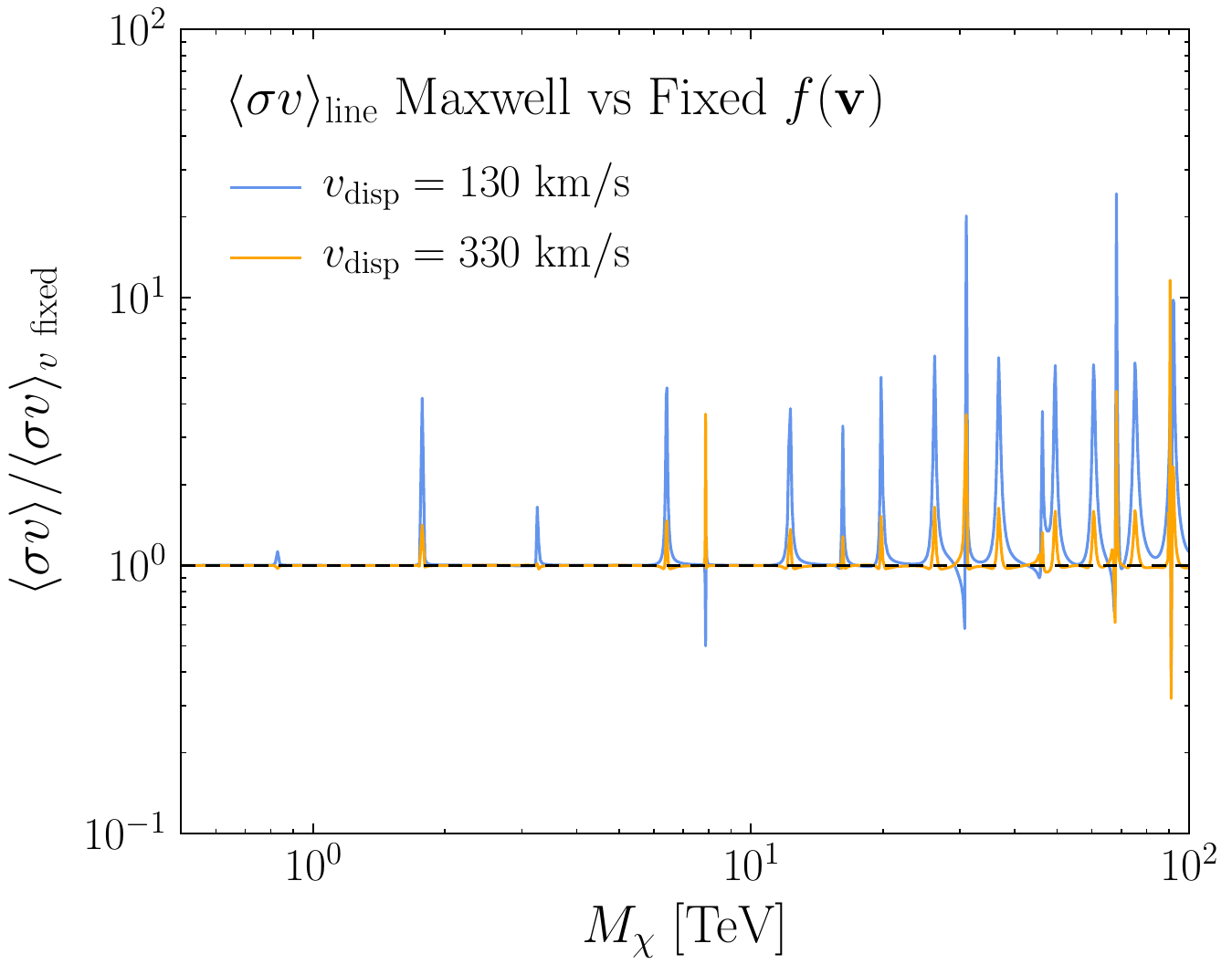}
\caption{An estimate of the range of uncertainty in our results associated with the DM velocity distribution for the dominant direct annihilation.
We use two Maxwell distributions with $v_{\rm disp}$ -- {\it cf.}~Eq.~\eqref{eq:vpdf} -- at the extremal values found by Ref.~\cite{Boddy:2018ike}.
We divide their resulting $\langle \sigma v\rangle$ by that of the simplified case of all quintuplets annihilating with $v = 10^{-3}$.}
\label{fig:arc}
\end{figure}

In Fig.~\ref{fig:arc}, we plot the leading contribution to endpoint photon production, direct annihilation from an $s$-wave initial state, for two different velocity distributions, normalized to the simple assumption of all quintuplets having a fixed $v=10^{-3}$.
Except on resonances, the uncertainty is typically negligible.
We also see that off resonance, and particularly at lower masses, the simple fixed-velocity assumption is a good approximation of either more realistic model.
The reason for this is simply that we are in the saturation regime, as seen in Fig.~\ref{fig:vscans}.
Therefore, we conclude that in general the fixed velocity assumption is a good one at low masses, although at higher masses one is generically underestimating the actual cross section, sometimes by more than an order of magnitude.
Accordingly, for an actual experimental analysis, completeness would require an appropriate weighting of the cross section according to the specific region of interest studied.

\begin{figure}
\centering
\includegraphics[width=0.49\linewidth]{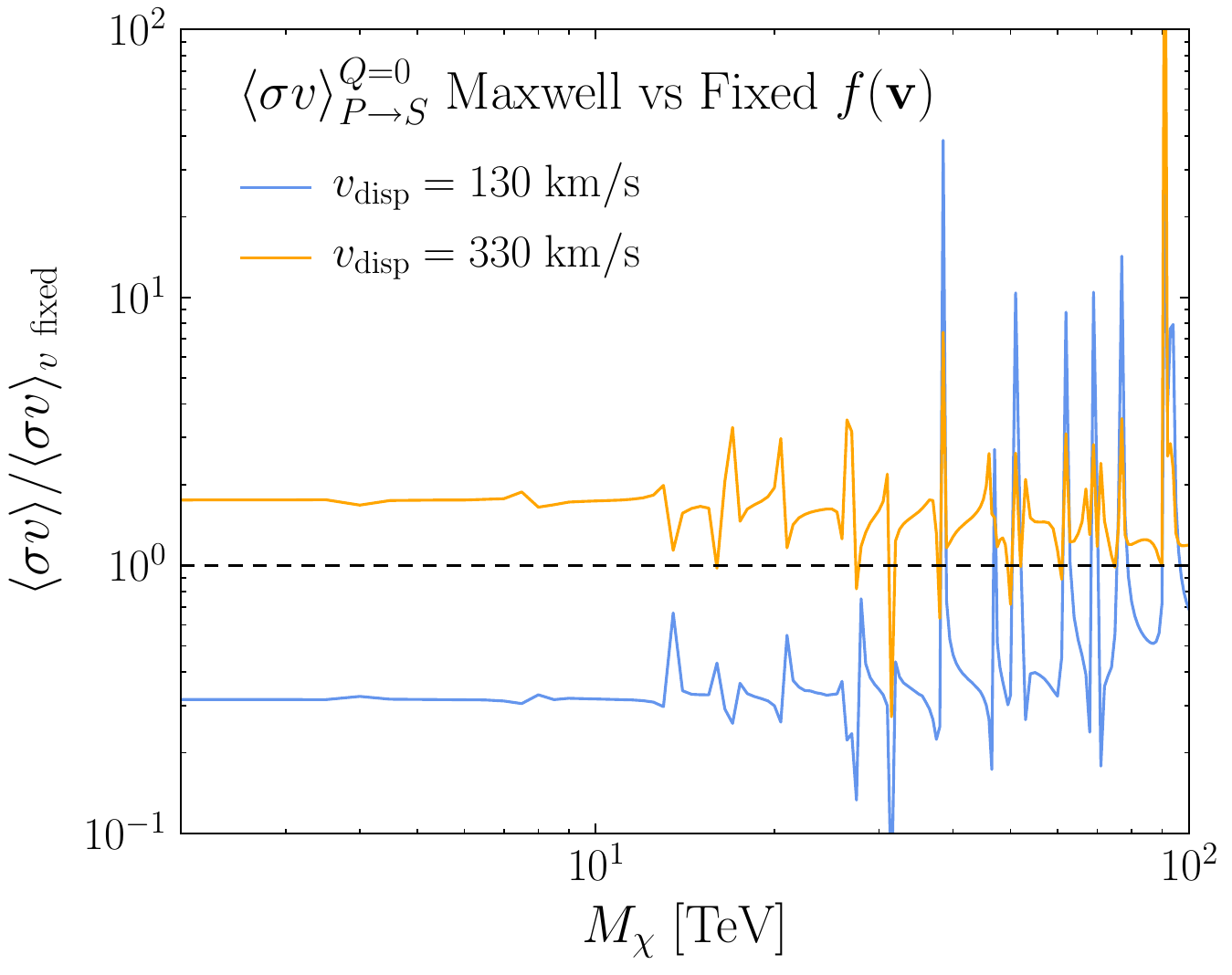}
\caption{Capture rate by emission of a dipole photon from $p$-wave initial state to $s$-wave quintuplet bound state for $Q=0$ states.
The three choices of velocity mirror those in Fig.~\ref{fig:arc}.  We see two different values of $v_\text{disp}$ plotted, plus a $v_\text{fixed}=10^{-3}$ whose capture rate we divide by.}
\label{fig:crc}
\end{figure}

For bound state capture, the off-resonance uncertainties are generally larger than for direct annihilation, as anticipated.
This is demonstrated in Fig.~\ref{fig:crc}, where we show $p$-to-$s$ capture, the leading single-photon dipole transition.
As we see by comparing the rates with those in Fig.~\ref{fig:Line} though, capture is generally far subdominant to direct annihilation.
In this channel, however, the simple assumption of all DM having $v=10^{-3}$ generally provides a result in the middle of the band given by the remaining two options.
Even where it does not, we see that it still provides a good approximation to the more realistic velocity profiles.
Accordingly, just as with direct annihilation, we will take this value as a representative approximation of this subleading contribution to endpoint photons when forecasting experimental sensitivity.

\subsection{Estimating the experimental sensitivity to quintuplet DM}

Finally we turn to an estimate of the experimental sensitivity to the quintuplet DM hypothesis using the spectra we have computed.
Using our definition of the spectrum in Eq.~\eqref{eq:sigmavline-def}, the average DM-generated flux an instrument observes from a region of interest (ROI) of solid angle $\Omega_{\rm ROI}$ is,
\be
\frac{d\Phi}{dE} = \frac{\langle \sigma v \rangle_{\rm line}}{8 \pi M_{\chi}^2} \left( \frac{dN}{dE} \right)_{\rm smeared} \left( \frac{1}{\Omega_{\rm ROI}} \int ds\,d\Omega\,\rho_{\chi}^2 \right)\!.
\label{eq:flux}
\ee
As defined, the flux has units of $[{\rm counts}/{\rm cm}^2/{\rm s}/{\rm TeV}/{\rm sr}]$ (for a detailed discussion of the units, see Ref.~\cite{Lisanti:2017qoz}).
The final term in parentheses here is often referred to as the $J$-factor, and is an integral over the DM density squared in the region being observed.
If the DM density $\rho_{\chi}$ was known exactly, then for a model like the thermal quintuplet the flux is fully determined, as we have computed both the cross section and spectrum (up to residual uncertainties from higher order terms in the theory prediction, and velocity distribution as discussed in the last subsection).

To test the quintuplet hypothesis, we need to compare the flux in Eq.~\eqref{eq:flux} to experimental measurements.
For this, we will estimate the sensitivity of H.E.S.S. to the endpoint photon signal using the ``mock analysis'' method described in Ref.~\cite{Baumgart:2017nsr}.
The approach in that work was to make use of the publicly available H.E.S.S. data in Ref.~\cite{HESS:2013rld}, where a search for $\chi \chi \to \gamma \gamma$ was performed using 112 hours of Galactic Center observations taken by the instrument between $2004-2008$.
This is a small fraction of the observations H.E.S.S. has taken towards the galactic center.
As emphasized in Ref.~\cite{Montanari:2022buj}, the collaboration has already collected roughly 800 hours of data in the region, and continues to collect roughly 150 hours each year.
In that sense, the dataset we consider represents a small fraction of what is available.
A further limitation of our approach is that the analysis in Ref.~\cite{HESS:2013rld} was a search purely for line photons, and therefore adopted a flexible background model that absorbed all smooth components.
The analysis is therefore unsuitable to consider continuum contributions, which can play an important role in these sorts of analyses, as emphasized in, for instance, Ref.~\cite{Rinchiuso:2020skh}.
Our rationale for adopting this mock analysis, however, is that Ref.~\cite{HESS:2013rld} provided enough information that the full analysis they undertook can be performed reconstructed, as was shown in Ref.~\cite{Baumgart:2017nsr}.
Later on we will provide a rough estimate of how sensitive more recent and upcoming analyses could be.

\begin{figure*}[!t]
\centering
\includegraphics[width=0.47\textwidth]{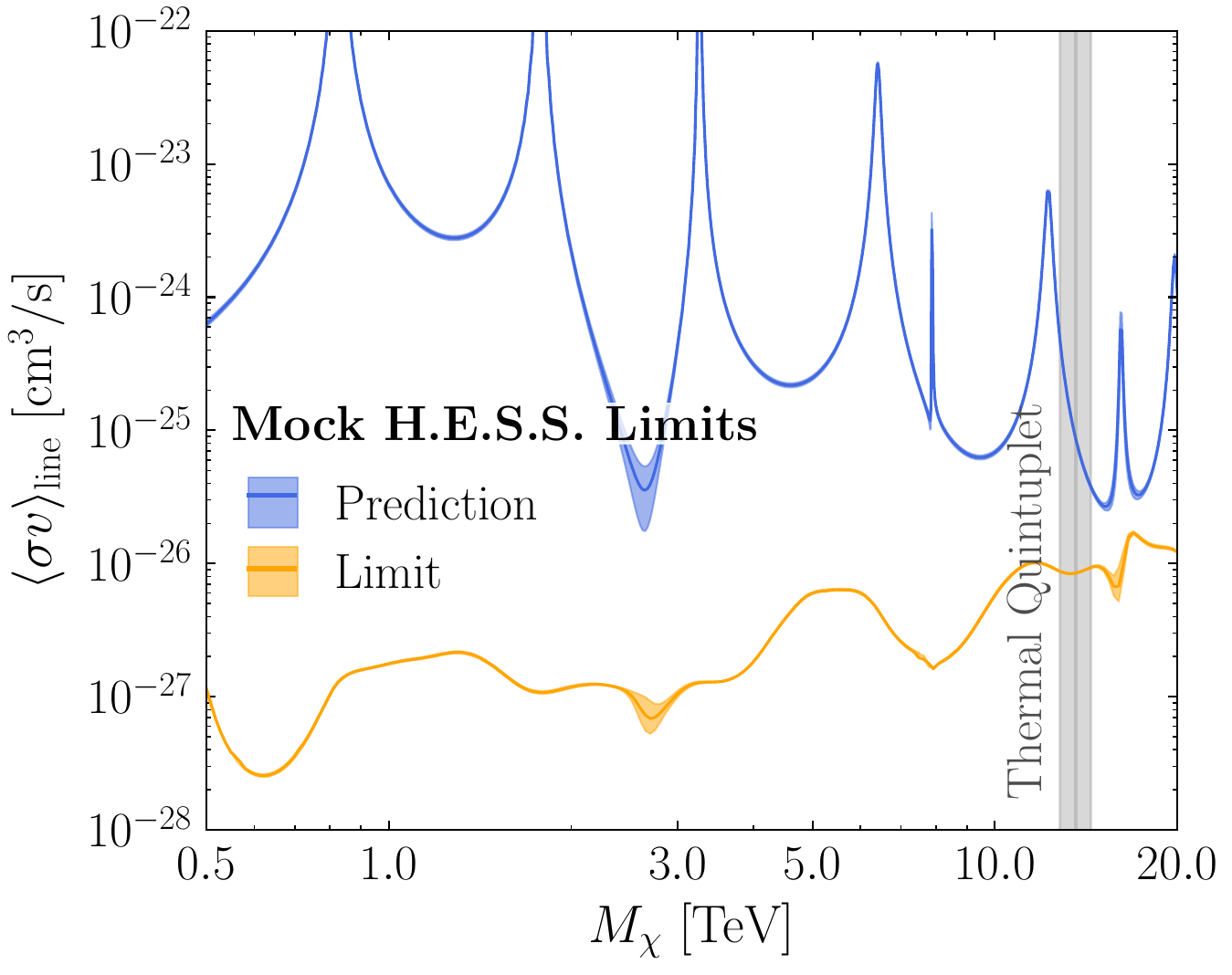} 
\hspace{0.5cm}
\includegraphics[width=0.47\textwidth]{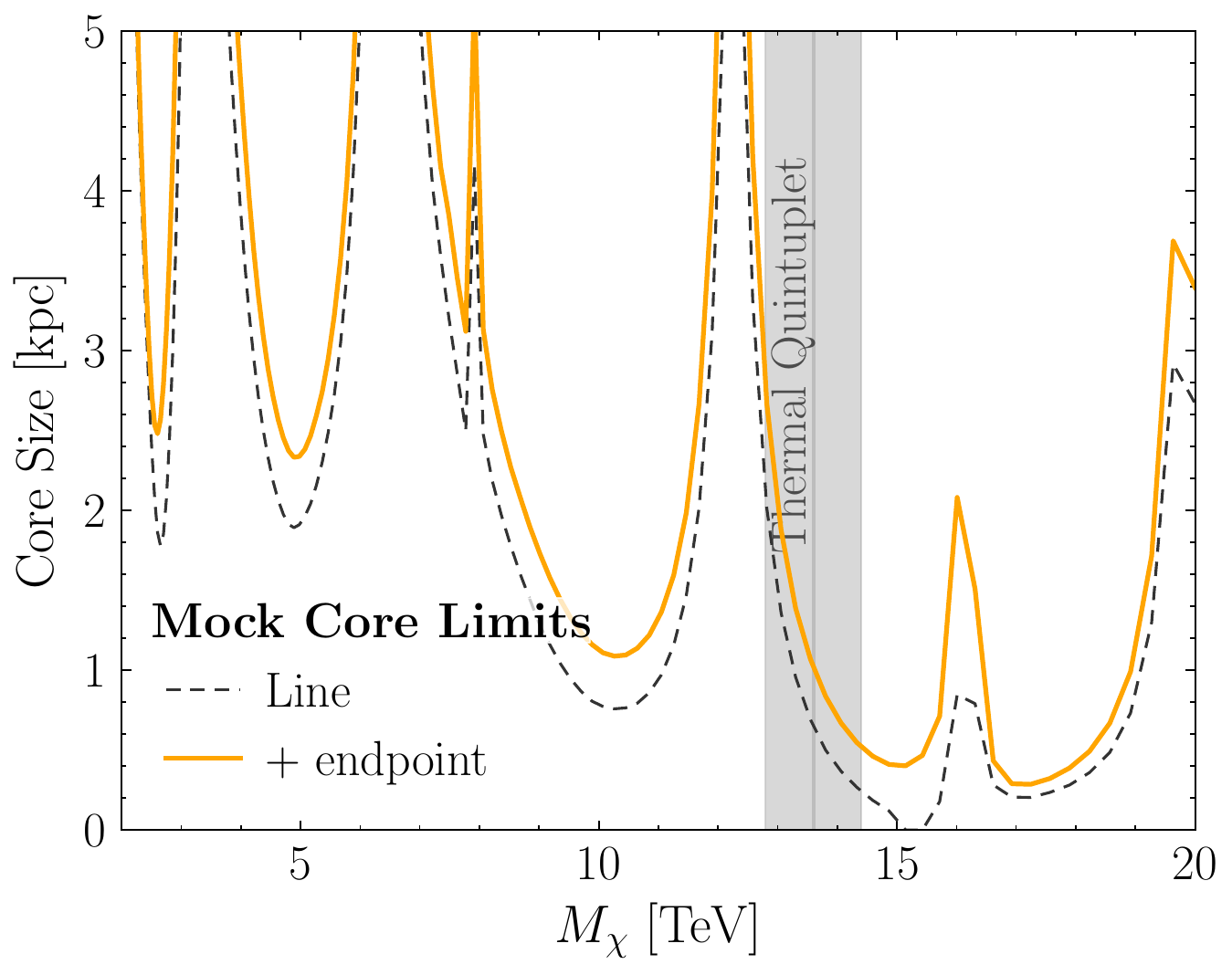}
\caption{
(Left) The estimated sensitivity of H.E.S.S.~I using 112 hours of galactic center (GC) data to the quintuplet.
Assuming an Einasto profile, we show sensitivity to $\langle \sigma v \rangle_{\rm line}$ as defined in Eq.~\eqref{eq:sigmavline-def}, which can then be compared to the equivalent theoretical prediction.
Across the entire mass range considered here, H.E.S.S. would be able to exclude the quintuplet assuming an Einasto profile.
(Right) If the DM profile is cored as in Eq.~\eqref{eq:core}, the core sizes that would be required to make a non-observation consistent with our quintuplet predictions.
At the thermal mass, 13.6 TeV, a core of 1 kpc is required, whereas at upper edge of the thermal band, 14.4 TeV, the results would be consistent with a 0.5 kpc core.
}
\label{fig:Limit}
\end{figure*}

For the mock analysis, we fit the data provided in Ref.~\cite{HESS:2013rld} to a combination of the flux in Eq.~\eqref{eq:flux} and a parametric background model adopted by the experiment---full details are provided in Ref.~\cite{Baumgart:2017nsr}.
If we have a prediction for $\rho_{\chi}$, this approach combined with our prediction for the spectrum can be used to obtain an estimated sensitivity for $\langle \sigma v \rangle_{\rm line}$.
For this we adopt the Einasto profile~\cite{1965TrAlm...5...87E} used by the H.E.S.S. analysis (based on Ref.~\cite{Pieri:2009je}),
\be
\rho_{\rm Einasto}(r) \propto \exp \left[ - \frac{2}{\alpha} \left( \left( \frac{r}{r_s} \right)^{\alpha} - 1\right) \right]\!,
\label{eq:Einasto}
\ee
with $\alpha=0.17$, $r_s = 20~{\rm kpc}$, and the normalization fixed to $0.39~{\rm GeV/cm}^3$ at the solar radius, $r = 8.5~{\rm kpc}$.
The resulting estimated constraint is shown in the left of Fig.~\ref{fig:Limit}.\footnote{One can find an early attempt to make projected $\gamma$-ray constraints from the H.E.S.S. galactic center data in Ref.~\cite{Cirelli:2015bda} (see also Refs.~\cite{Garcia-Cely:2015dda,Aoki:2015nza}).
Their Fig.~7 is analogous to our Fig.~\ref{fig:Limit}.
The earlier paper did not include LL (or NLL) resummation, nor the NLO corrections to the electroweak potential.
For similar Einasto parameters, their bounds on the line cross section are about an order of magnitude weaker than ours (however, note that our predicted line cross section is also smaller).}
We emphasize, that even though this is a constraint on $\langle \sigma v \rangle$ it is {\it not} based solely on the line prediction---the results are based on the estimated detectability of the entire spectrum resulting from line, endpoint, and bound state photons.
For the mass range considered, the bound state contribution is negligible ({\it cf.} Fig.~\ref{fig:Line}), however the endpoint is not: at 13.6 TeV, it enhances the sensitivity by a factor of 1.9.
As mentioned above, by default, the results do not include either continuum contribution considered in Fig.~\ref{fig:Spectra}.
The motivation for this is the particular background model adopted in Ref.~\cite{HESS:2013rld} was designed solely to search for a narrow line feature, and there can be considerable degeneracy with the continuum emission (see the discussion in Ref.~\cite{Baumgart:2017nsr}).
Nevertheless, we have tested adding the continuum emission from direct annihilation to $W$ and $Z$ final states, and found for most masses it has no impact on the estimated limits, although there is a slight fluctuation around the thermal mass, which increases the sensitivity by $\simeq$$20\%$.
We also note that there are several locations, such as just below $M_{\chi} = 3~{\rm TeV}$ and just above the thermal mass where there is a larger theoretical error.
This results from the sharp variations in the spectra observed in Fig.~\ref{fig:Spectra-var}.
In fact, the sensitivity to these features is reduced by the insensitivity of the background model used in this work to smooth features.
These results can also be compared to Ref.~\cite{Montanari:2022buj}, where an alternate H.E.S.S. analysis is performed using the spectra from this work, and much sharper variations are observed in their sensitivity.
(We note that the results of that work made use of the LO Sommerfeld potential calculations, not the NLO results we have adopted here.)

\begin{figure*}[!t]
\centering
\includegraphics[width=0.47\textwidth]{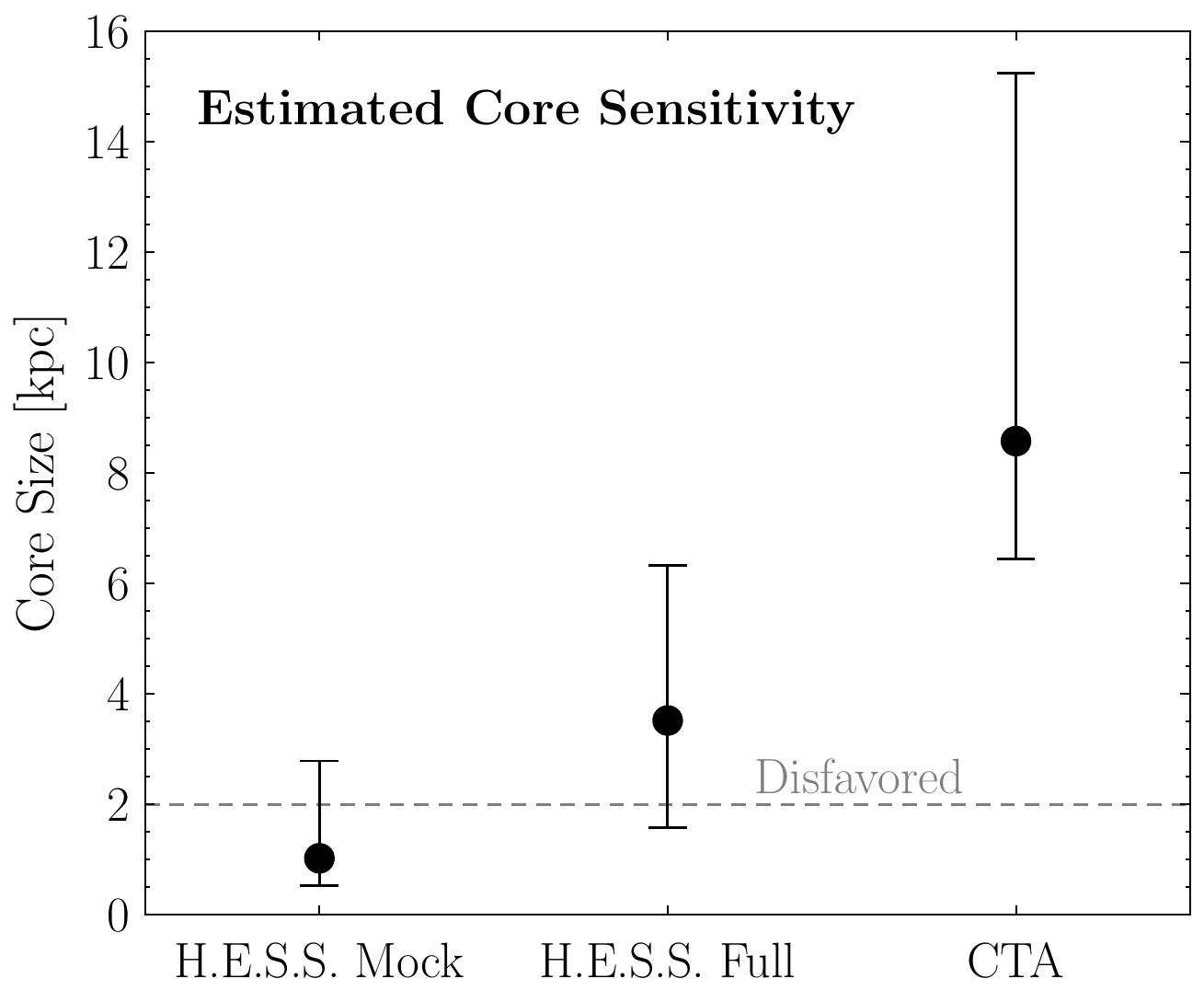} 
\caption{An estimate for the required core size of the Einasto profile, given no preference for a signal is seen in our H.E.S.S. mock analysis, and if no signal emerges in an analysis of the full H.E.S.S. dataset, or at CTA.
The central values correspond to 13.6 TeV, the central thermal mass, whereas the upper and lower error bars correspond to 12.8 and 14.4 TeV, the edges of the thermal mass window.
The dashed corresponds to the rough core size constraint from Ref.~\cite{Hooper:2016ggc}.
Our results suggest that H.E.S.S. can already considerably test the quintuplet, with the final word likely being left to CTA.
}
\label{fig:CoreProjection}
\end{figure*}

The results of the mock analysis suggest that for the central value of the thermal mass, even 112 hours of H.E.S.S. data can exclude the thermal prediction by a factor of 10.
Nevertheless, this varies considerably across the uncertainty band on the thermal mass: at 12.8 TeV, the exclusion factor is 55, whereas it is only 4 at 14.4 TeV.
The sharp variation is a result of the thermal window sitting near a Sommerfeld resonance, as shown in Fig.~\ref{fig:Limit}.
We emphasize once more that although we use the NLO potential in our computations, the thermal mass range was computed with the LO potential, and given the sensitivity of our findings to the exact mass, computing the thermal mass at NLO will be important for narrowing the fate of the thermal quintuplet.
If we relax the thermal cosmology assumption and consider a broader range of masses, we see the quintuplet is excluded across the full $0.5-20~{\rm TeV}$ mass range.
Yet this statement is contingent on the form of $\rho_{\chi}$ adopted, about which there is considerable uncertainty.
In particular, the density profile may flatten toward the inner Galaxy. 
As the annihilation signal is sensitive to $\rho_{\chi}^2$ flattening the profile has a marked impact on the flux, making this one of the dominant uncertainties in $\rho_{\chi}$ for our purposes.
We parameterize a possible flattening of the profile by replacing the Einasto density profile with a constant value for Galactocentric distances $r < r_c$, where we will refer to $r_c$ as the ``core size'':\footnote{To be explicit, we fix the normalization of the Einasto profile in Eq.~\eqref{eq:Einasto} to the solar radius {\it before} we impose the core restriction of Eq.~\eqref{eq:core}.
This implies that for a core size larger than the solar radius, the profile will predict less than 0.39 GeV/cm$^3$ at our location.
Of course, the more significant concern is that such a large core is not consistent with observations, as discussed in the text, and should solely be viewed as a proxy for how much the $J$-factor needs to be reduced.}
\be 
\rho(r) = 
\begin{cases}
\rho_{\rm Einasto}(r) & r > r_c, \\
\rho_{\rm Einasto}(r_c) & r < r_c.
\end{cases}
\label{eq:core}
\ee
We can then ask what choice of $r_c$ would raise the estimated constraint on $\langle \sigma v \rangle_\text{line}$ above the theoretical prediction.
This is plotted in the right-hand panel of Fig.~\ref{fig:Limit}, both employing our full endpoint spectrum and in the case where we (incorrectly) use only the line cross section in setting the bounds.
To provide an estimate for what core sizes are consistent with data, we note that simulations of Milky Way like galaxies can generate ${\cal O}(1~{\rm kpc})$ cores~\cite{Chan:2015tna}, however, measurements of stars in the Bulge seem to disfavor $r_c \gtrsim 2~{\rm kpc}$~\cite{2015MNRAS.448..713P,Hooper:2016ggc}. 
At the lower end of the thermal mass range, 12.8 TeV, the thermal Quintuplet would already be in tension with this, requiring a 2.8 kpc core.
At the central (13.6 TeV) and upper (14.4 TeV) end, however, the required core size is 1.0 and 0.5 kpc, respectively, and therefore not obviously in tension with our mock analysis.
A more recent study claims evidence for a few kpc core that could potentially saturate the earlier limits~\cite{Ou:2023adg}.
If confirmed, this could suppress the $J$-factor by nearly an order of magnitude.
That would challenge the indirect-detection community to set aggressive limits, but as we estimate, even cores of this size are in reach of CTA and possibly H.E.S.S ({\it cf.}~Fig.~\ref{fig:CoreProjection}).

As already mentioned, the mock analysis we consider makes use of only a very small amount of existing data, and with that we forecast a sensitivity to $\langle \sigma v \rangle_{\rm line} \simeq 8.5 \times 10^{-27}~{\rm cm^3/s}$ at the central thermal mass.
(The error on this value due to uncertainty on the spectrum from our NLL calculation is less than 1\%, far smaller than the variation across the thermal mass range, which is closer to 10\%.)
Using 500 hours of H.E.S.S. data and an identical Einasto profile, Ref.~\cite{Montanari:2022buj} forecast a sensitivity at the thermal mass of $\simeq$$9.3 \times 10^{-28}~{\rm cm^3/s}$, almost a factor of ten better than used here.
This is significantly more than the naive $\sqrt{5}$ the additional data would suggest, which can be primarily attributed to the fact that work used H.E.S.S. II observations, whereas we adopt the sensitivity from H.E.S.S. I, combined with a different analysis used in that work.
With such a sensitivity, we would require a core size slightly larger than $3.5~{\rm kpc}$ to save the thermal quintuplet, which would be in tension with observations.
Nevertheless, repeating the process at 14.4 TeV, the required core size would only be $1.6~{\rm kpc}$, and therefore not yet clearly excluded.
We can also give a crude estimate for the sensitivity the upcoming CTA could have for the quintuplet.
Although no dedicated forecast for the quintuplet has been performed using the spectra in our work, we can estimate the improved sensitivity as follows.
Reference~\cite{Baumgart:2018yed} performed an identical mock analysis to ours for the NLL wino spectrum, estimating sensitivity at $M_{\chi} = 13.6~{\rm TeV}$ of $\langle \sigma v \rangle_{\rm line} \simeq 8 \times 10^{-27}~{\rm cm^3/s}$, slightly stronger than the sensitivity to the quintuplet.
Using the identical NLL spectrum, Ref.~\cite{Rinchiuso:2020skh} then estimated that with 500 hours of data, CTA could reach $\simeq 1 \times 10^{-28}~{\rm cm^3/s}$, a factor of eighty improvement.
Assuming the same improvement for the quintuplet, CTA would be sensitive to $\langle \sigma v \rangle_{\rm line} \simeq 1.1 \times 10^{-28}~{\rm cm^3/s}$, excluding the thermal value by a factor of eight hundred.
To not have seen the thermal quintuplet, we would need to core $\rho_{\chi}$ out to almost $8.6~{\rm kpc}$ -- beyond the solar radius -- which is simply inconsistent with observations.
Even at the upper end of the mass range, a core of $6.4~{\rm kpc}$ would be required.
In this sense, CTA would provide the definitive word on the whether the thermal quintuplet is the DM of our Universe.

These results are summarized in Fig.~\ref{fig:CoreProjection}, where the point shows the core size required for the central thermal mass, 13.6 TeV, whereas the upper and lower error bars corresponds to the lower and upper ends of the thermal mass range.
We also show the core size disfavored by the analysis in Ref.~\cite{Hooper:2016ggc}.
The figure summarizes the conclusion reached above: CTA will seemingly have the final word of the thermal quintuplet, however if a full analysis of H.E.S.S. data sees no sign of the signal, the model would already begin to be disfavored.
There are two important caveats to this conclusion.
The first is that our findings are based off an extrapolation from a mock analysis of H.E.S.S. I data, and are no substitution for a full analysis or projection using the present and forecast H.E.S.S. and CTA instrumental responses.
To give one example of what could change, an analysis that accounts for the continuum emission could be able to even more strongly test the quintuplet.
Secondly, the range of masses we have considered is the thermal mass window of $13.6 \pm 0.8~{\rm TeV}$ that was determined using the LO potential, whereas the remainder of our calculations use the NLO results, as emphasized several times already.
Updating the thermal mass using the NLO potentials will be important.
To give a sense for the impact this could have, repeating the analysis in Fig.~\ref{fig:CoreProjection} for 13.6 TeV using the LO potential in our calculations, the core size would change from 1.0, 3.5, and 8.5 kpc, to 0.7, 2.5, and 7.1 kpc.

\section{Conclusions}
\label{sec:conclusions}

For all the vastness of the DM parameter space, a thermal WIMP has remained a constant focus for decades.
Minimal DM is an exemplar of thermal DM, and through indirect detection, many of the associated models are on the verge of being detected or firmly excluded, as we have shown for the quintuplet in the present work.
Either way, these are important times in the search for DM.

With this in mind, the present work has computed the quintuplet annihilation spectrum to NLL accuracy, and established the formalism to straightforwardly extend this to higher odd SU(2) representations.
We plot the spectrum along with projected limits from a simple extension of the H.E.S.S. I analysis in Fig.~\ref{fig:Limit}.
In doing so, we have demonstrated the power of the EFT of Heavy DM, and also extended this formalism to include the contribution from the rich set of bound states the model contains.
While the bound states can make a significant contribution to the continuum photon emission, their impact on the number of photons with $E_{\gamma} \sim M_{\chi}$ is minimal, except at isolated masses.
As seen in earlier studies of the wino, the same cannot be said for endpoint photons from direct annihilation, which again provide an ${\cal O}(1)$ correction to the line signal seen in IACTs.

Taken together, we estimated that with our spectra, H.E.S.S. should almost be able to probe the entire allowed range for the quintuplet once uncertainties on the DM density in the inner galaxy are accounted for.
Performing this analysis using the existing data, and the soon-to-be-collected data with CTA will be critical ({\it cf.}~Fig.~\ref{fig:CoreProjection}).
The use of background models which enhance sensitivity to smooth features such as the continuum as well as the full contribution from the endpoint-photon spectrum can provide an additional piece of experimental leverage beyond conventional line searches.

On the theory side, the thermal abundance should be recomputed using NLO potentials, as the sensitivity to the quintuplet depends strongly on what end of the predicted thermal mass range one sits.
Finally, it will be interesting to extend the techniques in this work to additional representations, such as a $\mathbf{7}$ of SU(2), where we expect key features of the quintuplet such as the strong variation in the spectrum as a function of mass, to appear and be even more pronounced.

\begin{acknowledgments}

The work we have presented benefited from useful discussions with Tobias Binder, Marco Cirelli, Tongyan Lin, Alessandro Montanari, Emmanuel Moulin, Nikhil Raghuram, and Diego Redigolo.
MB is supported by the DOE (HEP) Award DE-SC0019470.
VV is supported by startup funds from the University of South Dakota.
TRS' work is supported by the Simons Foundation (Grant Number 929255, T.R.S), by the National Science Foundation under Cooperative Agreement PHY-2019786 (The NSF AI Institute for Artificial Intelligence and Fundamental Interactions), and by the U.S. Department of Energy, Office of Science, Office of High Energy Physics of U.S. Department of Energy under grant Contract Number DE-SC0012567.
TRS thanks the Aspen Center for Physics, which is supported by National Science Foundation grant PHY-2210452, for hospitality during the completion of this work.

\end{acknowledgments}

\appendix
\addtocontents{toc}{\protect\setcounter{tocdepth}{1}}
\section*{Appendix}

\section{Quintuplet Dark Matter: A Brief Review}
\label{app:QDM}

Here we provide a brief review of quintuplet DM, also referred to as 5-plet electroweak DM.
In particular, we firstly outline the relevant group theory necessary to specify the interactions used for the calculations in the main text.
We will then review phenomenological aspects of the model beyond indirect detection.

\subsection{Interactions}

Quintuplet DM consists of adding to the SM five Majorana fermions that transform together in the $\mathbf{5}$ representation of SU(2), and as a singlet under the remaining SM forces.
Above the electroweak symmetry breaking scale, we collect the five fields into a multiplet $\chi = (\chi^1,\,\ldots,\,\chi^5)^T$, in terms of which the DM Lagrangian takes the following form
\bea
\mathcal{L}_{\scriptscriptstyle \textrm{DM}}
= \frac{1}{2} \bar{\chi} (i \slashed{D} - M_{\chi}) \chi
= \frac{1}{2} \bar{\chi} ([i \slashed{\partial} - M_{\chi}] \mathbb{1} + \gW T_{\mathbf{5}}^a \slashed{W}^a) \chi.
\label{eq:QLagrangian}
\eea
In the final expression, the first two contributions represent the kinetic terms for each of the fields, which are diagonal amongst the multiplets as indicated by $\mathbb{1}$.
The final term describes the interaction between the additional fields and the SM electroweak bosons.
Importantly, we emphasize that the interaction strength is specified by the SM SU(2) gauge coupling, $\gW$, and is not a free parameter.
Instead, $M_{\chi}$ is the unique free parameter in the theory.
If we assume a conventional thermal origin for the quintuplet, then even the mass can be fixed by the observed relic density to $M_{\chi} = 13.6 \pm 0.8~{\rm TeV}$~\cite{Mitridate:2017izz,Bottaro:2021snn}.\footnote{As emphasized above, this value was computed with the LO electroweak potential.  Redoing the analysis with the NLO potential would reduce an important theoretical uncertainty.}
The 13.6 TeV quintuplet is therefore a zero parameter DM model.

After electroweak symmetry breaking, the five Majorana fermions rearrange themselves into three mass eigenstates: a Majorana neutral fermion $\chi^0$, and two charged Dirac fermions $\chi^+$ and $\chi^{++}$.
At leading order, each of these states maintains a mass of $M_{\chi}$.  However, radiative corrections to the charged states break this degeneracy, raising the masses of the charged fermions, singling out $\chi^0$ as the lightest state and the DM candidate.
These corrections have been computed, and in detail $\delta_{0} = M_{\chi^+}-M_{\chi^0} \simeq 164~{\rm MeV}$, and $\delta_{+} = M_{\chi^{++}}-M_{\chi^+} = 3 \delta_{0}$~\cite{Cirelli:2005uq,Ibe:2012sx}.
For most aspects of our calculations these mass splittings will be irrelevant and we will take $\delta_{0} \simeq \delta_{+} \simeq 0$.  However, we do include them in the electroweak potential used to compute Sommerfeld factors, and scattering- \& bound-state wavefunctions.  This is done by adding $2\delta_0$ to the diagonal term in the potential matrix correspondingto the $\chi^+\chi^-$ component of any state, $8\delta_0$ to the diagonal term corresponding to the $\chi^{++}\chi^{--}$ component, and similar, appropriate shifts for diagonal elements in the potential matrix corresponding to components of $Q\neq 0$ states (the shift is given by the difference between the rest mass of the state constituents and $\chi^0\chi^0$).

We now turn to the interaction term in Eq.~\eqref{eq:QLagrangian}, $\tfrac{1}{2} \gW \bar{\chi} T_{\mathbf{5}}^a \slashed{W}^a \chi$.
Here $a=1,2,3$ indexes the electroweak gauge bosons, which transform together in an adjoint of SU(2).
In the broken theory, these are mapped to the charge and mass eigenstates in the usual way,
\bea
W^1_{\mu} = \frac{1}{\sqrt{2}} (W^+_{\mu}+W^-_{\mu}),\;\;
W^2_{\mu} = \frac{i}{\sqrt{2}} (W^+_{\mu}-W^-_{\mu}),\;\;
W^3_{\mu} = \sW A_{\mu} + \cW Z_{\mu}.
\label{eq:brokenbosons}
\eea
The only part of Eq.~\eqref{eq:QLagrangian} that remains undetermined is $T_{\mathbf{5}}^a$, the three generators of SU(2) in the quintuplet representation.
A convenient basis in which to specify $T_{\mathbf{5}}^a$ is the basis of charged states discussed above, where the DM can be cleanly identified.
We can determine the charged states through their couplings to the bosons in Eq.~\eqref{eq:brokenbosons}.
In particular, as $A_{\mu}$ couples to charge, we can read off the charges of the states as soon as we know $T^3$.
It will also be convenient to introduce
\be
T^{\pm} = \frac{1}{\sqrt{2}} (T^1 \pm i T^2),
\ee
in terms of which $T^a W^a = T^+ W^+ + W^- T^- + W^3 T^3$, independent of representation.

Before evaluating the generators for the quintuplet, let us review how the argument proceeds for the simpler case of the wino---a triplet or $\mathbf{3}$ of SU(2).
In this case, it is conventional to exploit the fact that the generators are given by the structure constants of SU(2), so that
\be
\bar{\chi} T_{\mathbf{3}}^a \gamma^{\mu} \chi = \bar{\chi}_b (T_{\mathbf{3}}^a)_{bc} \gamma^{\mu} \chi_c = - i \epsilon_{abc} \bar{\chi}_b \gamma^{\mu} \chi_c.
\ee
This approach depended on the fact we already knew a representation of the generators for the adjoint, and so does not simply generalise to larger representations.
However, we can derive a representation more systematically as follows.
Recall that one representation of $\mathbf{n}$ in SU(2) is an $n-1$ index symmetric tensor, with each index transforming in the fundamental.
In this representation we denote the adjoint as $\chi^{ij}$, and the quintuplet as $\chi^{ijkl}$, where the indices take values 1 and 2.
Beginning with the wino, $\chi^{ij}$ has three unique components, which we can embed into a vector as\footnote{The $\sqrt{2}$ ensures that $\chi$ as it appears in $\bar{\chi} \slashed{\partial} \chi/2 = \bar{\chi}^{ij} \slashed{\partial} \chi^{ij}/2$ is canonically normalized.
An identical argument explains the coefficients in Eq.~\eqref{eq:Qstvector}.}
\be
\chi = \begin{pmatrix} \chi^1 \\ \chi^2 \\ \chi^3 \end{pmatrix} 
= \begin{pmatrix} \chi^{11} \\ \sqrt{2} \chi^{12} \\ \chi^{22} \end{pmatrix}\!.
\ee
We can use this representation to explicitly construct $T^a_{\mathbf{3}}$ as follows.
The key is that we know exactly how the generators act on $\chi^{ij}$ as each index transforms in the fundamental.
Accordingly,
\be
T^a (\chi^{ij}) = [ (T^a_F)^i_k \delta^j_l + \delta^i_k (T^a_F)^j_n] \chi^{kl},
\ee
where $T^a_F = \sigma^a/2$, with $\sigma^a$ the Pauli matrices.
Consider a generic infinitesimal transformation, $U = 1 + i u$, with $u = u_a T^a$.
If we take $u^a = (0,0,\kappa)$ with $\kappa \ll 1$, then the components of $\chi$ transform as
\bea
\delta \chi^1 = i \kappa  \chi^1,\;\;
\delta \chi^2 = 0,\;\;
\delta \chi^3 = i \kappa \chi^3\!.
\eea
From this, we can read off the action of an infinitesimal $\mathbf{U}$ on $\chi$, and hence infer that we must have
\be
T^3_{\mathbf{3}} = {\rm diag}(+1,\,0,\,-1).
\ee
We can now identify $\chi^1$, $\chi^2$, and $\chi^3$ as having charges $+1$, $0$, and $-1$, yielding the expected spectrum in the broken phase.\footnote{The representation in the charge basis is not unique.
For instance, the transformation $\chi^{1,2} \to e^{\pm i\phi} \chi^{1,2}$ leaves the charge assignments unchanged, but will introduce a phase into the off-diagonal $W^{\pm}$ couplings.
The same will be true for the off-diagonal quintuplet couplings.}
The remaining components of $T^a_{\mathbf{3}}$ can be derived identically.

This approach readily generalizes to the quintuplet.
The five unique components of $\chi^{ijkl}$ can be embedded into a vector as follows,
\be
\chi = \begin{pmatrix} \chi^1 \\ \chi^2 \\ \chi^3 \\ \chi^4 \\ \chi^5 \end{pmatrix} 
= \begin{pmatrix} \chi^{1111} \\ 2 \chi^{1112} \\ \sqrt{6} \chi^{1122} \\ 2 \chi^{1222} \\ \chi^{2222} \end{pmatrix}
= \begin{pmatrix} \chi^{++} \\ \chi^+ \\ \chi^0 \\ \chi^- \\ \chi^{--} \end{pmatrix}\!.
\label{eq:Qstvector}
\ee
To justify the charge assignments, we repeat the above analysis to find
\be
T^3_{\mathbf{5}} = {\rm diag}(+2,\,+1,\,0,\,-1,\,-2).
\ee
Further, we can compute
\bea
T^+_{\mathbf{5}} =
\begin{pmatrix}
0 & \sqrt{2} & 0 & 0 & 0 \\
0 & 0 & \sqrt{3} & 0 & 0 \\
0 & 0 & 0 & \sqrt{3} & 0 \\
0 & 0 & 0 & 0 & \sqrt{2} \\
0 & 0 & 0 & 0 & 0
\end{pmatrix}\!,
\eea
and $T^-_{\mathbf{5}} = (T^+_{\mathbf{5}})^T$.
In the main body we will exclusively work in the charged basis of Eq.~\eqref{eq:Qstvector}, using the form of $T_{\mathbf{5}}^a$ as above whenever necessary.

\subsection{Phenomenology}

We end this section with a brief description of quintuplet phenomenology beyond indirect detection.
As already noted, the case where $M_{\chi} \simeq 13.6$ TeV is particularly appealing, as this is the mass singled out from a conventional cosmology via a WIMP-miracle-style argument.
The interactions of the quintuplets, reviewed in the previous subsection, are sufficient to keep it in thermal equilibrium in the early Universe.
As the Universe cools, eventually the quintuplet undergoes a conventional freeze-out.
When this occurs dictates the final abundance, and matching to the observed DM density fixes the single parameter of the theory, $M_{\chi}$.
For larger (smaller) values of $M_{\chi}$ than 13.6 TeV, one naively over (under) produces the observed DM density.
Nevertheless, it is entirely possible that there are effects in the early Universe that modify this simple picture.
The presence of additional beyond-the-SM states that either decay to the SM or directly to the quintuplet can dilute or increase its abundance, respectively, making a wider mass range viable.
Further, for $M_{\chi} \leq 13.6$ TeV, even with no additional states, the quintuplet would represent a well motivated (and predictable) fraction of DM.
For these reasons, in the main body we considered a wide range of quintuplet masses, although we emphasize once more that the scenario where $M_{\chi} \simeq 13.6$ TeV is compelling.

As with any DM candidate, one can also consider searching for the quintuplet with either direct detection or at a collider.
For direct detection, as the quintuplet carries no U(1) hypercharge, it does not couple to the $Z$ boson at tree level.
Nevertheless, couplings to SM nucleons can arise at loop level.
The spin-independent cross-section is $\sigma \simeq (1.0 \pm 0.3) \times 10^{-46}~{\rm cm}^2$~\cite{Bottaro:2021snn} (see also Ref.~\cite{Hisano:2015rsa,Chen:2023bwg}), beyond the reach of current searches at 13.6 TeV, and even next generation instruments such as LZ~\cite{Mount:2017qzi}.
Nevertheless, the cross-section is a factor of $\sim 4$ above the neutrino floor, and is in reach of generation-3 instruments such as DARWIN~\cite{DARWIN:2016hyl}.

Detection at colliders is similarly challenging, but also potentially within reach of future instruments.
Existing LHC searches reach masses of $\sim 270~\text{GeV}$; even the high-luminosity dataset will only reach to $\sim 520~\text{GeV}$~\cite{Ostdiek:2015aga}.
Even future hadron colliders are unlikely to reach the thermal mass.
A future 100 TeV hadron collider will just be able to reach the thermal masses for canonical neutralinos such as the higgsino and wino~\cite{Cirelli:2014dsa,Low:2014cba}.
Given these two candidates have significantly lower thermal masses of 1 and 2.9 TeV respectively~\cite{Hisano:2006nn,Cirelli:2007xd,Hryczuk:2010zi,Beneke:2016ync}, the prospects of probing the 13.6 TeV quintuplet appear discouraging.
Future muon colliders operating at lower center-of-mass energies could reach the quintuplet, although they would still need to obtain $\sqrt{s} \simeq 35~{\rm TeV}$~\cite{Bottaro:2021snn} (see also Refs.~\cite{Han:2020uak,Han:2022ubw}).
Taken together, in the short term indirect detection remains the most likely avenue for probing quintuplet DM.


\section{Operators for higher-$L$ Bound State annihilation}
\label{app:HigherL}

As argued in Sec.~\ref{sec:bsdecay}, the higher-$L$ bound states preferentially decay to the deeper bound states with lower $L$ instead of directly annihilating to SM particles.
Nevertheless, for completeness here we provide the complete set of relevant operators up to ${\cal O}(v)$.

We consider the structure of the operators which support up to $p$-wave bound states.
Thus we need to keep operators (at the amplitude  level)  suppressed by at most one power of the DM 3-momentum; the $\mathcal{O}(v^0)$ operators will support both the direct annihilation as well as $s$-wave  bound state annihilation.
By matching to the full tree level amplitude, we can obtain the structure that supports $S=0,\,L=1$ states,\footnote{Throughout this appendix we will work with 4-component DM fields and not reduce it to 2 components as was done for the $L=S=0$ operator in the main text.}
\begin{equation}
\mathcal{O}_1 = \mathbf{v}_{\chi} \cdot \mathbf{n}\left( \bar \chi \Big[T^a,T^b \Big]\gamma^0 \gamma^5 \chi \right)i \epsilon^{ijk}(n-\bar n)^k{\cal B}_{\perp n}^{i,a}{\cal B}_{\perp \bar n}^{j,b},
\end{equation}
where it is understood that the subscript $\chi$ on $\mathbf{v}_{\chi}$ indicates that the velocity vector is the velocity of the state created/annihilated by the $\chi$ field (as opposed to the $\bar \chi$ field).
As is evident, this operator supports a bound state with $L=1$ and $S=0$.
Likewise we can write down the next set of operators,
\begin{equation}\begin{aligned}
\mathcal{O}_2 &= \mathbf{v}_{\chi}\cdot \mathbf{n}\left( \bar \chi \Big\{T^a,T^b \Big\}\gamma^i \chi \right)(n-\bar n)^i  {\cal B}_{\perp n}^{\mu,a}{\cal B}^{b}_{\perp \bar n \mu},\\
\mathcal{O}_3 &= \mathbf{v}_{\chi} \cdot {\cal B}^a_{\perp n} {\cal B}^b_{\perp \bar n \mu} \left(\bar \chi T^bT^a \gamma^{\mu \perp}\chi \right)\!, \\
\mathcal{O}_4 &= \mathbf{v}_{\chi} \cdot {\cal B}^b_{\perp \bar n} {\cal B}^a_{\perp  n \mu} \left(\bar \chi T^aT^b \gamma^{\mu \perp}\chi \right)\!.
\label{basis}
\end{aligned}\end{equation}
From their Dirac structure, it is clear that these operators support $L=1, S=1$ bound states.
There is also another operator which supports an $L+S$ odd bound state, specifically the $L=1$, $S=0$ bound state which arises out of a correction to an ultra-soft gauge boson emission off the heavy DM particles.
The details of this operator are involved and hence are separately given further below.\footnote{A complete anaysis and resummation of this channel at NLL would require us to do a full two loop computation in order to recover the anomalous dimension as well matching to stage II of the EFT.
We leave this work for the future.}

We note that decays of SU(2)-singlet bound states with $L+S$ odd into two (transverse) gauge bosons are constrained by charge conjugation invariance and parity.
This is because the underlying Lagrangian in Eq.~\eqref{eq:QLagrangian} which couples the electroweak gauge sector and a Majorana quintuplet fermion is $C$ and $P$ invariant.
A fermion-antifermion bound state has $C$ eigenvalue $(-1)^{L+S}$ and $P$ eigenvalue $(-1)^{L+1}$.
A $\gamma \gamma$, $\gamma Z$, or $ZZ$ final state has $C$ eigenvalue $+1$, thereby forbidding an $L+S$ odd bound state decay into them by $C$ alone.
Decays to the $W^+W^-$ final state are allowed regardles of the $L,\,S$ quantum numbers of the bound state.\footnote{One might think the Landau-Yang theorem also forbids decay from $L+S$-odd initial states into two bosons at all orders, but the application of the theorem to non-Abelian theories is subtle; there is a generalized Landau-Yang theorem that holds so long as the decay products are in a color-singlet state \cite{Beenakker:2015mra}, but this is not the case for the decay of $L+S$-odd states.
In Ref.~\cite{Asadi:2016ybp}, one can see explicitly that for wino-onium, decays to $W^+W^-$ are allowed for all combinations of initial-state $L$ and $S$.}
In all cases, decays to longitudinal gauge bosons are potentially allowed, as well.

Finally, let us return to the additional operator which is sub-leading in velocity that we can write down by looking at the emission of another gauge boson which usually contributes to the $Y_v$ Wilson line.
We do not get any such correction from the $Y_n$ or the $Y_{\bar n}$ Wilson line since we do not wish to consider sub-leading terms in the SCET power counting parameter.

We start again with our ${\cal O}(v^0)$ operator at the amplitude level before we dress it with soft Wilson lines
\bea
\bar \chi \{T_{\chi}^a,T_{\chi}^b \} \Gamma \chi {\cal B}_{\perp n}^a {\cal B}^b_{\perp \bar n},
\eea
where $\Gamma$ is an arbitrary Lorentz structure. 
Consider the emission of an SCET ultra-soft gauge boson of momentum k off the initial $\chi$ or $\bar \chi$ particle.
First looking at the $\chi$ particle, 
\bea
u_e(\tilde{p})+ ig \frac{ i(\slashed{p}-\slashed{k}+M_{\chi})}{(p-k)^2-M_{\chi}^2+i\epsilon} \gamma^{\mu}\epsilon_{\mu}^a(T_{\chi}^a)_{ce} u_e(p) 
\eea
$\tilde p$, p is the momentum of the $\chi$ particle.
Let us now expand this result to $\mathcal{O}(v)$ (except inside the spinor).
We see that apart from the usual Wilson line contribution, we also get two other relevant terms,
\bea
u_e(\tilde p) - \gW (T_{\chi}^a)_{ce}\left( -\frac{\epsilon_0^a}{k_0}+ \frac{\mathbf{v} \cdot \mathbf{\epsilon}^a(k)}{k_0}- \frac{\epsilon_0^a(k)\mathbf{v} \cdot \mathbf{k}}{k_0^2} \right) u_e(M_{\chi}).
\eea
If we sum the infinite series of gauge bosons with one of the propagators expanded out to $\mathcal{O}(v)$, we then have a structure 
\bea
\chi \rightarrow  Y_v \mathbf{v} \cdot \mathbf{B_s}\chi
\eea
where we have defined the following Hermitian operator, 
\bea
\mathbf{B}_s =\frac{Y^{\dagger}_v \left(\mathbf{\mathcal{P}}-\gW\mathbf{A}_s \right)Y_v}{v \cdot \mathcal{P}},
\eea
where $\mathcal{P}$ is the momentum label operator.
If this term is included in a larger expression, the $v \cdot \mathcal{P}$ factor in the denominator only acts on the terms in the numerator while the one in the numerators acts only on the $Y_v$ Wilson line to the right.
We can now combine this with the emissions off $\bar \chi$ to give us an effective operator, now dressed with soft Wilson lines
\bea
\bar \chi \Big\{ Y_{v}^{\dagger}\{T^a,T^b \} \Gamma Y_v , \mathbf{v} \cdot \mathbf{B}_s\Big\}\chi {\cal B}_{\perp n}^{a'} {\cal B}^{b'}_{\perp \bar n}  Y_{n}^{aa'} Y_{\bar n}^{bb'}.
\eea
We can once again use our Wilson line identity to write this in terms of our usual soft function
\bea
\bar \chi \Big\{ \{T^a,T^b \} \Gamma , \mathbf{v} \cdot \mathbf{B}_s\Big\}\chi {\cal B}_{\perp n}^c {\cal B}^d_{\perp \bar n}  Y^{abcd}.
\eea
We can then formally define a new object $\mathbfcal{B}_s$ to explicitly separate out all the soft fields from the heavy DM fields
\bea
\mathbfcal{B}_s^a T^a = \mathbf{B}_s.
\eea
Then we see that 
\bea
\mathbfcal{B}_s^a = \mathrm{Tr}[ \mathbf{B}_s T^a],
\eea
where we have used $\mathrm{Tr}[T^aT^b] = \delta^{ab}$.
Our operator now becomes 
\bea
\bar \chi \Big\{ \{T^a,T^b \} \Gamma , T^e \Big\}\chi \, {\cal B}_{\perp n}^c {\cal B}^d_{\perp \bar n}  Y^{abcd}\mathbf{v} \cdot \mathbfcal{B}_s^e.
\eea
	
In summary, we have an additional soft function which is explicitly suppressed in velocity.
The Dirac structure for this operator is $\gamma^0\gamma^5$ so that this is the only operator that supports an L+S odd bound state.
However, it is clear that the soft operator only begins at one loop and hence we are always forced into at least a 3 gauge boson final state.

At the amplitude squared level, we simply have a square of this operator and there will no interference terms with other  bound state operators since all other operators support an L+S even bound state.
Let us consider the soft operator,
\bea
S^{abea'b'e'}=	\langle 0|( Y^{a'b'3d}\mathbf{v} \cdot \mathbfcal{B}_s^{e'})^{\dagger}\mathcal{M}|X_s\rangle \langle X_s|  Y^{ab3d}\mathbf{v} \cdot \mathbfcal{B}_s^{e}|0\rangle,
\eea
where $\mathcal{M}$ is the measurement performed on the soft operator and $X_s$ are the soft modes.
This soft operator now has 6 free indices which must be contracted into the DM wavefunction factor.
Since the soft final state is completely inclusive, we can simplify the operator as
\bea
S^{abea'b'e'}=	|\mathbf{v}|^2 \langle 0|( Y^{a'b'3d} \mathbfcal{B}_s^{e' i})^{\dagger}\mathcal{M}|X_s\rangle \langle X_s| Y^{ab3d}\mathbfcal{B}_s^{e i}|0\rangle.
\eea
We can now move the $|\mathbf{v}|^2$ factor to the DM wavefunction and instead redefine a soft operator 
\bea
S^{abea'b'e'} =	\langle 0|( Y^{a'b'3d} \mathbfcal{B}_s^{e' i})^{\dagger}\mathcal{M}|X_s\rangle \langle X_s|  Y^{ab3d}\mathbfcal{B}_s^{e i}|0\rangle.
\eea
	
The only term that contributes at one loop is the $ \mathbfcal{B}$ since it is 0 at ${\cal O}(\aW^0)$.
So we can set all other Wilson lines to their tree level values.
Explicitly,
\bea
Y^{ab3d} \big\vert_\textrm{tree} = (Y_v^{fa} Y_n^{f3})(Y_v^{gb} Y_{\bar n}^{gd}) \big\vert_\textrm{tree} = \delta^{a3} \delta^{bd}.
\eea
Thus, our operator at one loop becomes
\bea
S^{abea'b'e'}_{1\mhyphen\mathrm{loop}} = \delta^{a3}\delta^{a'3} \delta^{bb'} \langle 0|(\mathbfcal{B}_s^{e',i})^{\dagger}\mathcal{M}|X_s\rangle \langle X_s| \mathbfcal{B}_s^{e,i}|0\rangle.
\eea
We will only calculate this operator to one loop to elucidate its properties.
The one-loop integrand up to an overall factor take the form,
\begin{equation}\begin{aligned}
I =&2 \delta^{ee'}\gW^2 \int \frac{d^dk}{(2\pi)^{(d-1)}}\frac{\delta^+(k^2-M_{\chi}^2)\delta(q^+-k^+)}{k_0^2} \\
-  &\delta^{ee'}\gW^2\int \frac{d^dk}{(2\pi)^{(d-1)}}\frac{\delta^+(k^2-M_{\chi}^2)\delta(q^+-k^+)k^2}{k_0^4}.
\end{aligned}\end{equation}
where $q^+$ is the contribution to the final photon momentum from the soft function.
The second term is proportional to $M_{\chi}^2$ and gives a power correction in $M_{\chi}^2/(q^+)^2$ and hence can be ignored.
The first term  does not give a UV divergence, but will contribute a log which will be relevant for stage 2 of the EFT.
In detail,
\begin{equation}
2\delta^{ee'} \gW^2 \int \frac{d^dk}{(2\pi)^{(d-1)}}\frac{\delta^+(k^2-M_{\chi}^2)\delta(q^+-k^+)}{k_0^2}
= 2\delta^{ee'}\frac{\aW}{\pi} \frac{q^+}{(q^+)^2+M_{\chi}^2}.
\end{equation}
Going to Laplace space and expanding out in the limit $ M_{\chi}\rightarrow 0$, we have
\bea
I= -2\delta^{ee'} \frac{\aW}{\pi} \ln( M_{\chi} s e^{\gamma_E}).
\eea

This result has an IR divergence.
At first glance, this may not be that surprising since all our soft operators in the direct-channel annihilation also were IR divergent.
In those cases, we could trace back the IR divergence to the violation of the KLN theorem due to the semi-inclusive nature of the final state.
In this case however, the interesting point is that, even when the final state is completely inclusive, {\it i.e.}, we do not constrain the final state to be just a photon, our soft function is still IR divergent.
Here we can trace this to the exclusive nature of the initial state where we demand that our operator support an $L=1$, $S= 0$ state. This forces the emission of the 3rd gauge boson which has no virtual counterpart, leading to an IR divergence.
This is very similar to the IR divergence that appears in the computation of PDFs in QCD.

\section{Unstable Particle Effective Theory}
\label{app:UnstableET}

In this section, we justify our use of Eq.~\eqref{eq:Br} for computing the decay rate of bound states.
Let us now look at the effective theory for resonances systematically.
In the literature, this is referred to as unstable particle effective theory (for a review see Ref.~\cite{Beneke:2015vfa}).

To begin with, if we have an intermediate resonance state we expect a propagator in our amplitude of the form 
\bea 
D= \frac{i}{p^2-M_*^2},
\eea
where $M_*$ is a complex pole.
If we write $M_*= M+i\Gamma/2$, then we have the result 
\bea
D = \frac{i}{p^2-M^2-i\Gamma M +\Gamma^2/4}.
\eea
We wish to work in a regime of narrow width, {\it i.e.} $ p^2-M^2 \sim \Gamma M \ll M^2$, so that the propagator is well approximated by 
\bea
D \simeq \frac{i}{p^2-M^2-i\Gamma M},
\eea
and we want to develop an effective theory with an expansion in the small parameter $\lambda= \Gamma/M$.
For the case of inclusive decays of our DM bound state,  $\Gamma M_{\chi} \sim \aW^5 M_{\chi}^2$; we are interested in the inclusive decay rate, so there are no large logarithms and the perturbative cross section begins at $\aW^2$, and due to the nontrivial wavefunction, we have an additional factor of at least $\aW^3$.
The effective coupling thus scales as $\lambda \sim \aW^5 \ll 1$.

The hard scale in this process is just the resonance mass,  which in our case is simply $\sim$$M_{\chi}$.
We can now treat this as an HQET theory writing $p =M_{\chi}v+ k$, where $v$ is the four-velocity of the resonance with $v^2=1$ and $k$ is the residual momentum.
Given the scaling above of $p^2-M^2 \sim \Gamma M$, we can immediately see that $k \sim \Gamma$, so that we have a soft mode $k^{\mu} \sim (\Gamma, \Gamma, \Gamma) \equiv M_{\chi}(\aW^5, \aW^5, \aW^5)$.
Obviously, the question is how does this scale relate to the mass scale $\mW \sim v M_{\chi}$ that we already have.
Now the $\aW$ in the decay rate is evaluated at the scale $M_{\chi}$ while the $\aW$ in the HQET scaling is at the scale $\mW$; however, we can see by the equations relating $\aW$ at the two scales that they are parametrically of the same order.
If that is the case, then our mode has $k^{\mu} \sim  M_{\chi}(\aW^5, \aW^5, \aW^5)$ with $k^2 \ll \mW^2$ and hence can only be populated by a massless mode such as the photon.

Given the fact that in our case $\lambda \sim \aW^5$, any corrections of the order $\lambda^1$ are minute and will be sub-dominant in any error band at the accuracy we are aiming for.
So we can safely work at leading order in $\lambda$.
Following \cite{Beneke:2015vfa}, it is clear that at this order the only term that exists is the propagator with the 1PI self-energy corrections.
Any communication between the production and decay states via radiative emissions of our mode $k^{\mu}$  only occurs at $\mathcal{O}(\lambda)$ and hence is severely suppressed. Additionally, since $\Gamma \sim \alpha_W^5 M_\chi$, but the splitting between bound states, $\Delta E_n \sim \alpha_W^2 M_\chi$, any interference \textit{between} bound states is also subleading.
Therefore, we can safely ignore any radiative corrections by this mode.
This suppression is a manifestation of the length separation in space-time between the process of production and decay.

Now let us look at a cross-section for production of a resonance and its subsequent decay.
We assume that we are in a regime $p^2-M^2 \sim \Gamma M$, where $p$ is the momentum of the intermediate resonance state and $M$ is the real part of the pole.
Let $N$ be our initial scattering state that will create the resonance, and we focus on the cross section to produce an observed final state $f$ and an ultrasoft photon $\gamma_\textrm{us}$ indicating that a bound state was formed.
In detail, the differential cross section is
\bea
\frac{d\sigma}{dz} = \frac{1}{\mathcal{N}} \int d\Pi_{\gamma} d\Pi_f |{\cal M} ( N \rightarrow f+ \gamma_\textrm{us}) |^2 \delta^{(4)}( p_{\gamma}+p_f- p_N) \mathcal{M}_z(f),
\eea
where ${\cal M}_z$ is the measurement (in this case the photon energy) function on the final state particles, and ${\cal N}$ is a normalizing kinematic factor.
Since the photon emitted during the bound state formation is an ultrasoft photon, there is no measurement on it (via a multipole expansion of the measurement function) and its phase space is integrated over fully.
The amplitude squared will contain the squared propagator for the resonance,
\bea
J = \frac{1}{(p^2-M^2)^2+\Gamma^2M^2}.
\eea
Now, the key point is that if we are not interested in the details of  the variation of the cross section near resonance, and we are sufficiently inclusive over $p^2$ around the resonance (by a value $\gg \Gamma M$), then we can make the following substitution 
\bea
\lim_{\Gamma/M \rightarrow 0}J \rightarrow \frac{\pi}{\Gamma M} \delta(p^2-M^2).
\eea
This substitution is true only in the distribution sense, {\it i.e.} under the integral which at least encompasses the region of the size of the width about the resonance.
This is the narrow width approximation.
This substitution then puts the intermediate resonance on-shell.
The cross section can then be written as 
\begin{equation}\begin{aligned}
\frac{d\sigma}{dz} &=\frac{1}{\mathcal{N}} \frac{\pi}{\Gamma M}\int d\Pi_{\gamma} d\Pi_f |{\cal M} ( N \rightarrow B(p) + \gamma_\textrm{us}) |^2  |{\cal M} (B(p) \rightarrow f) |^2 \\
&\times \delta^{(4)}( p_{\gamma}+p_f- p_N) \delta(p^2-M^2) \mathcal{M}_z(f).
\end{aligned}\end{equation}
Here $B(p)$ represents a bound state with momentum $p$, and again this result is true at leading order in $\Gamma/M$, which forbids any communication between the production and decay states.
If we then insert a factor of unity, $1 =\int d^4p\, \delta^{(4)}(p - p_f)$, the result can be rearranged to yield,
\begin{equation}\begin{aligned}
\frac{d\sigma}{dz} = &\frac{1}{\mathcal{N}} \frac{\pi}{M}\Big[\int d\Pi_{\gamma} d\Pi_{R}| {\cal M} ( N \rightarrow B(p) + \gamma_\textrm{us}) |^2 \delta^{(4)}( p_{\gamma}+p- p_N)\Big]\\
\times &\frac{1}{\Gamma} \Big[\int d\Pi_f | {\cal M} (B(p) \rightarrow f) |^2\delta^{(4)}( p - p_f) \mathcal{M}_z(f) \Big] \\
= &\sigma( N \rightarrow B+ \gamma_\textrm{us}) \frac{1}{\Gamma} \frac{ d \Gamma_{B \rightarrow f}}{dz},
\end{aligned}
\label{eq:nwapprox}
\end{equation}
which is simply the product of the production cross section and the differential branching ratio.
The above separation holds where the process proceeds solely through the long lived bound state, but in practice the result should be summed over all bound states in the spectrum, as well as the direct annihilation case where no bound states are formed.
Applied to our specific case we arrive at Eq.~\eqref{eq:Br}.
%

\section{Proof of the Wilson Line Identity}
\label{app:WilsonLineProof}

In this section, we prove the identity involving soft Wilson lines used in Eq.~\eqref{eq:1hardop} that eventually leads to a universal factorization of the IR physics in terms of soft and jet functions independent of the representation.
The property that we wish to show is
\bea
S_v T^a S_v^{\dagger} = T^{a'} S_v^{a'a},
\label{eq:softid}
\eea
where $T$ is a generator in an arbitrary representation (we used $\mathbf{5}$ for the quintuplet), and on the left hand side we have two Wilson lines in the same representation, whereas on the right it is in the adjoint.
(In the main text we used $Y_v$ for the latter, we keep all as $S$ here for notational convenience.)
In the main text we actually used $S_v^{\dagger} T^a S_v = S_v^{aa'} T^{a'}$, although this follows from the above by applying various inverses.
In position space, the Wilson lines are defined as 
\begin{equation}\begin{aligned}
S_v(x) &= Pe^{ ig \int_{-\infty}^{x} ds\, v \cdot A_s(vs)} = Pe^{ ig \int_{-\infty}^{\bar v \cdot x} d \bar v \cdot y\, v \cdot A_s(\bar v \cdot y)}, \\
S_v^{\dagger}(x) &= \bar P e^{-ig  \int_{-\infty}^{x} ds\, v \cdot A_s(vs)}.
\end{aligned}\end{equation}
where $v \cdot A_s = v \cdot A_s^a T^a$, with $T^a$ in the appropriate representation for the Wilson line, whilst $P$ is path ordering and $\bar P$ indicates anti-path ordering.
The variable $s$ parametrizes the path along the light cone direction $n$ from $x$ to $-\infty$. 
The statement in Eq.~\eqref{eq:softid} is a generalization of the identity applied for QCD~\cite{Becher:2014oda} for other group representations. 
The soft Wilson line obeys the equation
\begin{equation}\begin{aligned}
\frac{d}{d \bar v \cdot x}S_v(x) &= ig\, v \cdot A_s( \bar v \cdot x) S_v(x), \\
\frac{d}{d \bar v \cdot x}S^{\dagger}_v(x) &=  -S_v^{\dagger}(x)ig\, v \cdot A_s( \bar v \cdot x).
\end{aligned}\end{equation}

Coming back to our question, let us define $U^a(x) = S_v(x) T^a S_v^{\dagger}(x)$.
We can then immediately see
\bea
\frac{d}{d \bar v \cdot x}U^a(x) = \left[  ig v \cdot A_s( \bar v \cdot x), U^a(x) \right]\!.
\eea
We will solve this equation by recursion, order by order in the coupling $g$ to build up the full solution.
The tree level result is simply $U^{a(0)}(x) = T^a = T^{a'} \delta^{a' a}$.
At the next order,
\begin{equation}\begin{aligned}
U^{a(1)}(x) &= \int_{-\infty}^{\bar v \cdot x} d \bar v \cdot y  \left[  ig n \cdot A_s( \bar v \cdot y ), T^a\right]\\
&= ig \int_{-\infty}^{\bar v \cdot x} d \bar v \cdot y\,  v \cdot A_s^b( \bar v \cdot y )if^{baa'}T^{a'} \\
&= T^{a'} ig \int_{-\infty}^{\bar v \cdot x} d \bar v \cdot y\,  v \cdot (A_s( \bar v \cdot y ))^{a'a}.
\end{aligned}\end{equation}
At ${\cal O}(g^2)$,
\begin{equation}\begin{aligned}
U^{a(2)}(x) &= \int_{-\infty}^{\bar v \cdot x} d \bar v \cdot y_1 \int_{-\infty}^{\bar v \cdot y_1}d \bar v \cdot y_2\left[ ig v \cdot A_s( \bar v \cdot y_1 ),\left[  ig v \cdot A_s( \bar v \cdot y_2 ), T^a \right]\right]\\
&= \frac{1}{2}\int_{-\infty}^{\bar v \cdot x} d \bar v \cdot y_1  \int_{-\infty}^{\bar v \cdot x} d \bar v \cdot y_2 P \Big\{\left[ ig v \cdot A_s( \bar v \cdot y_1 ),\left[  ig v \cdot A_s( \bar v \cdot y_2 ), T^a \right]\right] \Big\} \\
&= T^{a'}\frac{(ig)^2}{2}\int_{-\infty}^{\bar v \cdot x} d \bar v \cdot y_1  \int_{-\infty}^{\bar v \cdot x} d \bar v \cdot y_2 P \Big\{ \left(v \cdot A_s( \bar v \cdot y_1 ) v \cdot A_s( \bar v \cdot y_2 )\right)^{a'a}\Big\}.
\end{aligned}\end{equation}
From here, we can see that the $n^{th}$ term will be
\begin{equation}\begin{aligned}
U^{a(n)}(x) &= \frac{1}{n!} \int_{-\infty}^{\bar v \cdot x} d \bar v \cdot y_1\int_{-\infty}^{\bar v \cdot x} d \bar v \cdot y_2  \ldots\int_{-\infty}^{\bar v \cdot x} d \bar v \cdot y_n \\
&\times P \Big\{\left[ ig v \cdot A_s( \bar v \cdot y_1 ),\left[  ig v \cdot A_s( \bar v \cdot y_2 ),\left[ \ldots\left[ ig v \cdot A_s( \bar v \cdot y_n ), T^a \right] \right]\right]\right] \Big\} \\
&= T^{a'} \frac{(ig)^n}{n!} \int_{-\infty}^{\bar v \cdot x} d \bar v \cdot y_1 \int_{-\infty}^{\bar v \cdot x} d \bar v \cdot y_2  \ldots\int_{-\infty}^{\bar v \cdot x} d \bar v \cdot y_n \\
&\times P \Big\{\left( v \cdot A_s( \bar v \cdot y_1 ) v \cdot A_s( \bar v \cdot y_2 ) \ldots v \cdot A_s( \bar v \cdot y_n )\right)^{a'a}\Big\}.
\end{aligned}\end{equation}
Summing to all orders then proves our result.

\section{Subtle Signs in the Bound-state Formation and Decay Calculations}
\label{app:signs}

In the bulk of the paper, we have freely used results from Ref.~\cite{Harz:2018csl} that are written in terms of two-body states of the form $|ij\rangle$.
However, in general our 2-body states for non-identical particles will in fact be combinations of the form $\frac{1}{\sqrt{2}} (|ij\rangle + (-1)^{L+S} |ji\rangle)$.
(This convention choice is also discussed in the context of Sommerfeld enhancement in Ref.~\cite{Beneke:2014gja}, where these two approaches are labeled ``method-1'' and ``method-2''; we largely adopt ``method-2'', where we treat $|ij\rangle$ and $|ji\rangle$ as components of a single state, rather than tracking them separately.)
The factor of $(-1)^{L+S}$ arises from a factor of $(-1)^{S+1}$ from the behavior of the spin configuration under particle exchange, a factor of $(-1)^L$ from the parity of the spatial wavefunction, and a factor of $(-1)$ from the exchange of two fermions.

This means that when considering a transition of the form $|ij\rangle \rightarrow |i^\prime j^\prime\rangle$, we also need to include transitions between the $|ji\rangle$ and $|j^\prime i^\prime\rangle$ states.
In many cases this does not make a difference and it is adequate to represent states purely by one component $|ij\rangle$.
For example, if the $|ij\rangle \rightarrow |i^\prime j^\prime\rangle$ and $|ji\rangle \rightarrow |j^\prime i^\prime\rangle$ processes have equal rates, but $|ij\rangle \rightarrow |j^\prime i^\prime\rangle$ and $|ji\rangle \rightarrow |i^\prime j^\prime\rangle$ are forbidden (for example, this occurs if $|ij\rangle = |0,++\rangle$), then the combined rate is the same as what one would obtain from purely considering the $|ij\rangle \rightarrow |i^\prime j^\prime\rangle$ process.
However, this behavior is not universal.

As an example of a case where this matters, consider the transition between $Q=1$ bound state components $|\!+0 \rangle \rightarrow |\!+0 \rangle$.
Writing out the individual components of these states (labeled as $23$ and $32$, following the notation of Sec.~\ref{sec:BSF}),\footnote{One might be tempted to write the 23 state as $|\!+0 \rangle$ and 32 as $|0+ \rangle$.  However, we reserve $|\!+0 \rangle$ for the full quantum state such that $|\!+0 \rangle = (|2\,3\rangle + (-1)^{L+S} |3\,2\rangle)/\sqrt{2}$.} the full matrix element should be:
\begin{equation}\begin{aligned}
\mathcal{M} = \frac{1}{2} &\left(\mathcal{M}_{22,33} + (-1)^{(L+S)_i} \mathcal{M}_{32,23} +  (-1)^{(L+S)_f} \mathcal{M}_{23,32} \right.\\
&\left.+ (-1)^{(L+S)_i + (L+S)_f} \mathcal{M}_{33,22} \right)\!.
\end{aligned}\end{equation}
Now we can write $(L+S)_f = (L+S)_i + 1\, \textrm{(mod 2)}$ for dipole transitions, and consequently:
\begin{equation} \mathcal{M} = \frac{1}{2} \left(\mathcal{M}_{22,33} - \mathcal{M}_{33,22} + (-1)^{(L+S)_i} \left( \mathcal{M}_{32,23} - \mathcal{M}_{23,32} \right)  \right)\!.
\end{equation}

Now as calculated in Sec.~\ref{sec:BSF}, if $\psi_i$ and $\psi_f$ denote the initial- and final-state wavefunctions, we have:
\begin{align} 
\mathcal{M}^3_{22,33} & =  i \sqrt{2^6 \pi \alpha_\text{rad} M_\chi} \left[(T^3)_{22} - (T^3)_{33} \right] \int d^3\mathbf{r} \, \psi_f^*  \nabla \psi_i \\
& = i \sqrt{2^6 \pi \alpha_\text{rad} M_\chi} \int d^3\mathbf{r} \, \psi_f^*  \nabla \psi_i, \nn \\
\mathcal{M}^3_{33,22} & =  i \sqrt{2^6 \pi \alpha_\text{rad} M_\chi} \left[(T^3)_{33} - (T^3)_{22} \right] \int d^3\mathbf{r} \, \psi_f^*  \nabla \psi_i \nn \\
& = i \sqrt{2^6 \pi \alpha_\text{rad} M_\chi} \int d^3\mathbf{r} \, (- \psi_f^*  \nabla \psi_i), \nn \\
\mathcal{M}^3_{23,32} & =  i \sqrt{2^6 \pi \alpha_\text{rad} M_\chi} \left\{\left[-i ((T^1)_{32} (T^2)_{23} - (T^2)_{32} (T^1)_{23}  \right]  M_\chi \alpha_\text{NA} \int d^3\mathbf{r} \, \hat{\mathbf{r}} \psi_f^* \psi_i  \right\} \nn \\
& = i \sqrt{2^6 \pi \alpha_\text{rad} M_\chi} \left\{ -3 M_\chi \alpha_\text{NA} \int d^3\mathbf{r} \, \hat{\mathbf{r}} \psi_f^* \psi_i  \right\}\!,\nn \\
\mathcal{M}^3_{32,23} & =  i \sqrt{2^6 \pi \alpha_\text{rad} M_\chi} \left\{\left[-i ((T^1)_{23} (T^2)_{32} - (T^2)_{23} (T^1)_{32}  \right]  M_\chi \alpha_\text{NA} \int d^3\mathbf{r} \,\hat{\mathbf{r}} \psi_f^* \psi_i  \right\} \nn \\
& = i \sqrt{2^6 \pi \alpha_\text{rad} M_\chi} \left\{ 3 M_\chi \alpha_\text{NA} \int d^3\mathbf{r} \, \hat{\mathbf{r}} \psi_f^* \psi_i  \right\}\!, \nn
\end{align}
recalling that the ``3'' superscript is for $\gamma,\,Z$ emission, depending on the value of $\alpha_\text{rad}$, the coupling of the emitted boson to the charged particle that radiated it.  We also recall that the terms with $\alpha_\text{NA}$ correspond to emission off the virtual particles sourcing the potential.  The $\alpha_\text{NA}$ factor is thus the coupling of the virtual line to the WIMPs.  For the case of capture by a radiated $\gamma$, $\alpha_\text{rad} = \alpha_\text{em}$ and $\alpha_\text{NA} = \alpha_W$. 

Thus, overall we have $\mathcal{M}_{32,23} = - \mathcal{M}_{23,32}$ and $\mathcal{M}_{33,22} = - \mathcal{M}_{22,33}$, and consequently:
\begin{align} \mathcal{M} & = \mathcal{M}_{22,33} + (-1)^{(L+S)_i}  \mathcal{M}_{32,23} \nn \\
& = i \sqrt{2^6 \pi \alpha_\text{rad} M_\chi} \left[ \int d^3\mathbf{r}\, \psi_f^*  \nabla \psi_i + (-1)^{(L+S)_i} 3 M_\chi \alpha_\text{NA} \int d^3\mathbf{r}\, \hat{\mathbf{r}} \psi_f^* \psi_i\right]\!.
\end{align}
The $(-1)^{(L+S)_i}$ factor is obtained by treating the different components correctly, and is required to ensure the correct behavior of the matrix element under time reversal.

\section{Analytic Approximate Results for Annihilation and Bound-state Formation}
\label{app:analytic}

In this final appendix we provide analytic estimates for the annihilation and bound state formation rate of DM.
We consider the quintuplet case of interest first, followed by providing equivalent results for a general representation.

\subsection{Results for the quintuplet}

In the limit of unbroken SU(2), the wavefunctions and their integrals, and hence the bound-state capture rate, can be computed analytically in the low-velocity limit.
In this regime the Sommerfeld enhancement can also be computed analytically. These results can also be applied (approximately, and with caveats we will discuss below) to the case where SU(2) is broken but the DM mass is very heavy relative to the symmetry breaking scale.
These calculations can be useful both as a cross-check on our numerical results, and to develop intuition for which channels are likely to dominate the overall annihilation signal.
As such, we present the details of these analytic calculations below, beginning with the DM in the quintuplet representation.

\subsubsection{Capture and annihilation rates}

As an opening example, let us estimate the spin-averaged capture rate into the spin-triplet ground state via photon emission.
The total cross section for this process is given by (see App.~C of Ref.~\cite{Asadi:2016ybp}),\footnote{This expression as written includes capture only from the components of the incoming state that experience an attractive potential; we expect the contribution from repulsed incoming states to be suppressed, due to their small overlap with the bound states.}
\begin{align} 
\sigma v & = \frac{3}{2} \times \frac{2^8 \pi \alpha \,k}{3} \left|\sum_i ({\bf I} \cdot \eta_i) \aW (\aW \lambda_f M_\chi/2)^{-3/2} e^{-2 \lambda_i/\lambda_f} e^{\pi \alpha^{}_{\scaleto{W}{3.5pt}} \lambda_i /(2 v)} \Gamma(1 - i \aW \lambda_i /v) \right. \nn \\
&\left. \times \eta_f^\dagger \left[\lambda_i \hat{C}_1 + \hat{C}_2 \lambda_i/\lambda_f  \right] \eta_i \right|^2\!,
\label{eq:analytic}
\end{align}
The notation we employ follows Ref.~\cite{Asadi:2016ybp}; $\eta_i$ and $\eta_f$ are potential eigenvectors which for the quintuplet are given by (in our basis) $\eta_f=\{-2,1,0\}/\sqrt{5}$, $i=1,2$ with $\eta_1 = \{\sqrt{2}, -\sqrt{2}, 1 \}/\sqrt{5}$, $\eta_2 = \{-2, -1, \sqrt{2}\}/\sqrt{7}$, with corresponding attractive eigenvalues $\lambda_f=5$, $\lambda_1 = 6$, $\lambda_2=3$.
The $I$ vector describes the fraction of the incoming plane wave in each state; we will choose $I = \{0,0,1\}$, as the state asymptotes to two noninteracting, neutral DM particles.
The energy of the outgoing photon is $k$, which at low velocities can be approximated as the binding energy of the ground state, $\lambda_f^2 \aW^2 M_\chi/4$.
Lastly, the $\hat{C}_1$ and $\hat{C}_2$ matrices describe the couplings between the different components of the initial and final states; for capture via photon (or $Z$) emission, they take the form:
\begin{equation} 
\hat{C}_1 = 
\begin{pmatrix} 2 & 0 & 0 \\ 0 & 1 & 0 \\ 0 & 0 & 0 \end{pmatrix}\!, \quad
\hat{C}_2 = 
\begin{pmatrix}0 & 2 & 0 \\ -2 & 0 &3 \sqrt{2} \\ 0 &  -3\sqrt{2} & 0 \end{pmatrix}\!.
\end{equation}

In the wino and positronium cases studied in Ref.~\cite{Asadi:2016ybp}, there was only one eigenstate that experienced an attractive initial-state potential, and so the sum in Eq.~\eqref{eq:analytic} was trivial. 
There is a simple expression for $|e^{\pi \alpha^{}_{\scaleto{W}{3.5pt}} \lambda_i /(2 v)} \Gamma(1 - i \aW \lambda_i /v) |^2 \simeq 2 \pi  \aW \lambda_i /v$ in the limit of small $v$ (for positive $\lambda_i$), which scales purely as $1/v$, and consequently in those cases (wino and positronium) $\sigma v$ had a simple $1/v$ scaling at low relative velocities.
However, in the quintuplet case, we see there are multiple terms in the sum that can interfere with each other, and so even in the limit of unbroken SU(2), we expect there to be a non-trivial velocity dependence in the capture cross section.

Similarly, the Sommerfeld factors can be read off from the components of the scattering-state wavefunction at the origin, and in the unbroken limit this wavefunction has the form $\sum_i (I \cdot \eta_i) \eta_i \, \phi(\lambda_i \aW, r)$, where $\phi(\alpha, r)$ is the solution to the scalar Schrodinger equation with an attractive Coulomb potential with coupling $\alpha$.
In principle this sum runs over both positive and negative eigenvalues of the potential (corresponding to both attracted and repulsed eigenstates), but for low velocities we expect the contribution of the eigenstates experiencing a repulsive interaction to be very small. 
Nonetheless, where (as in the quintuplet case) there are two eigenstates experiencing an attractive potential, the value of the wavefunction at the origin (and hence the Sommerfeld factors) will experience a non-trivial interference between the two contributions.
This can give rise to a velocity dependence differing from the simpler case where there is only one attractive eigenstate.
Similar interference effects can be seen in the form of rapid changes in spectrum with respect to $M_\chi$ in the case of broken SU(2) symmetry, where the interference occurs between the various Sommerfeld factors with resonances at different positions (as discussed in the context of Fig.~\ref{fig:Spectra-var} and App.~\ref{app:massvariation}).
In contrast, the manifestation of the eigenstate interference identified here persists in the SU(2)-symmetric limit and does not require any resonance structure, only differing (velocity-dependent) phases between the interfering contributions.

However, as noted in Ref.~\cite{Schutz:2014nka, Asadi:2016ybp}, at low velocities the system is often in an ``adiabatic'' regime where the incoming particle wavefunction evolves such that at short distances it has complete overlap with the eigenvector with the largest-magnitude attractive eigenvalue.
The criterion for this behavior is roughly $v \lesssim \delta/\mW$, where $\delta$ is the mass splitting between the states; for $\delta=164$ MeV, we expect this behavior to hold roughly for $v \lesssim 2\times 10^{-3}$, {\it i.e.}~for Milky-Way-scale velocities and lower.
Note that this criterion is independent of the DM mass, so even when the DM is very heavy and the ratio $\mW/M_\chi$ is small, the effect of SU(2) breaking can still be seen in the presence of this adiabatic regime.
In this case, the interference will be suppressed for both bound-state formation and Sommerfeld enhancement, with only the $i=1$, $\lambda_i=6$ term contributing significantly, and with the coefficient $I\cdot \eta_i$ replaced with $\delta_{i1}$.

This is an important simplifying approximation within its regime of validity. Note that the presence of this regime relies on SU(2) being broken, and also on low velocity ($v \lesssim \delta / \mW$); it will not appear if an unbroken symmetry ensures the degeneracy of the mass eigenstates, and it will also not generally be relevant in the early universe ({\it e.g.}~for relic density calculations) where velocities are much higher. However, it is well-suited to the case of indirect detection in the Milky Way halo.

For example, within this approximation, we obtain the spin-averaged capture rate to the ground state as:
\begin{equation}
\sigma v \simeq \frac{2^8}{5} \frac{\pi \alpha \aW }{M_\chi^2}  \times \frac{3^3 \cdot 2^9}{5^2} e^{-24/5} \frac{\pi \aW}{v} = \frac{2^{17} \cdot 3^3}{5^3} e^{-24/5} \frac{\pi^2 \alpha \aW^2}{M_\chi^2 v} \simeq \frac{233 \pi^2}{v} \frac{\alpha \aW^2}{M_\chi^2}.\label{eq:p2s}
\end{equation}
Here we have employed the low-velocity approximation $|e^{\pi \alpha^{}_{\scaleto{W}{3.5pt}} \lambda_i /(2 v)} \Gamma(1 - i \aW \lambda_i /v) |^2 \simeq 2 \pi  \aW \lambda_i /v$.

In the same regime, where the initial state rotates into the most-attracted eigenstate, the $s$-wave direct annihilation cross section to gauge bosons can be computed as,
\begin{equation}
\sigma v 
\simeq \frac{720 \pi^2 \aW^3}{M_\chi^2 v}.
\end{equation}
In this unbroken limit, the effective branching ratio to the line ({\it i.e.} to $\gamma \gamma$ + half the branching ratio to $\gamma Z$) should be given by $(\sW^4 + \sW^2 \cW^2)/3 = \sW^2/3$, so the line cross section should be:
\begin{equation}
(\sigma v)_\text{line} \simeq \frac{240 \pi^2 \aW^2 \alpha}{M_\chi^2 v}.
\end{equation}
As for the bound-state formation, at higher velocities ($v \gtrsim 2\times 10^{-3}$), we expect to see the onset of interference between the contributions from the two attracted eigenstates, resulting in a non-monotonic dependence of the cross section on velocity even in the $s$-wave case.
This behavior, and its onset at roughly Milky Way-scale velocities, can be observed in Fig.~\ref{fig:vscans}.
(Note that the velocity dependence can also be suppressed if the DM mass is small enough that the Sommerfeld enhancement is fully saturated, such that the velocity dependence of the individual Sommerfeld factors is very different from the case of unbroken SU(2) symmetry.)

This would suggest the cross-sections for capture (to the spin-triplet ground state) and for annihilation producing a line should be very similar, at least when this adiabatic approximation holds (numerical calculations indicate the non-adiabatic cross section for bound-state capture, as estimated in Eq.~\ref{eq:analytic}, can range between larger than the adiabatic result by a factor of $\sim$$2$ and smaller by a factor of $\sim$$5$, as $v$ is varied). 
However, for $v \ll \mW/M_\chi$, we have the SU(2)-breaking effect that $p$-wave processes should be parametrically suppressed by a factor of order $(v M_\chi/\mW)^2$, which at a 13.6 TeV mass can suppress the $p$-wave capture cross section by $\sim$$2$ orders of magnitude.

We can also study the cross-section for spin-singlet $s \rightarrow p$ capture to an $n=2$, $l=1$ state.
In this case we have $k = (25/16) \aW^2 M_\chi$, and
\begin{equation}\begin{aligned}
\sigma v & = \frac{2^{13} \pi \alpha k}{3^3} \frac{1}{\aW} M_\chi^{-3} \frac{1}{\lambda_f^3} \left|\sum_i ({\bf I} \cdot \eta_i) e^{-4 \lambda_i/\lambda_f} e^{\pi \alpha^{}_{\scaleto{W}{3.5pt}} \lambda_i /(2 v)} \Gamma(1 - i \aW \lambda_i /v) \eta_f^\dagger \right. \\
& \left. \times \left[\lambda_i \left(\frac{4\lambda_i}{\lambda_f} - 3 \right) \hat{C}_1  + \hat{C}_2 \left(3 - 12 \frac{\lambda_i}{\lambda_f} + 8 \frac{\lambda_i^2}{\lambda_f^2} \right) \right] \eta_i \right|^2 \\
& =  \frac{2^{9} \pi \alpha \aW }{5 \cdot 3^3  M_\chi^{2}}  \Bigg|\frac{3}{25} \sqrt{\frac{2}{5}} e^{-24/5} e^{3 \pi \alpha^{}_{\scaleto{W}{3.5pt}}/(2 v)} \left( 37 e^{12/5} \Gamma(1 - 3 i \aW/v) \right. \\
&\hspace{6.5cm}\left.+ 89 e^{3\pi \alpha^{}_{\scaleto{W}{3.5pt}}/(2 v)} \Gamma(1 - 6 i \aW/v)\right) \Bigg|^2,
\end{aligned}\end{equation}
or in the adiabatic regime,
\begin{equation}\begin{aligned}
\sigma v & \rightarrow  \frac{2^{9} \pi \alpha \aW }{5 \cdot 3^3  M_\chi^{2}}  \left|\frac{267}{25}\sqrt{2} e^{-24/5} e^{3\pi \alpha^{}_{\scaleto{W}{3.5pt}}/(2 v)} \Gamma(1 - 6 i \aW/v) \right|^2 \\
& = \frac{2^{12} \cdot 89^2 \pi^2}{5^5\, v}  e^{-48/5} \frac{\alpha \aW^2}{M_\chi^2} \\
& \simeq  \frac{0.70 \pi^2}{v} \frac{\alpha \aW^2}{M_\chi^2}. \label{eq:s2p} \end{aligned}\end{equation}

We see that the scale for this cross-section is naturally two orders of magnitude smaller than the previous ones, which arises from the various numerical prefactors, primarily the factor of $e^{-48/5}$ compared to $e^{-24/5}$ for the capture to the $n=1$ state, corresponding to a suppression factor of $8\times 10^{-3}$.
If these exponential terms were removed, the other prefactors would differ by less than a factor of 3.
(Numerical calculations indicate the non-adiabatic cross section is larger than the adiabatic one in this case, by factors between 1 and 3.6.)
This is suggestive that only the capture to the ground-state is likely to be comparable to direct annihilation, and the capture from the $p$-wave initial-state component suffers from a $v^2$ suppression once $v$ drops below $\mW/M_\chi$, which renders it subdominant at the $\mathcal{O}(1\%)$ level for our 13.6 TeV benchmark point.

This suppression for higher-$n$ capture also suggests that the contribution to the {\it endpoint} hard photon spectrum from bound state formation and decay will be suppressed, as these contributions are dominated by capture into states with odd $L$ (and thus $n>1$) and $S=0$ that decay to $L=S=0$ states before annihilating (see Sec.~\ref{sec:BSA} for a more in-depth discussion).
Capture to the ground-state via emission of a dipole gauge boson changes $L$ by 1, thus requiring an initial $L=1$ state (which must then have $S=1$ if it contains identical DM particles), and $S=1$ states do not produce a leading-power contribution to the endpoint spectrum when they decay.

For the $Q=1$ sector, let us again consider capture from the spin-triplet $p$-wave incoming wave to the spin-triplet $s$-wave state.
In the unbroken limit the potential matrix for the final state takes the form:
\begin{equation} 
V = \begin{pmatrix} -2 & \sqrt{6} \\ \sqrt{6} & -3 \end{pmatrix}\!.
\end{equation}
Here the first row refers to the $++-$ state and the second to the $+\,0$ state.
The transition matrices are now,
\begin{equation}
\hat{C}_1 = \begin{pmatrix} - \frac{1}{\sqrt{2}} & \frac{1}{\sqrt{2}} & 0 \\
0 & -\frac{\sqrt{3}}{2} & \sqrt{\frac{3}{2}} \end{pmatrix}\!, \quad \hat{C}_2 = \begin{pmatrix} 2\sqrt{2} & \sqrt{2} & 0 \\
0 & \sqrt{3} & 0\end{pmatrix}\!.
\end{equation}

In this case we must also replace the $\alpha$ prefactor in the cross section with $\aW$.
The attractive eigenvalue for the final state has $\lambda_f=5$, $\eta_f = \{-\sqrt{2},\sqrt{3}\}/\sqrt{5}$.
Again, the binding energy of the ground state is $k=\aW^2 \lambda_f^2 M_\chi/4 = (25/4)\aW^2 M_\chi$ (and since in the unbroken limit $\mW=0$, we do not need to include a kinematic suppression for the $W$ mass).
Then we obtain for the unbroken limit:
\begin{equation}\begin{aligned}
\sigma v  = \frac{2^{7} 3^2  \pi \aW^2}{5^4 M_\chi^2} &\left| 16  e^{-12/5} e^{6 \pi \alpha^{}_{\scaleto{W}{3.5pt}} /(2 v)} \Gamma(1 - 6 i \aW /v)  \right. \\
&\left.+  7 e^{-6/5} e^{3 \pi \alpha^{}_{\scaleto{W}{3.5pt}} /(2 v)} \Gamma(1 - 3 i \aW /v)  \right|^2\!,
\end{aligned}\end{equation}
When we assume the adiabatic regime, the result reduces to,
\begin{equation} 
\sigma v = \frac{2^{17} \cdot 3^3 \pi^2}{5^3 }  e^{-24/5} \frac{ \aW^3}{v M_\chi^2} \simeq  233  \frac{ \pi^2 \aW^3}{v M_\chi^2}.
\end{equation}
This is for capture to the $Q=+1$ state; there is an equal rate for capture to the $Q=-1$ state.
Note the $\aW$ prefactor (rather than $\alpha$); including formation of the $Q=0$ state through $Z$ emission (as well as photon emission) would similarly promote that capture rate to have a prefactor of $\aW$ rather than $\alpha$, for an overall capture rate (summing across all three channels) of $\sim 700 \pi^2 \aW^3/(M_\chi^2 v)$, similar to the full $s$-wave direct annihilation rate.
The primary difference between this $p\rightarrow s$ capture rate and the inclusive direct annihilation rate will arise from velocity suppression of the $p\rightarrow s$ capture cross section in the broken-SU(2) case (with this suppression being lifted in the truly unbroken limit).

\subsubsection{Presence of metastable bound states}
\label{subsubsec:metastable}

The unbroken-SU(2) limit is also helpful for studying the question of whether there could be $L>0$ states in the spectrum whose decays to more deeply bound states are highly suppressed, leading them to decay through annihilation to SM particles with a substantial branching ratio.
States which are degenerate in the unbroken limit are likely to remain close in energy as we reduce $M_{\chi}$, and therefore decays between them will be suppressed (although if this is decisive in determining whether a state is metastable, a more careful analysis will be required).
The $L+S$-even potential for the quintuplet has two attractive eigenvalues, $Z=6$ and $Z=3$, whereas the $L+S$-odd potential has a single attractive eigenvalue, $Z=5$. Thus for spin-singlet states ($S=0$) we expect $L$-even bound states with energies $E_n / \aW^2 M_\chi = - 9/n^2, -2.25/n^2$, and $L$-odd bound states with energies $E_n / \aW^2 M_\chi = -6.25/n^2$.
For spin-triplet states ($S=1$) we expect $L$-odd states with energies $E_n / \aW^2 M_\chi = -9/n^2, -2.25/n^2$, and $L$-even states with energies $E_n / \aW^2 M_\chi = -6.25/n^2$.

We first consider the case of $L$-odd states.
Given a spin-singlet $L$-odd state with $n > L > 0$ (binding energy $6.25/n^2$), there should always be a more deeply bound state with $n^\prime = n$, $L^\prime = L-1$ (binding energy $9/n^2$), which is accessible through a dipole transition.
So in the spin-singlet case and Coulomb limit there should be no metastable states with $L > 0$. 
For the spin-triplet, the case is slightly more complicated, as given an $L$-odd state with $n > L > 0$ (binding energy $9/n^2$ or $2.25/n^2$), the accompanying state with $n^\prime = n$, $L^\prime = L-1$ has binding energy $6.25/n^2$.
We see that the $L$-odd states with binding energies $2.25/n^2$ will always have an accompanying more-deeply-bound $L$-even spin-triplet state, to which they can decay, but this is not necessarily true for the states with binding energies $9/n^2$. A state with $L^\prime = L-1$ will be available if $9/n^2 < 6.25/m^2$ for some $m$ with $L \le m < n$, {\it i.e.} if $m < n \sqrt{6.25/9} = n/1.2$ is consistent with $m \ge L$.
This will be true for $n > 1.2 L$, so the dangerous range is states with $L < n \le 1.2 L$. In order for this range to include an integer, we must have $L > 5$. For example, consider the spin-triplet state with $L=7$ and $n=8$, with dimensionless binding energy $9/8^2 \simeq 0.14$ in the Coulombic limit.
The lowest-lying $L=6$ spin-triplet state that is accessible via $\Delta L=1$, $\Delta S=0$ transitions has $n=7$, and consequently binding energy $6.25/7^2 = 0.13$ in the Coulombic limit; thus the $L=7$ state cannot decay through such a transition.
For the $L=5$ case, in the Coulombic limit the states are degenerate, and so we would need to perform a more careful calculation.

If we now consider the case of even-$L$ states, the situation is reversed between the spin-singlet and the spin-triplet; in the spin-triplet case we expect the even-$L$ states will always be able to decay to their accompanying, more deeply bound state with $L^\prime = L-1$, with the exception of the $L=0$ case where no such state exists (we will consider the $L=0$ case below).
In the spin-singlet case, the same argument as previously tells us that for $L > 0$, a state with $L^\prime=L-1$ will be available except in the case where $L < n \le 1.2 L$, which in the case of even $L$ is potentially relevant for $L \ge 6$.

Therefore, based on the Coulombic limit, we would predict that the only possible (meta)stable states with $L>0$ are $L+S$-even states (spin-triplet with $L$ odd or spin-singlet with $L$ even) with $L \ge 5$, $L < n \le 1.2 L$, and the eigenstate structure corresponding to the $Z=6$ eigenvalue.
These high-$L$ states may not even be bound for masses of interest to us, and in any case the capture rate into them is likely to be very small.

\subsection{Results for general representations}

Let us now consider the more general situation where the DM is the lightest component of an SU(2) multiplet in a real representation of odd dimension $N$.
Larger representations require higher DM masses to obtain the correct relic density ({\it e.g.} Ref.~\cite{Bottaro:2021snn}), and hence the unbroken-SU(2) approximation is likely to be better at their thermal masses.
Recall, however, that the condition to be in the adiabatic regime is mass-independent, $v \lesssim \delta/\mW$, so while this is a feature of the broken SU(2) symmetry, we expect it to be retained at sufficiently low velocities even for very heavy DM.

In the unbroken-SU(2) limit we can use the results of Ref.~\cite{Mitridate:2017izz} for general representations.
They proceed by decomposing the two-particle state into  eigenstates of isospin $I$ ($I=1, 3, \cdots, 2N-1$); this corresponds to identifying the eigenstates of the potential in our language.
They find the eigenvalue associated with the state with isospin $I$ is $\lambda = (2N^2 - 1 - I^2)/8$ (where positive eigenvalues correspond to attracted states, as per our previous convention) and so the most-attracted channel is the singlet, where $\lambda = (N^2 - 1)/4$.
The isospin singlet corresponds to an $L+S$-even state with total charge $Q=0$, and for the quintuplet $\lambda=6$, as discussed above.
In general, states with $I < \sqrt{2 N^2 - 1}$ can support bound states; for the quintuplet this means we have $I=1,3,5$ bound states.

In the adiabatic regime, SU(2)-breaking effects cause the lowest-energy state at large distances ({\it i.e.}~the $L+S$-even, $Q=0$ state of two identical DM particles) to smoothly transition into the isospin-singlet state at short distances.
Thus in this regime, we expect both the Sommerfeld-enhanced direct annihilation and the bound state capture rates to be consistent with an initial isospin-singlet state.
Since the dominant bound-state capture process (dipole emission of a gauge boson) changes isospin by 2, the final state must then have $I=3$, {\it i.e.} it is a SU(2) adjoint (this state is $L+S$-odd and the three components have $Q=0,\pm 1$).

Consequently, within the adiabatic approximation, we only need to concern ourselves with (Sommerfeld enhanced) direct annihilation from the isospin-singlet state, and capture from an isospin-singlet initial state to an isospin-triplet final state, with the relevant eigenvalues being $\lambda_i = (N^2-1)/4$ and $\lambda_f = (N^2 - 5)/4$.
(Transitions amongst bound states can involve higher-$I$ states; in particular $I=5$ for the quintuplet, with $\lambda = 3$.)

We will thus focus in this appendix on singlet-to-adjoint transitions. We reiterate that this approximation is {\it not} appropriate if the gauge symmetry is actually unbroken or at the high velocities associated with thermal freezeout in the early universe (as studied in e.g.~Ref.~\cite{Binder:2023ckj}), where other transitions can also contribute significantly and may dominate. The quality of this approximation---{\it i.e.}~the degree to which the incoming state retains non-singlet components at small $r$, which could contribute significantly to the capture rate---is an interesting question, but we ignore it here, as our main purpose is simply to develop some simple intuition for the importance of bound-state effects for the gamma-ray endpoint signal. The corresponding approximation for the quintuplet appears to do a reasonable job of estimating the relative size of bound-state capture and annihilation, as we see in Fig.~\ref{fig:analytic-comparison}.

For these singlet-to-adjoint transitions, we can write the group theory coefficients for bound state formation from Ref.~\cite{Mitridate:2017izz} in the simplified form:
\begin{equation}\begin{aligned}
C^{a1b}_\mathcal{J} & = \frac{1}{\sqrt{T_R d_R}} \text{Tr}(T^b T^a), \\
C^{a1b}_\mathcal{\tau} &= i \frac{1}{\sqrt{T_R d_R}} \text{Tr}(T^b T^c T^d) f^{acd}= - \frac{1}{\sqrt{T_R d_R}} \text{Tr}(T^b T^c T^d) (T^a_\text{adj})^{cd}.
\end{aligned}\end{equation}
We can now use $\text{tr}(T^a T^b) = T_R \delta^{ab}$, and also
\begin{align}
\text{Tr}(T^b T^c T^d) (T^a_\text{adj})^{cd} 
& = \frac{1}{2} T_R T_\text{adj}  \delta^{ab}.
\end{align}
Thus, finally we obtain:
\begin{align}
C^{a1b}_\mathcal{J} & = \sqrt{\frac{T_R}{d_R}}  \delta^{ab}, \quad C^{a1b}_\mathcal{\tau} = - \frac{1}{2}  \sqrt{\frac{T_R}{d_R}} T_\text{adj}\delta^{ab},
\end{align}
where we will show how the $C^{a1b}_\mathcal{J}$ and $C^{a1b}_\mathcal{\tau}$ coefficients enter the bound state capture rate in Secs.~\ref{subsubsec:cgs} and \ref{subsubsec:ces}.

Now if $R$ is the representation of size $N$, for SU(2) we have $T_R = N(N^2 - 1)/12$, and so $T_\text{adj} = 2$, while $d_R=N$ (and in particular $d_\text{adj}=3$).
Thus for SU(2) we obtain the coefficients:
\begin{align} 
C^{a1b}_\mathcal{J} & = \sqrt{\frac{N^2-1}{12}}  \delta^{ab}, \quad C^{a1b}_\mathcal{\tau} = -  C^{a1b}_\mathcal{J}. \end{align}

Let us also note that we can now extend the argument given in App.~\ref{subsubsec:metastable} to general representations.
A bound state of isospin $I$ will generally have open decay channels to states with lower isospin (by 2 units) which are hence more deeply bound due to the larger eigenvalue $\lambda$.
The exception is $I=1$ states, which must decay to $I=3$ states which are more shallowly bound for the same principal quantum number.
For this reason, excited $I=1$ states can be metastable if they have sufficiently large $L$ that all the $I=3$ states differing by only one unit in $L$ are more shallowly bound.
This can occur for a general representation if $L < n \le \frac{N^2 - 1}{N^2-5} L = \left(1 + \frac{4}{N^2-5}\right) L$; this range will contain an integer if $L > (N^2 - 5)/4$. 
Thus the threshold $L$ at which this effect can occur increases as the representation size goes up.

\subsubsection{Direct annihilation}

In this case, if we can evaluate the tree-level cross section for annihilation from an isospin-singlet initial state to any desired SM final state, we can account for the Sommerfeld enhancement by simply multiplying the tree-level cross section by $S=2 \pi \aW \lambda_i/v$, in the low-velocity limit.
This cross section is given for Majorana fermion DM by Ref.~\cite{Mitridate:2017izz} as:
\begin{align} 
(\sigma v)_\text{tree,I=1} & = \frac{\pi \aW^2}{M_\chi^2} \frac{T_R^2 d_\text{adj}}{d_R} \nn \\
& = \frac{\pi \aW^2}{2^4 \times 3 \times M_\chi^2} N (N^2 - 1)^2. 
\end{align}
Multiplying by the Sommerfeld factor gives:
\begin{align}
(\sigma v)_\text{I=1} 
& = \frac{\pi^2 \aW^3}{2^5 \times 3 \times M_\chi^2 v} N (N^2 - 1)^3 \rightarrow \frac{\pi^2 \aW^3}{ M_\chi^2 v} \frac{N^7}{96},
\end{align}
where in the final step we have assumed $N \gg 1$.

Checking, for the wino and quintuplet this yields:
\begin{align} 
(\sigma v)_\text{I=1} 
& = \frac{\pi^2 \aW^3}{M_\chi^2 v}\begin{cases} 16, & N=3, \\ 720, & N=5. \end{cases} 
\end{align}
For the quintuplet this agrees with our calculation above.
This also agrees with the wino result from Ref.~\cite{Asadi:2016ybp}, accounting for our assumption that the adiabatic condition holds.

\subsubsection{Capture to the ground state}
\label{subsubsec:cgs}

At small velocities, from Ref.~\cite{Mitridate:2017izz} we can read off the low-velocity bound state capture cross section to the $n=1$ state as:
\begin{align} 
(\sigma v)^{n=1,l=0}_\text{bsf} = \frac{\pi \aW^2}{M_\chi^2} \frac{2S+1}{g_\chi^2} \frac{2^{11} \pi}{3} \sum_{ab} |C^{a1b}_\mathcal{J}+(1/\lambda_f) C^{a1b}_\mathcal{\tau}|^2 \frac{\lambda_i^3 \aW}{\lambda_f v} e^{-4\lambda_i/\lambda_f}
.
\end{align}
This expression involves an average over initial states, with degrees of freedom denoted by $g_\chi$; since we are interested in the case where $100\%$ of the DM captures from the singlet state, we only need to average over spin degrees of freedom, so $g_\chi=2$.
Our initial state must be $L+S$-even and thus for capture to an $L=0$ final state, it must have $S=1$.
Thus we obtain:
\begin{align} 
(\sigma v)^{n=1,l=0}_\text{bsf} & = \frac{\pi \aW^2}{M_\chi^2} \frac{3}{4} \frac{2^{11} \pi}{3} d_\text{adj} \left(\frac{N^2-1}{12} \right) |1 - (1/\lambda_f)|^2 \frac{\lambda_i^3 \aW}{\lambda_f v} e^{-4\lambda_i/\lambda_f} \nn \\
& =\frac{8 \pi^2 \aW^3}{M_\chi^2 v}   \frac{(N^2 - 9)^2 (N^2-1)^4}{ (N^2 - 5)^3} e^{-4 (N^2 - 1)/(N^2 - 5)}.
\end{align}
where $d_{\text{adj}}=3$ counts the number of generators and arises from $\sum_{ab} |\delta^{ab}|^2 = d_\text{adj}$.

In the limit of $N \gg 3$, which is helpful for comparison against direct annihilation, we obtain the simplified result:
\begin{align} 
(\sigma v)^{n=1,l=0}_\text{bsf} 
& \rightarrow \frac{\pi^2 \aW^3}{M_\chi^2 v} 8 N^6 e^{-4} \simeq \frac{\pi^2 \aW^3}{M_\chi^2 v} \frac{N^6}{6.8}, \quad N \gg 3.
\end{align}
(Note in the quintuplet case this approximate value is larger than the truth by about a factor of 3; it will be a better approximation for larger $N$.)

\subsubsection{Capture to the $n=2$ states}
\label{subsubsec:ces}

Now let us consider capture to the $n=2$, $L=1$ states, as their subsequent decays and annihilations can give rise to endpoint photons, unlike capture directly to the $n=1$ state. 
In this case the final state must have $S=0$ (so the initial state has $L+S$ even). 
From Ref.~\cite{Mitridate:2017izz} in the low-velocity limit the $s$-wave (1st line) and $d$-wave (2nd line) contributions are:
\begin{align} 
(\sigma v)^{n=2,l=1}_\text{bsf} & =\frac{\pi \aW^2}{M_\chi^2} \left(\frac{2S+1}{g_\chi^2} \right) \frac{2^{12} \pi\lambda_i}{9 \lambda_f^5} \frac{\aW}{v} \left[ \sum_{ab} |C^{a1b}_\mathcal{J}(\lambda_f \lambda_i (3\lambda_f - 4 \lambda_i) +C^{a1b}_\mathcal{\tau} (-3\lambda_f^2 + 12 \lambda_f \lambda_i - 8 \lambda_i^2)|^2  \right. \nn \\
& \left. + \sum_{ab} 2^5 \lambda_i^4 |C^{a1b}_\mathcal{J} \lambda_f + 2 C^{a1b}_\mathcal{\tau} |^2\right]e^{-8\lambda_i/\lambda_f}   \nn \\
& = \frac{\pi^2 \aW^3}{M_\chi^2 v} \frac{2^{8}\lambda_i}{9 \lambda_f^5} (N^2 -1) \left[ |(\lambda_f \lambda_i (3\lambda_f - 4 \lambda_i) - (-3\lambda_f^2 + 12 \lambda_f \lambda_i - 8 \lambda_i^2)|^2 \right. \nn \\
& \left. + 2^5 \lambda_i^4 | \lambda_f - 2 |^2\right] e^{-8\lambda_i/\lambda_f}  \nn \\
& = \frac{\pi^2 \aW^3}{M_\chi^2 v} \frac{2^4 (N^2 -1)^2 }{9 (N^2-5)^5} \left[ (N^6 + 9 N^4 - 165 N^2 - 37)^2 \right. \nn \\
& \left. +   2^5 (N^2-1)^4 (N^2 - 13)^2 \right] e^{-8 (N^2-1)/(N^2-5)}.
\end{align}
The first line here agrees with the $s\rightarrow p$ quintuplet result computed in Eq.~\eqref{eq:s2p}, once we multiply that result (which was for photon-mediated capture into a specific $n=2$ $L=1$ bound state) by a factor of 3 to account for $W$- and $Z$-mediated capture and a second factor of 3 to account for the $m=0,\pm 1$ states.

In the limit of large $N$, this expression has the scaling:
\begin{align} 
(\sigma v)^{n=2,l=1}_\text{bsf} 
& \rightarrow \frac{\pi^2 \aW^3}{M_\chi^2 v} \frac{2^4 N^6}{9 } \left[ 1 +   2^5 \right] e^{-8} \simeq  \frac{\pi^2 \aW^3}{M_\chi^2 v} \frac{N^6}{1700} \left[ 1 +   32 \right], \quad N \gg 5 
\end{align}
So we see that compared to direct annihilation, in the large-$N$ limit we expect the various contributions to scale as:
\begin{itemize}
\item $p\rightarrow s$ capture to the $n=1, L=0, S=1$ state (contributions to endpoint photons are power-suppressed): direct annihilation rate $\times 14/N$, in addition to (when SU(2) is broken) any kinematic suppression of $W/Z$ emission (by a factor as small as $\alpha/(3\aW)$) or velocity suppression due to the $p$-wave initial state (parametrically $\mathcal{O}(M_\chi^2 v^2/\mW^2)$ for $v \lesssim \mW/M_\chi$),\footnote{As mentioned above, the large-$N$ approximation overestimates this ratio for the quintuplet by about a factor of 3, and so the rates are actually very comparable.}
\item $s\rightarrow p$ capture to the $n=2, L=1$ states collectively: direct annihilation rate $\times 0.06/N$, in addition to any kinematic suppression of $W/Z$ emission (up to a factor of $3\aW/\alpha$),
\item $d\rightarrow p$ capture to the $n=2, L=1$ states collectively: direct annihilation rate $\times 1.8/N$, in addition to any kinematic suppression of $W/Z$ emission (up to a factor of $3\aW/\alpha$) or velocity suppression due to the $d$-wave initial state (parametrically $\mathcal{O}(M_\chi^4 v^4/\mW^4)$ for $v \lesssim \mW/M_\chi$).
\end{itemize}
We see that the only contribution that is not suppressed at low velocities and that gives rise to leading-power contributions to the endpoint photon spectra (via its decay and subsequent annihilation) is generically expected to have a cross section 2 or more orders of magnitude below direct annihilation.
Consequently, it is quite plausible for bound state formation to be a large or even dominant contribution to the inclusive annihilation rate when the velocity suppression for higher partial waves is mild or absent and the gauge bosons are massless (as in the case of freezeout), while simultaneously having a generically small effect on the endpoint spectrum for indirect detection, especially at low velocities ($v \ll \mW/M_\chi$) or where the $n=2$ states are too loosely bound to allow $W$- or $Z$-mediated capture.

One might ask about the contribution from capture into states with $n > 2$. While in the unbroken limit bound states with large $L$ and $n$ may play a large role in the capture rate ({\it e.g.}~\cite{Binder:2023ckj}), for the parameter space we have considered in this paper, the number of bound states is always truncated by the non-zero SU(2) breaking scale, preventing large enhancements from the proliferation of high-$n$ states. Furthermore, we expect velocity suppressions (of order $(M_\chi v/\mW)^{2L}$) for all capture rates with initial $L > 0$. Finally, within our adiabatic approximation both initial and final states always experience attractive interactions, with couplings that obey $\lambda_i/\lambda_f > 1$, leading to increasingly strong exponential suppression for large-$n$ states and avoiding the potentially unitarity-violating region of parameter space identified in Ref.~\cite{Binder:2023ckj}.

\section{Understanding sharp variations in the spectrum as a function of mass}
\label{app:massvariation}

\begin{figure*}[!t]
\centering
\includegraphics[width=0.47\textwidth]{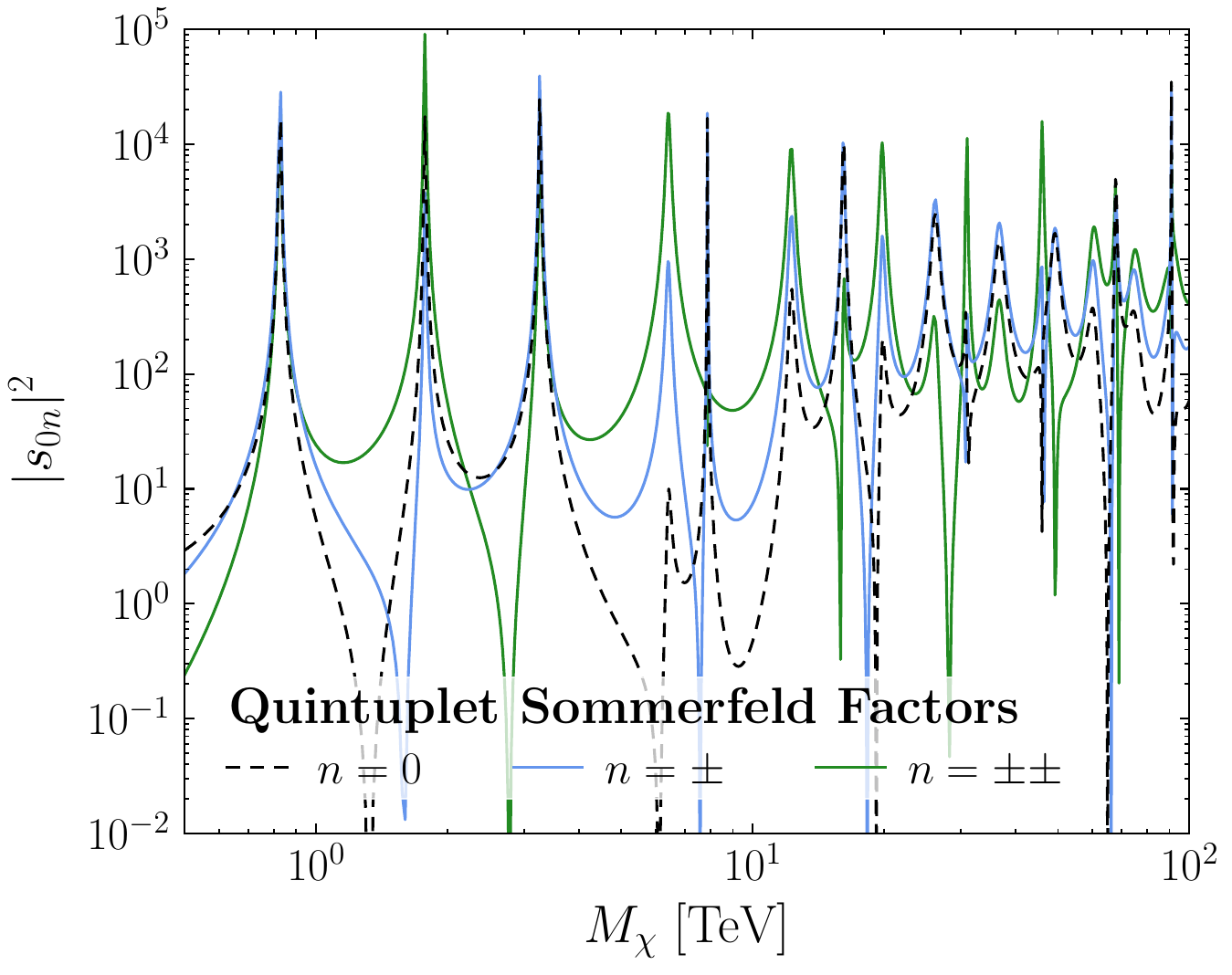} 
\hspace{0.5cm}
\includegraphics[width=0.47\textwidth]{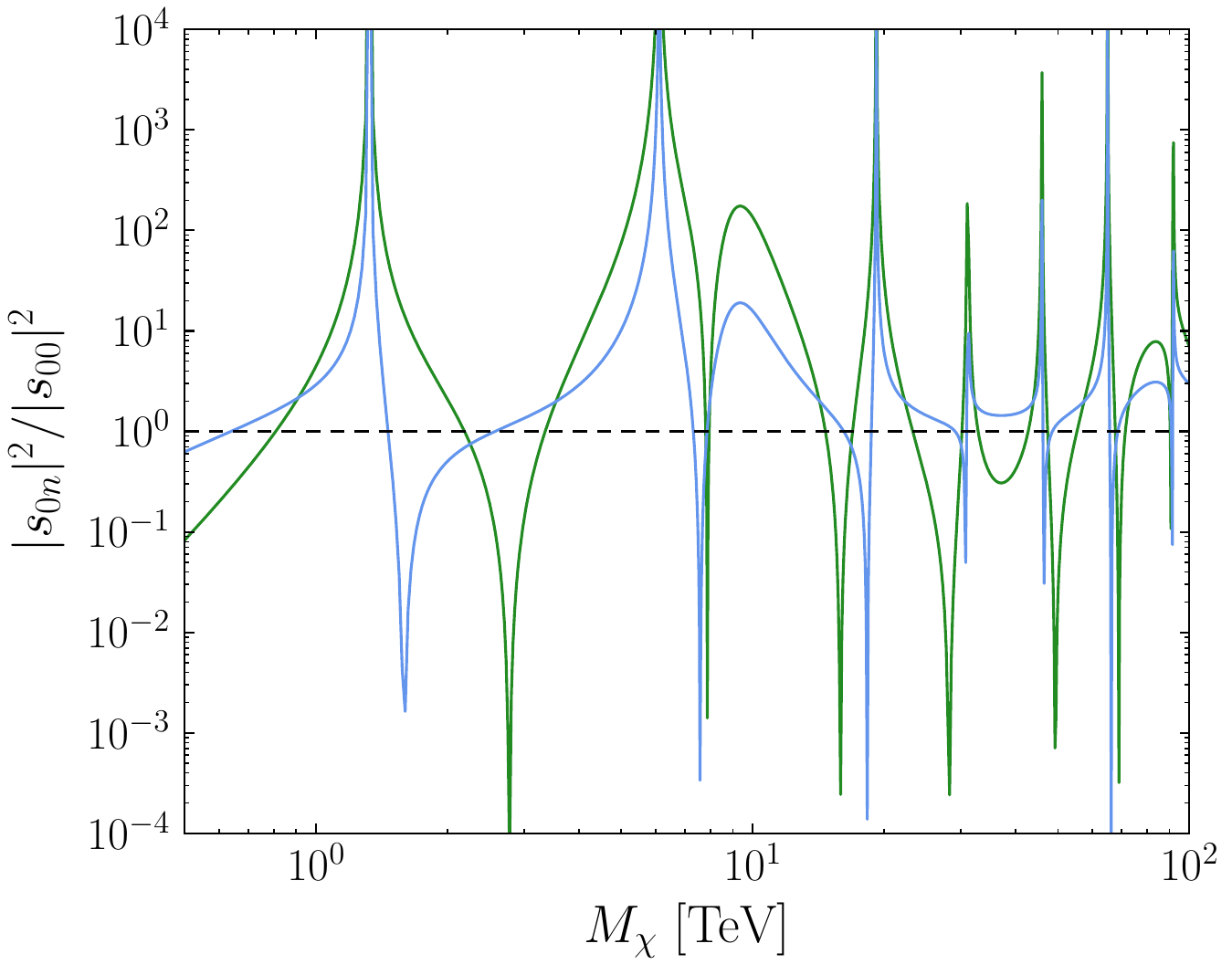}
\includegraphics[width=0.47\textwidth]{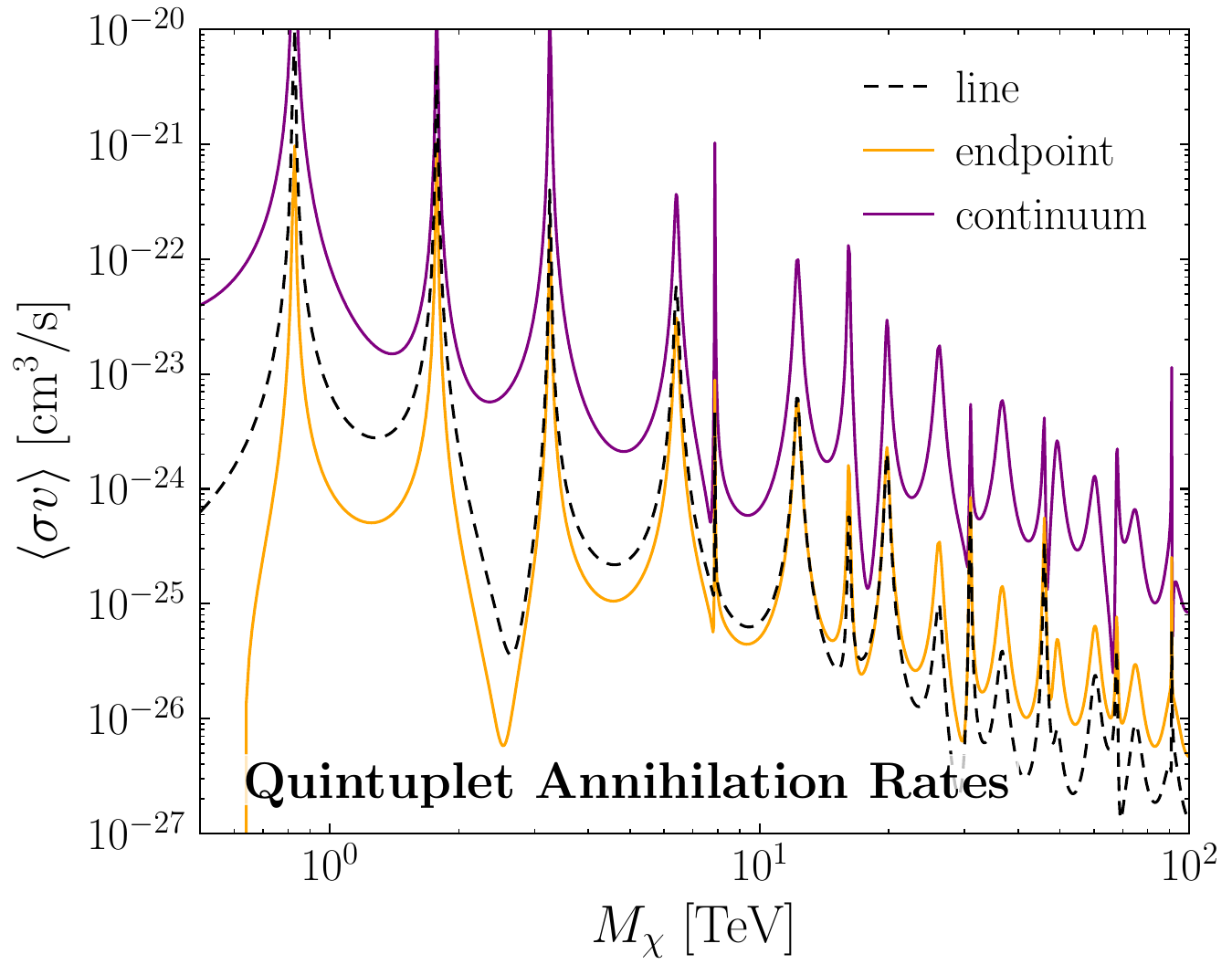} 
\hspace{0.5cm}
\includegraphics[width=0.47\textwidth]{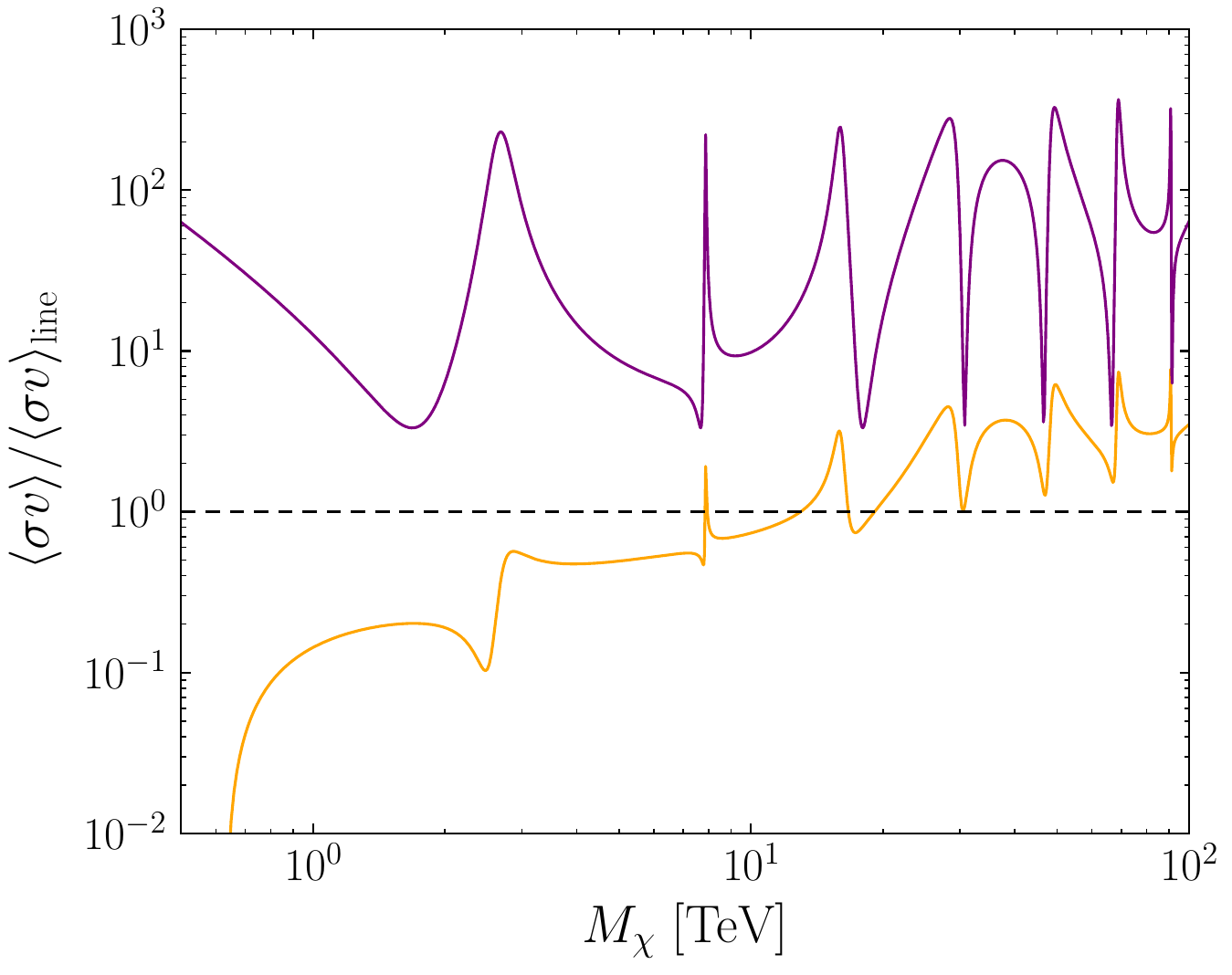}
\caption{The quintuplet Sommerfeld factors (top) and annihilation rates to line, endpoint, and continuum photons (bottom).
The right hand plots show the equivalent ratios.
The three rates depend on different combinations of the Sommerfeld factors, whose ratio varies rapidly as a function of $M_{\chi}$, which is the ultimate origin of the rapid variation of the spectrum seen, for instance, in Fig.~\ref{fig:Spectra-var}.
See the text for additional details.
}
\label{fig:rapidvariationQuint}
\end{figure*}

\begin{figure*}[!t]
\centering
\includegraphics[width=0.47\textwidth]{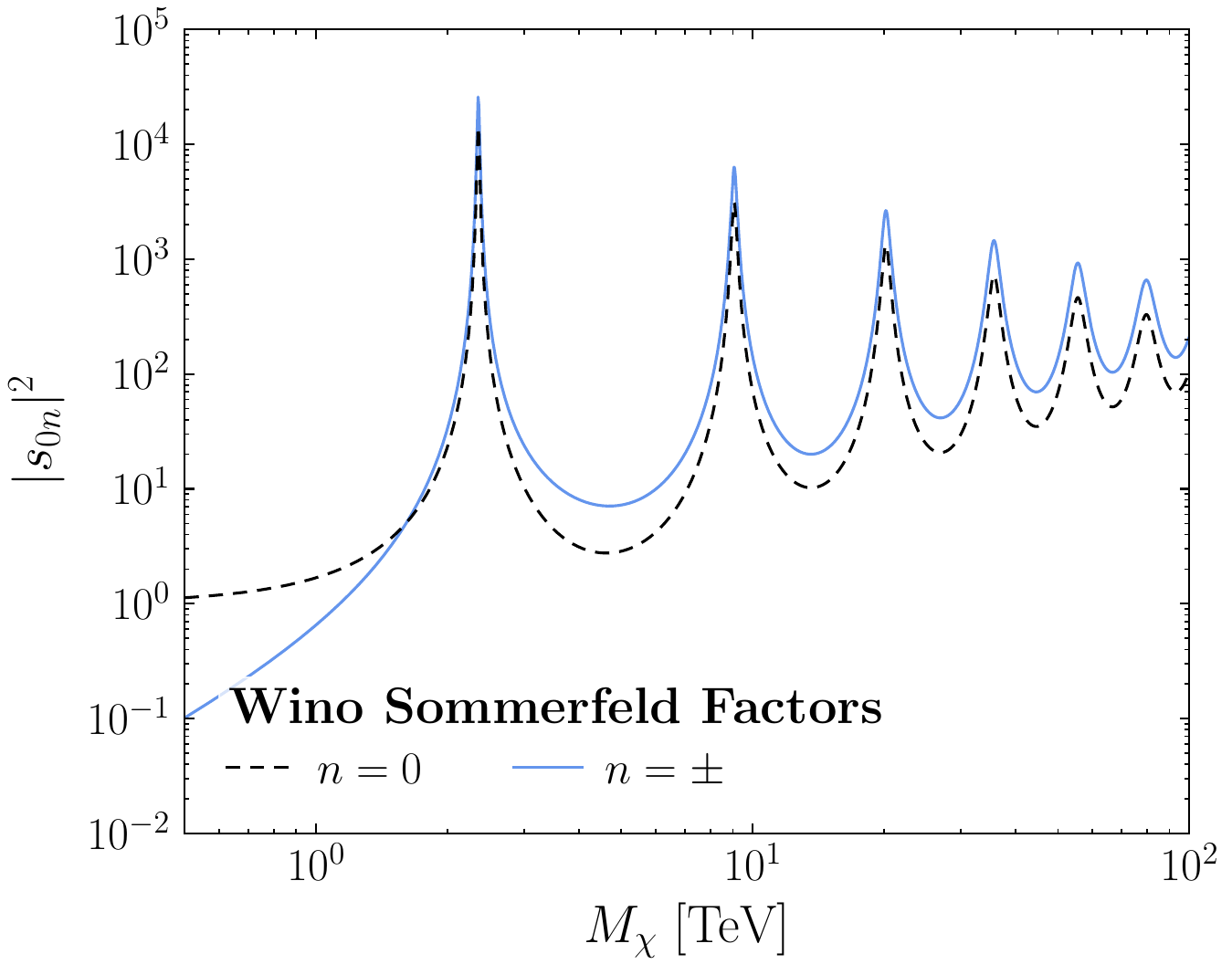} 
\hspace{0.5cm}
\includegraphics[width=0.47\textwidth]{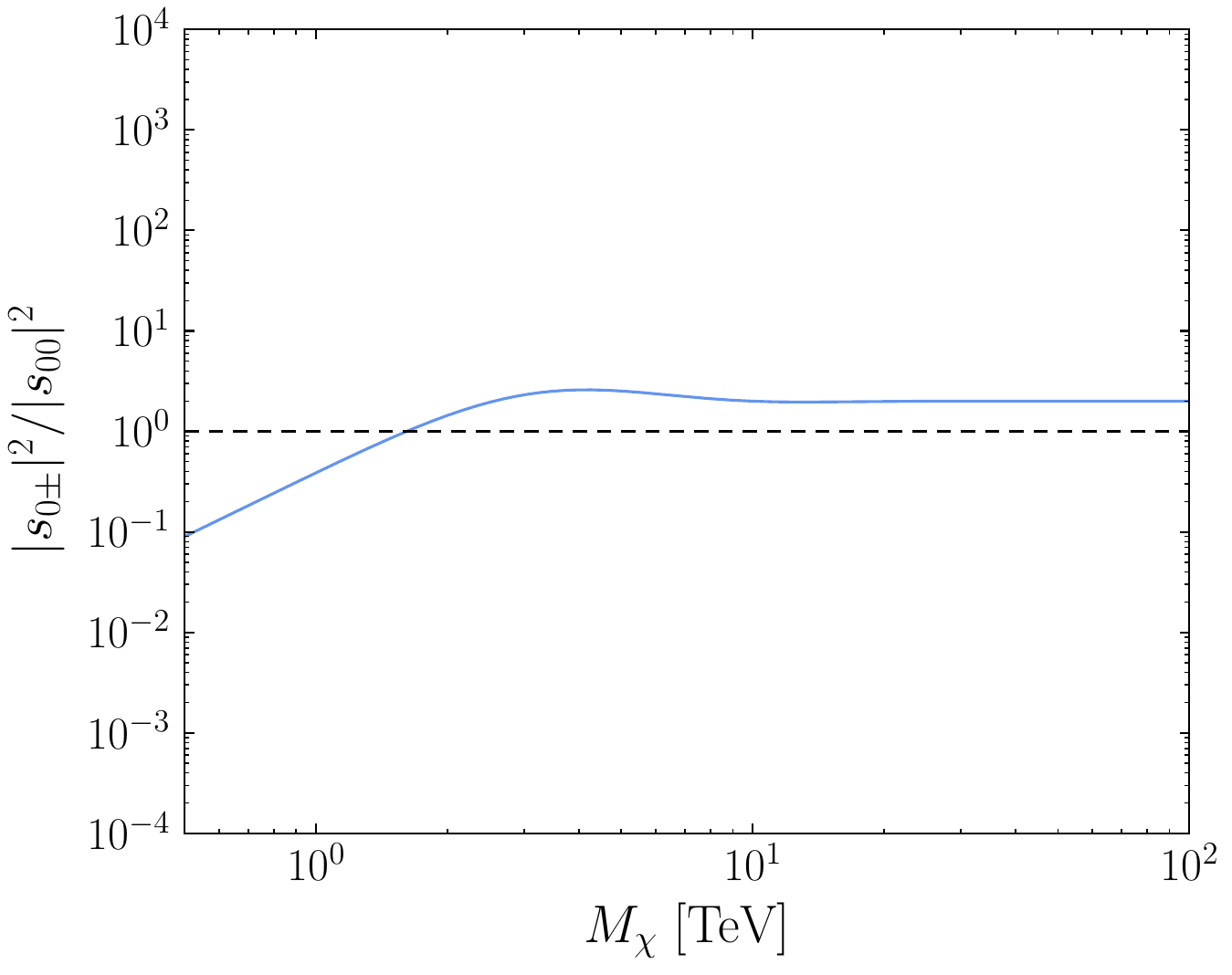}
\includegraphics[width=0.47\textwidth]{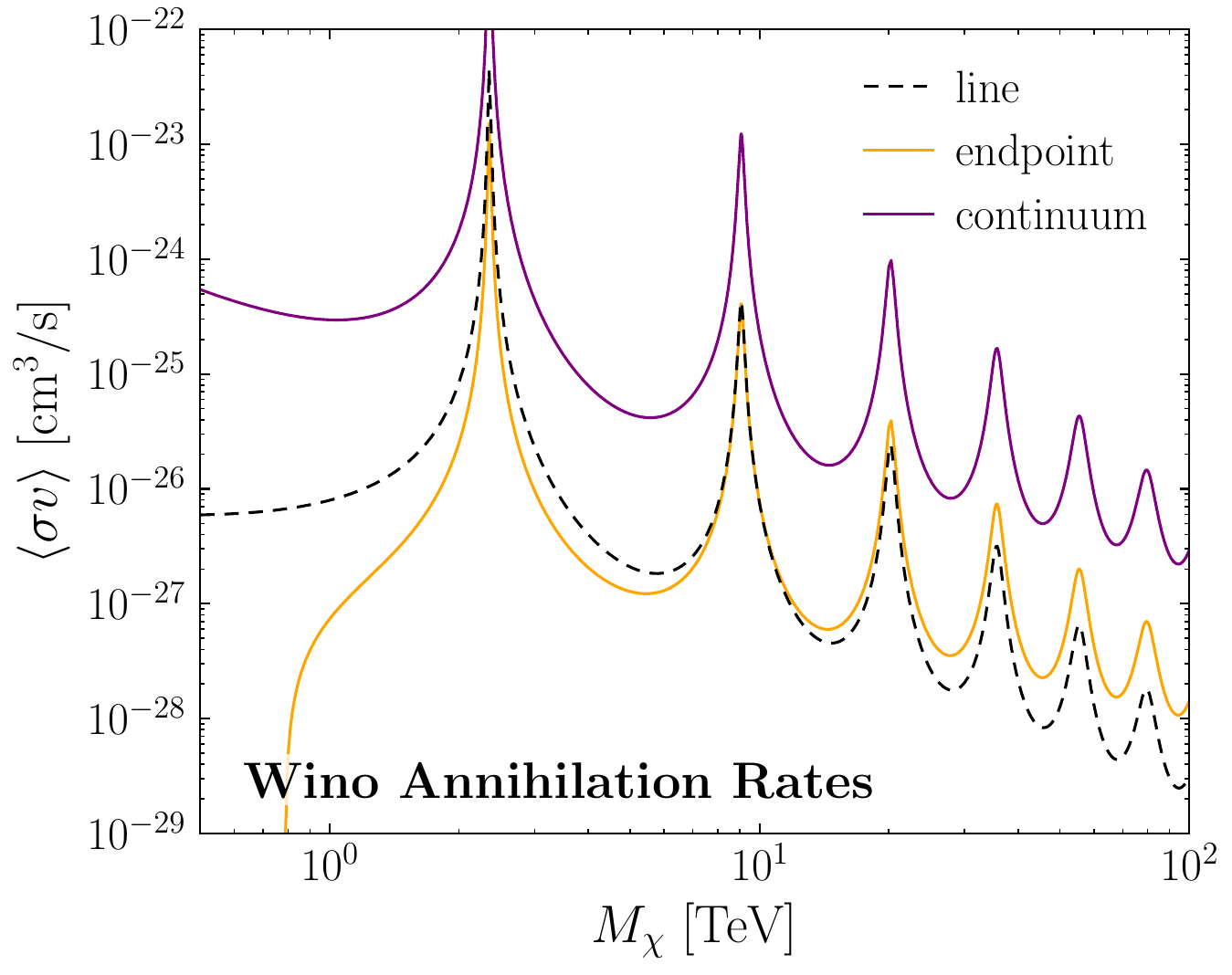} 
\hspace{0.5cm}
\includegraphics[width=0.47\textwidth]{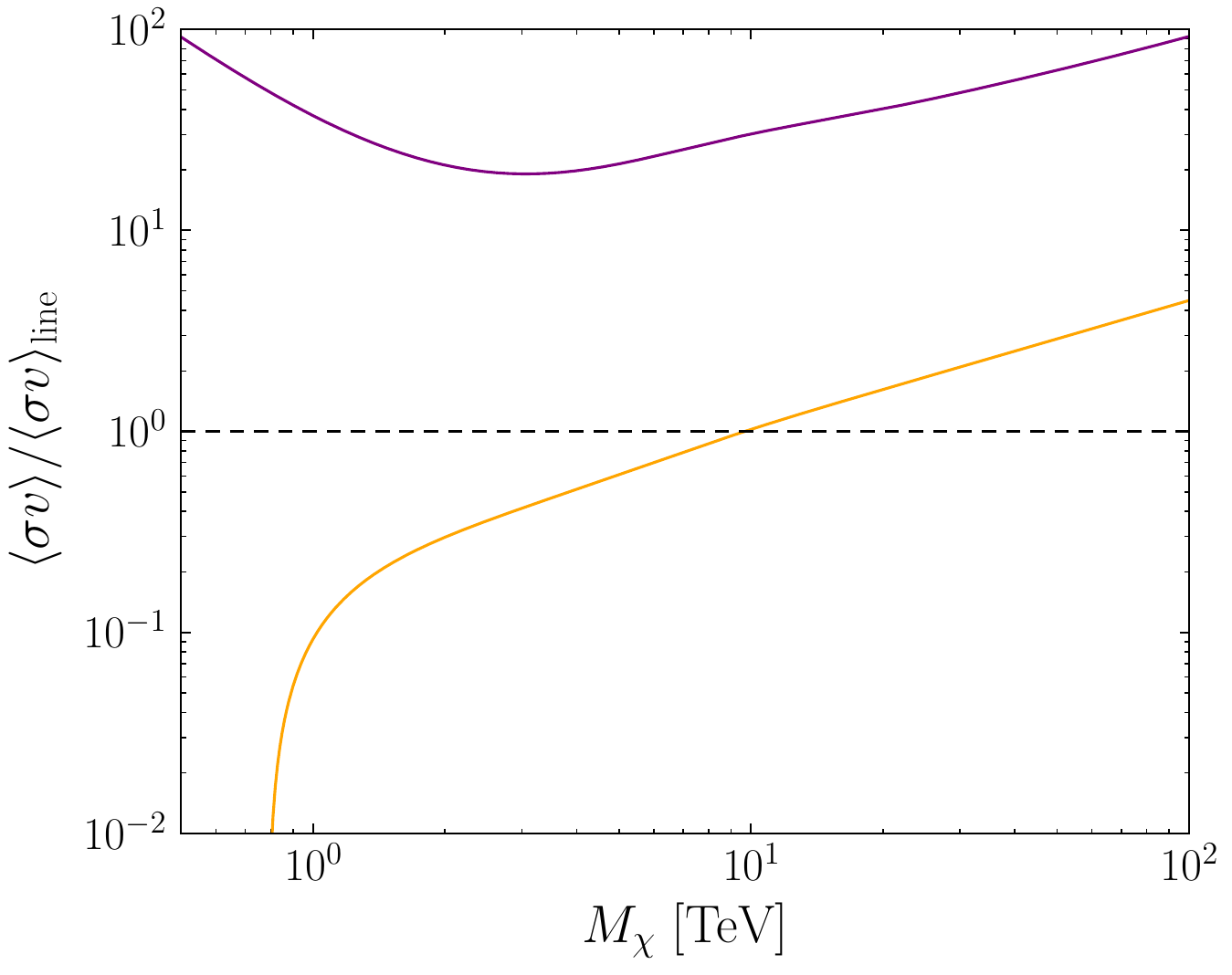}
\caption{A replica of Fig.~\ref{fig:rapidvariationQuint} for the case of the wino.
No sharp variation in the spectrum as a function of mass is observed for the wino, and indeed in the ratio plots we see that all variations are smooth functions of $M_{\chi}$.
}
\label{fig:rapidvariationWino}
\end{figure*}

As discussed in the main text, the ratio of the continuum and endpoint spectra to the line signal can vary sharply as a function of mass in the quintuplet case.
We attribute this behavior to the fact that each of these contributions to the spectrum depends on different linear combinations of the Sommerfeld factors $s_{00}$, $s_{0\pm}$, and $s_{0\pm\pm}$.
In this appendix we briefly expand on this point, and contrast the quintuplet to the wino where a similar rapid variation in the spectrum does not occur.

In Fig.~\ref{fig:rapidvariationQuint} we plot the absolute values of the three quintuplet Sommerfeld coefficients as a function of mass, and the annihilation rates to produce a line of photons, endpoint photons, and continuum photons.
To be explicit, the three cross sections are determined as follows: the line is specified by $\langle \sigma v \rangle_{\gamma\gamma} + \tfrac{1}{2} \langle \sigma v \rangle_{\gamma Z}$, the endpoint is the cumulative cross section evaluated at $z_{\rm cut}=0.9$ (see Fig.~\ref{fig:Thermal-Spectrum}) minus the line, and the continuum is $\langle \sigma v \rangle_{WW} + \langle \sigma v \rangle_{ZZ} + \tfrac{1}{2} \langle \sigma v \rangle_{\gamma Z}$.
We observe that the mass dependence of the Sommerfeld enhancement differs for the three Sommerfeld factors, and this in turn leads to differing mass dependencies for the three relevant annihilation rates (note that the phases of the Sommerfeld factors may differ with mass as well as the amplitudes, contributing to interference; the upper panels of Fig.~\ref{fig:rapidvariationQuint} only show the absolute values).
To make this point clearer, the right-hand panels of Fig.~\ref{fig:rapidvariationQuint} show ratios of Sommerfeld enhancement factors and annihilation rates as a function of mass.
We observe sharp features in the annihilation rate ratios, which will give rise to rapid mass-dependent variation in the continuum and endpoint spectra when they are normalized to the line cross section.

One might ask why this behavior was not seen for the wino, which also has multiple Sommerfeld factors that can interfere with each other.
For the line cross section, one might suspect that the issue is that only the $\chi^+\chi^-$ state can annihilate to photons at tree-level, so only the $s_{0\pm}$ factor is relevant (not $s_{00}$); however, we also do not see sharp quintuplet-like features in annihilation of wino DM to $W$ bosons, which is allowed at tree-level from both the $\chi^0\chi^0$ and $\chi^+\chi^-$ states.
The explanation is instead that the mass dependence of the $s_{00}$ and $s_{0\pm}$ factors is closely aligned for the wino, such that their ratio is a smooth function with no resonant features, as shown in Fig.~\ref{fig:rapidvariationWino}.
Thus in the wino case, the strong mass dependence in the Sommerfeld enhancement is fully described by measuring one of the two Sommerfeld factors, and we expect all relevant annihilation cross sections to scale similarly with mass.
This expectation is borne out in the right-hand panels of Fig.~\ref{fig:rapidvariationWino}.

More insight can be gained by working in the basis of eigenstates of the potential at small $r$, rather than mass eigenstates; this corresponds to the basis of potential eigenstates in the limit of unbroken SU(2), as discussed in App.~\ref{app:analytic}.
For the quintuplet case, there are two attractive eigenvalues and one repulsive eigenvalue; for the wino, there is only one attractive and one repulsive eigenvalue. 
Our expectation is that any linear combination of states that evolves under a repulsive potential at small $r$ will be highly suppressed as $r\rightarrow 0$, and will effectively not contribute to the Sommerfeld enhancement.
In the quintuplet case, we have confirmed numerically that this suppression is quite pronounced, typically several orders of magnitude at the masses and velocities considered in this study.

For this reason, we expect that the number of significant and physically distinct contributions to the Sommerfeld factors, corresponding to different short-distance potentials which can source different mass dependences (in both phase and amplitude), will correspond to the number of attractive eigenvalues.
This is consistent with the fact that we observe such interference in the case of the quintuplet, with two attractive eigenvalues, and not in the case of the wino, with a single attractive eigenvalue.
Based on this argument, we expect the presence of interfering Sommerfeld factors with differing resonance structures, leading to sharp variations in the spectrum with mass, to be generic for higher SU(2) representations.

\bibliography{HDMA}
\bibliographystyle{JHEP}

\end{document}